\documentclass[a4paper,11pt]{article}

\usepackage{caption,booktabs,array}
\usepackage{adjustbox}
\usepackage{pdflscape}

\usepackage{gensymb}
\pdfoutput=1 

\usepackage{jcappub} 

\usepackage[T1]{fontenc} 

\usepackage{graphicx}
\usepackage{epsfig}
\usepackage{amsmath,amssymb}
\usepackage{natbib}

\usepackage{subfigure}
\usepackage{longtable}
\usepackage{textcomp}
\usepackage{color}
\usepackage{url}
\usepackage{relsize}

\usepackage{epstopdf}

\usepackage{multirow}
\usepackage[table,usenames,dvipsnames]{xcolor}

\usepackage[greek,english]{babel}
\languageattribute{greek}{polutoniko}
\usepackage{enumitem}

\def\mission{CORE\ }

\newcommand{\alm}{$a_{\ell m}$}
\newcommand{\dd}{\textrm{d}}

\def\simlt{\mathrel{\rlap{\lower 3pt\hbox{$\sim$}}\raise 2.0pt\hbox{$<$}}}
\def\simgt{\mathrel{\rlap{\lower 3pt\hbox{$\sim$}} \raise 2.0pt\hbox{$>$}}}
\def\lsim{\,\lower2truept\hbox{${<\atop\hbox{\raise4truept\hbox{$\sim$}}}$}\,}
\def\gsim{\,\lower2truept\hbox{${> \atop\hbox{\raise4truept\hbox{$\sim$}}}$}\,}
\def\dalemb#1#2{{\vbox{\hrule height.#2pt
  \hbox{\vrule width.#2pt height#1pt \kern#1pt \vrule width.#2pt}
    \hrule height.#2pt}}}

\def\ba{\begin{eqnarray}}
\def\ea{\end{eqnarray}}
\def\be{\begin{equation}}
\def\ee{\end{equation}}

\def\gtorder{\mathrel{\raise.3ex\hbox{$>$}\mkern-14mu
             \lower0.6ex\hbox{$\sim$}}}
\def\ltorder{\mathrel{\raise.3ex\hbox{$<$}\mkern-14mu
             \lower0.6ex\hbox{$\sim$}}}

\title{\boldmath Exploring cosmic origins with CORE: effects of observer peculiar motion}

\author[*,1,2,3]{C. Burigana,\footnote{$^*$ Corresponding author. E-mail: burigana@iasfbo.inaf.it}}
\author[4]{C.S. Carvalho,}
\author[1,2,3]{T. Trombetti,}
\author[5]{A. Notari,}
\author[6,7]{\\M. Quartin,}
\author[8,9]{G. De Gasperis,}
\author[10,9]{A. Buzzelli,}
\author[8,9]{N. Vittorio,}
\author[12]{\\G. De Zotti,}
\author[10,11]{P. de Bernardis,}
% ...

\affiliation[1]{\scriptsize INAF--Istituto di Astrofisica Spaziale e Fisica Cosmica di Bologna, Via Piero Gobetti 101, I-40129 Bologna, Italy}
\affiliation[2]{\scriptsize Dipartimento di Fisica e Scienze della Terra, Universit\`a degli Studi di Ferrara, Via Giuseppe Saragat 1, I-44122 Ferrara, Italy}
\affiliation[3]{\scriptsize INFN, Sezione di Bologna, Via Irnerio 46, I-40126, Bologna, Italy}
\affiliation[4]{\scriptsize Institute of Astrophysics and Space Sciences, University of Lisbon, Tapada da Ajuda, 1349-018 Lisboa, Portugal}
\affiliation[5]{\scriptsize Departament de F\'isica Qu\`antica i Astrof\'isica i Institut de Ci\`encies del Cosmos, Universitat de Barcelona, Mart\'i i Franqu\`es 1, 08028 Barcelona, Spain.}
\affiliation[6]{\scriptsize Instituto de F\'\i sica, Universidade Federal do Rio de Janeiro, 21941-972, Rio de Janeiro, Brazil}
\affiliation[7]{\scriptsize Observat\'orio do Valongo, Universidade Federal do Rio de Janeiro, Ladeira Pedro Ant\^onio 43, 20080-090, Rio de Janeiro, Brazil}
\affiliation[8]{\scriptsize Dipartimento di Fisica, Universit\`a di Roma ``Tor~Vergata'', Via della Ricerca Scientifica 1, I-00133, Roma, Italy}
\affiliation[9]{\scriptsize INFN, Sezione Roma~2, Via della Ricerca Scientifica 1, I-00133, Roma, Italy}
\affiliation[10]{\scriptsize Dipartimento di Fisica, Universit\`a di Roma ``La Sapienza'', P.le Aldo Moro 2, 00185, Rome, Italy}
\affiliation[11]{\scriptsize INFN, Sezione di Roma, P.le A. Moro 2, 00185 Roma, Italy}
\affiliation[12]{\scriptsize INAF--Osservatorio Astronomico di Padova, Vicolo dell'Osservatorio 5, I-35122 Padova, Italy}
% ...

\abstract{
We discuss the effects on the cosmic microwave background (CMB), cosmic
infrared background (CIB), and thermal Sunyaev-Zeldovich effect due to the
peculiar motion of an observer with respect to the CMB rest frame, which
induces boosting effects.  After a brief review of the current observational and
theoretical status, we investigate the scientific perspectives opened by future
CMB space missions, focussing on the Cosmic Origins Explorer (CORE) proposal.
The improvements in sensitivity offered by a mission like CORE, together with
its high resolution over a wide frequency range, will provide a more accurate
estimate of the CMB dipole.  The extension of boosting effects to polarization
and cross-correlations will enable a more robust determination of purely
velocity-driven effects that are not degenerate with the intrinsic CMB dipole,
allowing us to achieve an overall signal-to-noise ratio of 13;
this improves on the {\it Planck\/} detection and essentially equals that of
an ideal cosmic-variance-limited experiment up to a multipole $\ell\simeq2000$.
Precise inter-frequency calibration will offer the opportunity to constrain or
even detect CMB spectral distortions, particularly from the cosmological
reionization epoch, because of the frequency dependence of the dipole spectrum,
without resorting to precise absolute calibration.  The expected improvement
with respect to COBE-FIRAS in the recovery of distortion parameters (which
could in principle be a factor of several hundred for an ideal experiment with
the CORE configuration) ranges from a factor of several up to about 50,
depending on the quality of foreground removal and relative calibration.
Even in the case of $\simeq1$\,\% accuracy in both foreground removal and
relative calibration at an angular scale of $1^\circ$, we find that dipole
analyses for a mission like CORE will be able to improve the recovery of the
CIB spectrum amplitude by a factor $\simeq 17$ in comparison with current
results based on COBE-FIRAS.  In addition to the scientific potential of a
mission like CORE for these analyses, synergies with other planned and ongoing
projects are also discussed.
}

\graphicspath{{figures_newrev/}}

\begin{document}
\maketitle
\flushbottom

\section{Introduction}\label{sect:Intro}

The peculiar motion of an
observer with respect to the cosmic microwave background (CMB)
rest frame gives rise to boosting effects (the largest of which is the
CMB dipole, i.e., the multipole $\ell=1$ anisotropy in the Solar System
barycentre frame), which can be explored by future CMB missions.
In this paper, we focus on peculiar velocity effects and their relevance to the Cosmic Origins Explorer (CORE) experiment. CORE is a satellite proposal dedicated to microwave polarization and submitted to the European Space Agency (ESA) in October 2016 in response to a call for future medium-sized space mission proposals for the M5 launch opportunity of ESA's Cosmic Vision programme.

This work is part of the {\it Exploring Cosmic Origins\/} (ECO) collection of articles, aimed at describing different scientific objectives achievable with the data expected from a mission like CORE.
We refer the reader to the CORE proposal \cite{CORE2016} and to other dedicated ECO papers for more details, in particular the mission requirements and design paper \cite{delabrouille_etal_ECO}
and the instrument paper \cite{debernardis_etal_ECO}, which provide a comprehensive discussion of the key parameters of CORE adopted in this work. We also refer the reader to the paper on extragalactic sources \cite{2016arXiv160907263D}
for an investigation of their contribution to the cosmic infrared background (CIB), which is one of the key topics addressed in the present paper, as well as
the papers on $B$-mode component separation \cite{baccigalupi_etal_ECO} for a stronger focus on polarization, and mitigation of systematic effects \cite{ashdown_etal_ECO} for further discussion of potential residuals included in some analyses presented in this work.
Throughout this paper we use the CORE specifications summarised in Table~\ref{tab:CORE-bands}.

\begin{table}[ht!]
\begin{center}
\scalebox{0.93}{\begin{tabular}{|c|c|c|c|c|c|c|c|}
\hline
Channel & Beam  &  $N_{\rm det}$  &  $\Delta T$  &  $\Delta P$  & $\Delta I$& $\Delta I$  & $\Delta y\times 10^6$  \\ \relax
[GHz]  & [arcmin] &  & [$\mu$K.arcmin] & [$\mu$K.arcmin] & [$\mu$K$_{\rm RJ}$.arcmin]  & [kJy$\,$sr$^{-1}$.arcmin] & [$y_{\rm SZ}$.arcmin] \\
\hline
\hline
60  &  17.87  &  48  &  7.5  &  10.6  &  6.81  &  0.75  &  $-$1.5  \\
70  &  15.39  &  48  &  7.1  &  10  &  6.23  &  0.94  &  $-$1.5    \\
80  &  13.52  &  48  &  6.8  &  9.6  &  5.76  &  1.13  &  $-$1.5   \\
90  &  12.08  &  78  &  5.1  &  7.3  &  4.19  &  1.04  &  $-$1.2  \\
100  &  10.92  &  78  &  5.0  &  7.1  &  3.90  &  1.2  &  $-$1.2   \\
115  &  9.56  &  76  &  5.0  &  7.0  &  3.58  &  1.45  &  $-$1.3  \\
130  &  8.51  &  124  &  3.9  &  5.5  &  2.55  &  1.32  &  $-$1.2  \\
145  &  7.68  &  144  &  3.6  &  5.1  &  2.16  &  1.39  &  $-$1.3  \\
160  &  7.01  &  144  &  3.7  &  5.2  &  1.98  &  1.55  &  $-$1.6  \\
175  &  6.45  &  160  &  3.6  &  5.1  &  1.72  &  1.62  &  $-$2.1  \\
195  &  5.84  &  192  &  3.5  &  4.9  &  1.41  &  1.65  &  $-$3.8  \\
220  &  5.23  &  192  &  3.8  &  5.4  &  1.24  &  1.85  &  \dots   \\
255  &  4.57  &  128  &  5.6  &  7.9  &  1.30  &  2.59  &  3.5  \\
295  &  3.99  &  128  &  7.4  &  10.5  &  1.12  &  3.01  &  2.2  \\
340  &  3.49  &  128  &  11.1  &  15.7  &  1.01  &  3.57  &  2.0  \\
390  &  3.06  &  96  &  22.0  &  31.1  &  1.08  &  5.05  &  2.8  \\
450  &  2.65  &  96  &  45.9  &  64.9  &  1.04  &  6.48  &  4.3  \\
520  &  2.29  &  96  &  116.6  &  164.8  &  1.03  &  8.56  &  8.3  \\
600  &  1.98  &  96  &  358.3  &  506.7  &  1.03  &  11.4  &  20.0  \\
\hline
\hline
Array         &              & 2100 & 1.2  & 1.7 &  &  & 0.41   \\
\hline
\end{tabular}}
\end{center}
\caption{\small \textbf{Proposed CORE-M5 frequency channels.}
The sensitivity is estimated assuming $\Delta \nu/\nu=30\,\%$ bandwidth, 60\,\% optical efficiency, total noise of twice the expected photon noise from the sky and the optics of the instrument being at 40\,K. The second column gives the FWHM resolution of the beam.
This configuration has 2100 detectors, about 45\,\% of which are located in CMB channels between 130 and 220\,GHz. Those six CMB channels yield an aggregated CMB sensitivity of $2\,\mu$K.arcmin ($1.7\,\mu$K.arcmin for the full array).
}
\label{tab:CORE-bands}
\end{table}

The analysis of cosmic dipoles is of fundamental relevance in cosmology, being related
to the isotropy and homogeneity of the Universe at the largest scales.
In principle, the observed
dipole is a combination of various contributions, including observer
motion with respect to the CMB rest frame, the intrinsic primordial
(Sachs-Wolfe) dipole and the Integrated Sachs-Wolfe dipole
as well as dipoles from astrophysical (extragalactic
and Galactic) sources.  The interpretation that the CMB dipole is mostly
(if not fully) of kinematic origin
has strong support from independent studies of the galaxy and cluster distribution, in particular via the measurements of the so-called \textit{clustering dipole}.
According to the linear theory of cosmological perturbations, the peculiar velocity of an observer (as imprinted in the CMB dipole) should be related to the observer's peculiar gravitational acceleration via $\vec{v}_{\rm lin} = \beta_{\rm rd} \vec{g}_{\rm lin}$, where
$\beta_{\rm rd} \simeq \Omega_{\rm m}^{0.55} / b_{\rm g}$ is also know as the redshift-space distortion parameter
($b_{\rm g}$ and $\Omega_{\rm m}$ being, respectively, the bias of the particular galaxy sample and the matter density parameter at the present time).
The peculiar velocity and acceleration of, for instance, the Local Group treated as one system,
i.e., as measured from its barycentre, should thus be aligned and have a specific relation between amplitudes. The former fact has been confirmed from analyses of many surveys over the last three decades, such as IRAS \cite{1987MNRAS.228P...5H,1992ApJ...397..395S}, 2MASS \cite{2003ApJ...598L...1M,2006MNRAS.368.1515E}, or galaxy cluster samples \cite{1996ApJ...460..569B,1998ApJ...500....1P}. As far as the amplitudes are concerned, the comparison has been used to place constraints on the $\beta_{\rm rd}$ parameter
\cite{1996ApJ...460..569B,2000MNRAS.314..375R,2006MNRAS.368.1515E,2006MNRAS.373.1112B,2011ApJ...741...31B}, totally independent of those from redshift-space distortions observed in spectroscopic surveys.
In this context, confirming the kinematic origin of the CMB dipole, through a comparison accounting for
our Galaxy's motion in the Local Group and the Sun's motion in the Galaxy (see e.g., Refs.~\cite{1999AJ....118..337C,2012MNRAS.427..274S}), would provide support for the standard cosmological model,
while finding any significant deviations from this assumption could open up the possibility for other interpretations
(see e.g., Refs.~\cite{2013PhRvD..88h3529W,2016JCAP...06..035B,2016JCAP...12..022G,Roldan:2016ayx}).

Cosmic dipole investigations of more general type have been carried out in several frequency domains
\cite{2012MNRAS.427.1994G}, where the main signal comes from various types of astrophysical sources differently weighted in different shells in redshift.
An example are dipole studies in the radio domain, pioneered by Ref.~\cite{1984MNRAS.206..377E} and recently revisited by Ref.~\cite{2013A&A...555A.117R} performing a re-analysis of the NRAO VLA Sky Survey (NVSS) and the Westerbork Northern Sky Survey, as well as by Refs.~\cite{2015APh....61....1T, 2015MNRAS.447.2658T} using NVSS data alone.
Prospects to accurately measure the cosmic radio dipole with the Square Kilometre Array have been studied by Ref.~\cite{2015aska.confE..32S}.
Perspectives on future surveys jointly covering microwave/millimeter and far-infrared wavelengths aimed at comparing CMB and CIB dipoles have been presented in Ref.~\cite{2011ApJ...734...61F}. The next decades will see a continuous improvement of cosmological surveys in all bands. For the CMB, space observations represent the best, if not unique, way to precisely measure this large-scale signal. It is then important to consider the expectations from (and the potential issues for) future CMB surveys beyond the already impressive results produced by {\it Planck}.

In addition to the dipole due to the combination of observer velocity and Sachs-Wolfe and intrinsic (see ref.~\cite{2017arXiv170400718M} for a recent study) effects,
a moving observer will see velocity imprints on the CMB due to Doppler and aberration effects
\cite{Challinor:2002zh,Burles:2006xf}, which manifest themselves in correlations between the power at subsequent multipoles of both temperature and polarization anisotropies.
Precise measurements of such correlations \cite{Kosowsky:2010jm,Amendola:2010ty} provide important consistency checks of fundamental principles in cosmology, as well as an alternative and general way to probe observer peculiar velocities \cite{2014A&A...571A..27P,Roldan:2016ayx}. This type of analysis can in principle be extended to thermal Sunyaev-Zeldovich (tSZ) \cite{2005A&A...434..811C} and CIB signals.
We will discuss how these investigations could be improved when applied to data expected from a next generation of CMB missions, exploiting experimental specifications in the range of those foreseen for LiteBIRD \cite{2016SPIE.9904E..0XI} and CORE.

Since the results from COBE~\cite{1992ApJ...397..420B}, no substantial improvements have been achieved in the observations of the CMB spectrum at $\nu \gsim 30$\,GHz.\footnote{For recent observations at long wavelengths,
see the results from the ARCADE-2 balloon \cite{2011ApJ...730..138S,2011ApJ...734....6S} and from the TRIS experiment \cite{2008ApJ...688...24G}.}
Absolute spectral measurements rely on ultra-precise absolute calibration. FIRAS \cite{1996ApJ...473..576F} achieved an absolute calibration precision of 0.57\,mK, with a typical inter-frequency calibration
accuracy of $0.1$\,mK in one decade of frequencies around 300\,GHz. The amplitude and shape of the CIB spectrum, measured by FIRAS \cite{Fixsen:1998kq}, is still not well known. Anisotropy missions, like CORE, are not designed to have an independent absolute calibration, but nevertheless can investigate the CMB and CIB spectra by looking at the frequency spectral behaviour of the dipole amplitude \cite{DaneseDeZotti1981, Balashev2015, 2016JCAP...03..047D, Planck_inprep}. Unavoidable spectral distortions are predicted as the result of energy injections in the radiation field occurring at different cosmic times, related to the origin of cosmic structures and to their evolution, or to the different evolution of the temperatures of matter and radiation (for a recent overview of spectral distortions within standard $\Lambda$CDM, see Ref.~\cite{2016MNRAS.460..227C}).  For quantitative forecasts we will focus on well-defined types of signal, namely
Bose-Einstein (BE) and Comptonization distortions \cite{1969Ap&SS...4..301Z, 1970Ap&SS...7...20S}; however, one should also be open to the possible presence of
unconventional heating sources, responsible in principle for imprints larger than (and spectral shapes different from) those mentioned above, and having parameters that could be constrained through analysis of the CMB spectrum. Deciphering such signals will be a challenge, but holds the potential for important new discoveries and for constraining unexplored processes that cannot be probed by other means.  At the same time, a better determination of the CIB intensity greatly contributes to our understanding of the dust-obscured star-formation phase of galaxy evolution.

The rest of this paper is organised as follows.
In Sect.~\ref{sec:dipideal} we quantify the accuracy of a mission like CORE for recovering the dipole direction and amplitude separately at a given frequency, focussing on a representative set of CORE channels. Accurate relative calibration and foreground mitigation are crucial for analysing CMB anisotropy maps. In Sect.~\ref{sec:res_mod} we describe a parametric approach to modelling the pollution of theoretical maps with potential residuals. The analysis in Sect.~\ref{sec:dipideal} is then extended in Sect.~\ref{sec:dipsyst} to include a certain level of residuals. The study throughout these sections is carried out in pixel domain.

In Sect.~\ref{sect:aberration} we describe the imprints at $\ell>1$ due to Doppler and aberration effects, which can be measured in harmonic space. Precise forecasts based on CORE specifications are presented and compared with those expected from LiteBIRD. The intrinsic signature of a boost in Sunyaev-Zeldovich and CIB maps from CORE is also discussed in this section.

In Sect.~\ref{sect:DiffCMB} we study CMB spectral distortions and the CIB spectrum through the analysis of the frequency dependence of the dipole distortion; we introduce a method to extend predictions to higher multipoles, coupling higher-order effects and geometrical aspects. The theoretical signals are compared with sensitivity at different frequencies, in terms of angular power spectrum, for a mission like CORE.
In Sect.~\ref{sec:dip_spec_sim} we exploit the available frequency coverage through
simulations to forecast CORE's sensitivity to the spectral distortion parameters and the CIB spectrum amplitude, considering the ideal case of perfect relative calibration and foreground subtraction; however, we also parametrically quantify the impact of potential residuals, in order to define the requirements for substantially improve the results beyond those from FIRAS.

In Sect.~\ref{sec:end} we summarise and discuss the main results. The basic concepts and formalisms are introduced in the corresponding sections, while additional information and technical details are provided in several dedicated appendices for sake of completeness. 

\section{The CMB dipole: forecasts for CORE in the ideal case}
\label{sec:dipideal}

A relative velocity, $\beta\equiv v/c$, between an observer and the CMB rest frame induces a dipole (i.e., $\ell=1$ anisotropy) in the temperature of the
CMB sky through the Doppler effect. Such a dipole is likely dominated by the velocity of the Solar System, $\vec{\beta_{\rm S}}$, with respect to the CMB (Solar dipole), with a seasonal modulation
due to the velocity of the Earth/satellite, $\vec{\beta_{\rm o}}$, with respect to the Sun (orbital dipole). In this work we neglect the orbital dipole (which may
indeed be used for calibration), thus hereafter
we will denote with $\vec{\beta}$ the relative velocity of the Solar dipole.

In this section we forecast the ability to recover the dipole parameters (amplitude 
and direction)
by performing a Markov chain Monte Carlo (MCMC) analysis in the ideal case (i.e., without calibration errors or sky residuals). Results including
systematics are given in Sect.~\ref{sec:dipsyst}.
We test the amplitude of the parameter errors against the chosen sampling resolution and we probe the impact of both instrumental noise and
masking of the sky.
We consider the ``{\it Planck\/} common mask 76'' (in temperature), which is
publicly available from the Planck Legacy Archive (PLA)\footnote{http://pla.esac.esa.int/pla/} \cite{PLArefESA},
and keeps 76\,\% of the sky, avoiding the Galactic plane and regions at higher Galactic latitudes contaminated by
Galactic or extragalactic sources.  We exploit here an extension of this mask that excludes all the pixels at
$|b| \le 30^\circ$.\footnote{When we degrade the {\it Planck\/} common mask 76
to lower resolutions we apply a threshold of 0.5 for accepting or excluding pixels, so that the exact sky coverage not excluded by each mask
(76--78\,\%) slightly increases at decreasing $N_{\rm side}$.
In the case of the extended masks, typical sky coverage values are 47--48\,\%.}

Additionally, we explore
the dipole reconstruction ability for different frequency channels, specifically 60, 100, 145, and 220\,GHz. We finally investigate the impact of
spectral distortions (see Sects.~\ref{sect:DiffCMB} and \ref{sec:dip_spec_sim}), treating the specific case of a BE spectrum (with chemical potential $\mu_{0}=1.4\times10^{-5}$, which is several times smaller than FIRAS upper limits).

\begin{figure}[ht!]
\centering
  \includegraphics[angle=90,width=0.45\textwidth]{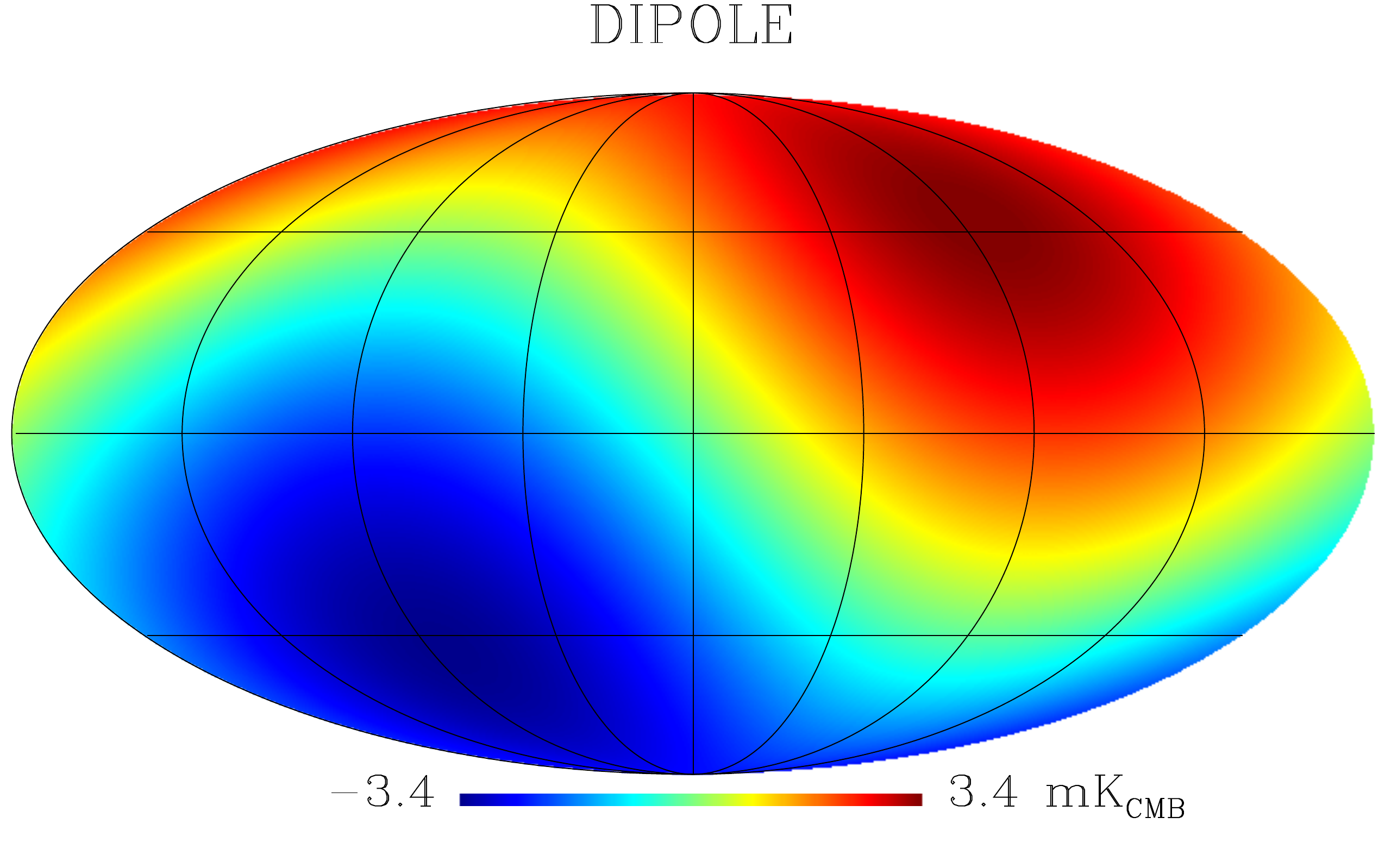}
  \caption{Map of the CMB dipole used in the simulations, corresponding to an
amplitude $A=3.3645$\,mK
  and a dipole direction defined by the Galactic coordinates $b_{0} =48.24^\circ$ and $l_{0}=$ $264.00^\circ$.
  The map is in Galactic coordinates
  and at a resolution of $\simeq 3.4$\,arcmin, corresponding to {\tt HEALPix} $N_{\rm side}=1024$.}%
\label{fig:dipolemap}
  \end{figure}
\begin{figure} 
\centering
  \includegraphics[angle=90,width=0.45\textwidth]{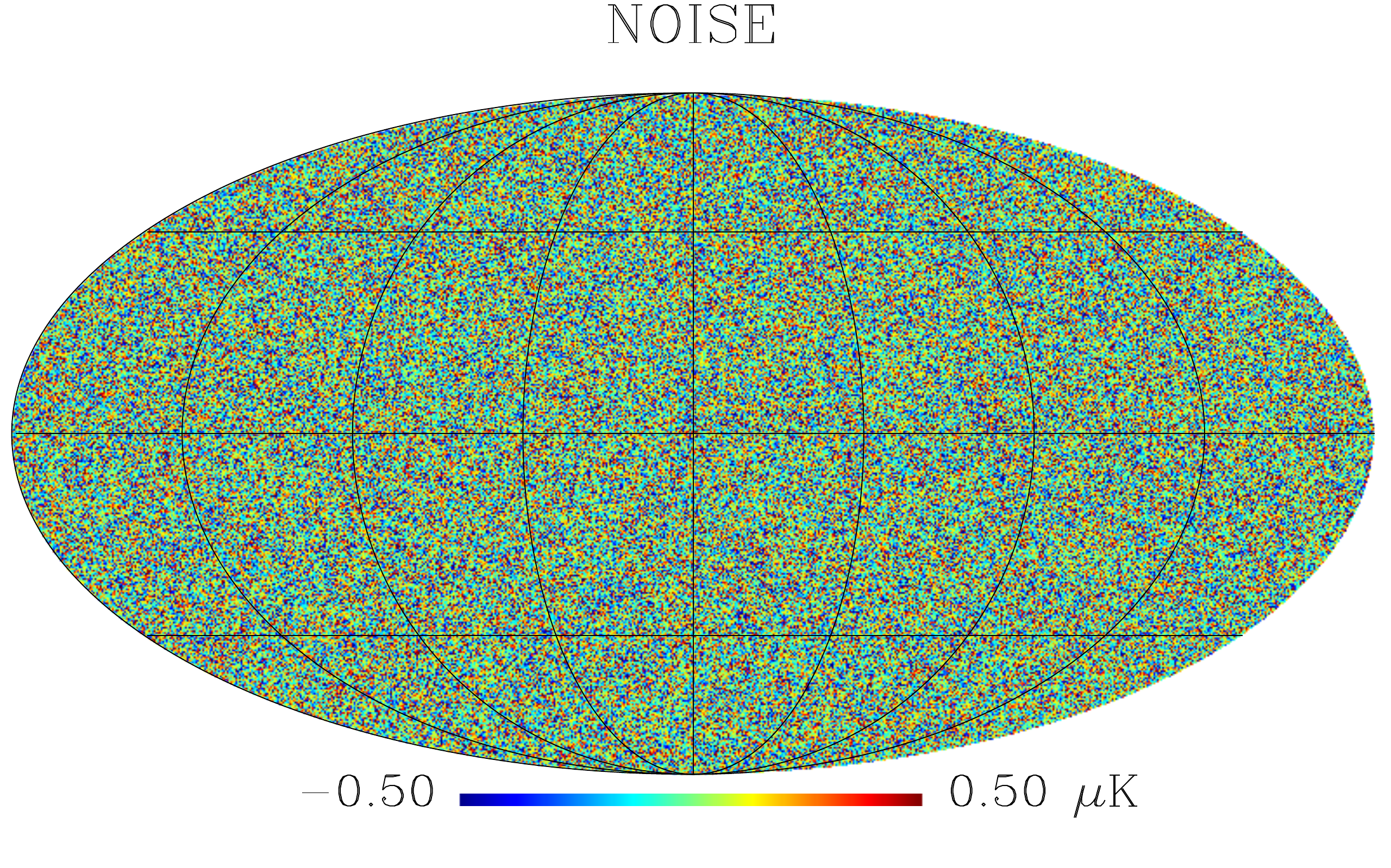}
  \includegraphics[angle=90,width=0.45\textwidth]{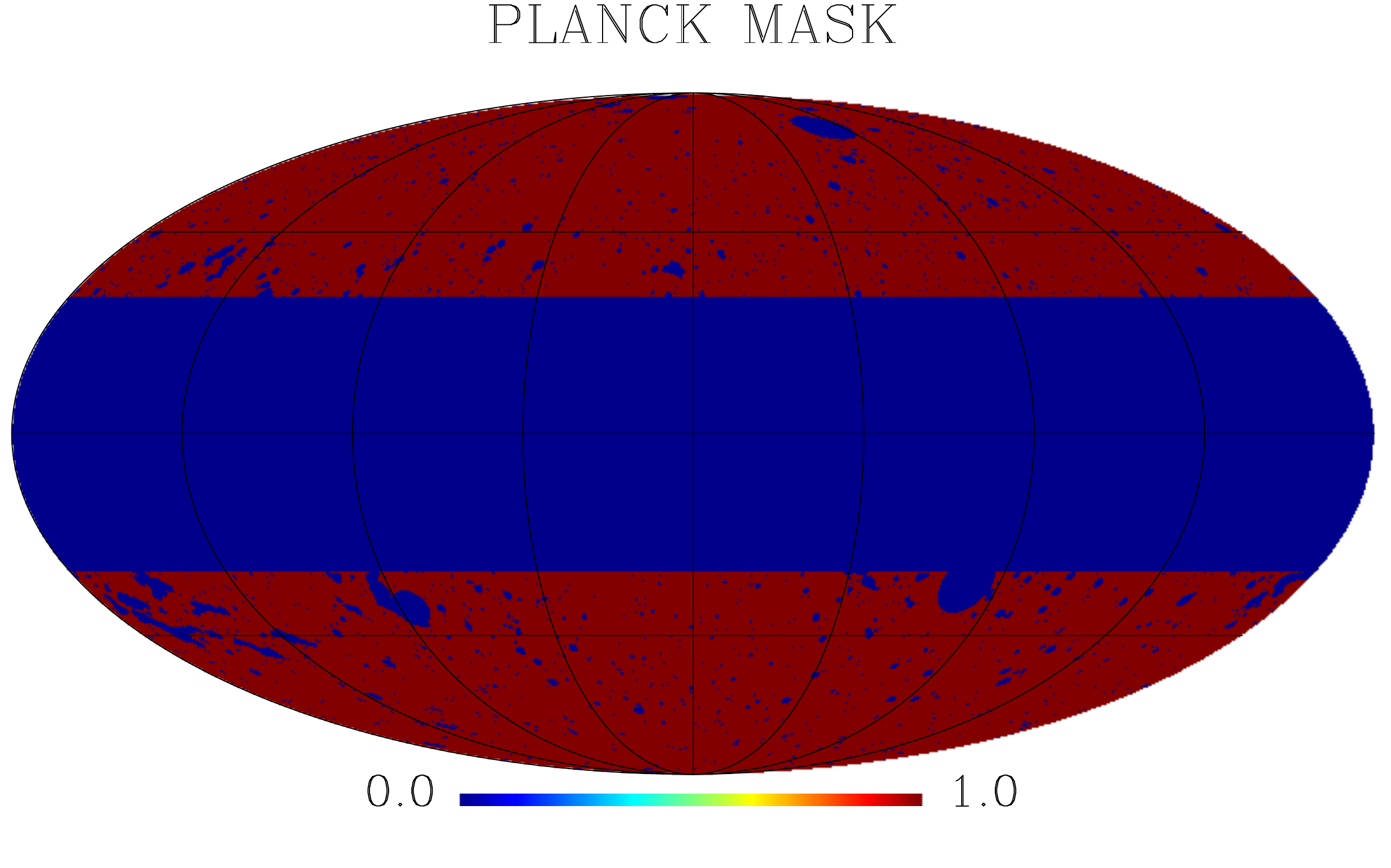}
  \caption{Instrumental noise map 
  and {\it Planck\/} Galactic
mask (extended to cut out $\pm30^\circ$ of the Galactic plane) employed in the simulations.
  The noise map corresponds to 7.5\,$\mu$K.arcmin, as expected for the 60-GHz band.
  The Map is in Galactic coordinates
  and at resolution of $\simeq 3.4$\,arcmin, corresponding to {\tt HEALPix} $N_{\rm side}=1024$.}%
\label{fig:instnoisemap}
\end{figure}

We write the dipole in the form:
\begin{equation}
    d(\hat{n}) = A\, \hat{n} \cdot \hat{n}_{0} + T_{0},
\end{equation}
\noindent
where $\hat{n}$ and $\hat{n}_{0}$ are the unit vectors defined respectively by the Galactic longitudes and latitudes $(l,b)$ and $(l_{0},b_{0})$.
In Fig.~\ref{fig:dipolemap} we show the dipole map we have used in our simulations, generated assuming the best-fit values
of the measurements of the dipole amplitude, $A=(3.3645\pm0.002)$\,mK, and direction, $l_{0}=264.00^\circ \pm 0.03^\circ$ and $b_{0}=48.24^\circ \pm 0.02^\circ$, found in the {\it Planck\/} (combined result from the High Frequency Instrument, HFI, and Low Frequency Instrument, LFI) 2015 release \cite{2016A&A...594A...1P,2016A&A...594A...5P,2016A&A...594A...8P}. Assuming the dipole to be due to velocity effects only, its amplitude corresponds to $\beta \equiv |\vec{\beta}| \equiv v/c = A / T_{0} = 1.2345 \times 10^{-3}$, with $T_{0}=2.72548 \pm 0.00057\,$K being the present-day temperature of the CMB \cite{2009ApJ...707..916F}. In Fig.~\ref{fig:instnoisemap} we show the instrumental noise map and the {\it Planck\/} Galactic mask employed in the simulations. The noise map corresponds to 7.5 $\mu$K.arcmin, as expected for the 60-GHz band.

We calculate the likelihoods for the parameters $A$, $l_{0}$, $b_{0}$
and $T_{0}$ using the publicly available {\tt COSMOMC} generic sampler package \citep{2002PhRvD..66j3511L,Metropolis53,Hastings70}.
While the monopole $T_{0}$ is not an observable of interest in this context, we include it as a free parameter, to verify any degeneracy with the other parameters and for internal consistency checks.

To probe the dependence of the parameter error estimates on the sampling resolution, we investigate the dipole reconstruction at {\tt HEALPix}
\cite{2005ApJ...622..759G} $N_{\rm side}=128$, 256, 512, and 1024,
eventually including the noise and the Galactic mask.
The reference frequency channel for this analysis is the 60-GHz band. The corresponding likelihoods are collected in Appendix~\ref{sec:Likelihoods} (see Fig.~\ref{fig:conflevel128+1024}) for the same representative values of $N_{\rm side}$ (see also Table \ref{table:conflevel128+1024} for the corresponding 68\% confidence levels).

In Fig.~\ref{fig:diperrors} we plot the 1\,$\sigma$ uncertainties on the parameter estimates as functions of the {\tt HEALPix} $N_{\rm side}$ value. We find that the pixelization error due to the finite resolution is dominant over the instrumental noise at any $N_{\rm side}$. This means that we are essentially limited
by the sampling resolution. As expected, the impact of noise is negligible, although the effect of reducing the effective sky fraction is relevant.
In fact, the presence of the Galactic mask results in larger errors (for all parameters) and introduces a small correlation between the
parameters $A$ and $b_{0}$, as clearly shown in these plots.

The likelihood results for
some of 
the different frequencies under analysis are collected in 
Figs.~\ref{fig:conflevel100220} of Appendix~\ref{sec:Likelihoods} (see also Table \ref{table:conflevel60220} for the 68\% confidence levels at the four considered frequencies).
Here we keep the resolution fixed at
{\tt HEALPix} $N_{\rm side}=1024$ and consider both noise level and choice of Galactic mask. We find that the dipole parameter estimates do not significantly change
among the frequency channels, which is clearly due to the sub-dominant effect of the noise.

As a last test of the ideal case, we compare the dipole parameter reconstruction between the cases of a pure blackbody (BB) spectrum and a BE-distorted
spectrum. The comparison of the likelihoods is presented in Fig.~\ref{fig:conflevelBE} of Appendix~\ref{sec:Likelihoods} (see also Table \ref{table:conflevelBE} for the corresponding 68\% confidence levels). 
This analysis shows that the parameter errors are not affected by the spectral distortion and that the direction of the dipole is successfully recovered. The difference found in the amplitude value is consistent
with the theoretical difference of about 76\,nK.


\section{Parametric model for potential foreground and calibration residuals in total intensity}
\label{sec:res_mod}

In the previous section we showed that in the ideal case of pure noise, i.e., assuming perfect foreground subtraction and calibration (and the absence of systematic effects) in the sky region being analysed, pixel-sampling
limitation dominates over noise limitation.

Clearly, specific component-separation and calibration methods (and implementations) introduce specific types of residuals. Rather than trying to accurately characterise them (particularly in the view of great efforts carried out in the last decade for specific experiments and the progress that is expected over the coming years), we implemented a simple toy model to parametrically estimate the potential impact of imperfect foreground
subtraction and calibration in total intensity (i.e., in temperature).  This includes using some of the {\it Planck\/} results and products made publicly available through the PLA.

The PLA provides maps in total intensity (or temperature) at high resolution ($N_{\rm side} = 2048$) of global foregrounds at each {\it Planck\/} frequency (here we use those maps based on the {\tt COMMANDER} method).\footnote{Adopting this choice or one of the other foreground-separation methods is not relevant for the present purpose.}
It provides also suitable estimates of the zodiacal light emission (ZLE)
maps (in temperature) from {\it Planck}-HFI. Our aim is to produce templates of potential foreground residuals that are simply scalable in amplitude according to a tunable parameter.
In order to estimate such emission at CORE frequencies, without relying on particular sky models,
we simply interpolate linearly
(in logarithmic scale, i.e., in ${\rm log} (\nu)$--${\rm log} (T))$
pixel by pixel the foreground maps and the ZLE maps, and linearly extrapolate the ZLE maps at $\nu < 100$\,GHz. We then create a template of signal sky amplitude at each CORE frequency,
adding the absolute values in each pixel of these foreground and ZLE maps\footnote{Since we are not interested here in the separation of the diffuse Galactic emission and ZLE, this assumption is
in principle slightly conservative. In practice, separation methods will at least distinguish between these diffuse components, which are typically treated
with different approaches, e.g., analysing multi-frequency maps in the case of Galactic emission, and different surveys (or more generally, data taken at different times) for the ZLE.}
and of the CMB anisotropy map available at the same resolution in the PLA (we specifically use that based on {\tt COMMANDER}).
Since for this analysis we are not interested in separating CMB and astrophysical emission
at $\ell \ge 3$, we then generate templates from these maps, extracting the $a_{lm}$ modes for $\ell \le 2$ only. These templates are then degraded to the desired resolution.
Finally, we generate maps of Gaussian random fields at each CORE frequency, with rms amplitude given by these templates, $T_{\rm amp,for}$, multiplied by a tunable parameter, $E_{\rm for}$, which globally characterizes
the potential amplitude of foreground residuals after component separation. Clearly, the choice of reasonable values of $E_{\rm for}$ depends on the resolution being considered (or on the adopted pixel size), with
the same value of $E_{\rm for}$ but at smaller pixel size implying less contamination at a given angular scale.

{\it Planck\/} maps reveal, at least in temperature, a greater complexity in the sky than obtained by previous experiments. The large number of frequencies of CORE is in fact designed to accurately model
foreground emission components with a precision much better than {\it Planck}'s. Also, at least in total intensity, ancillary information will come in the future from a number of other surveys, ranging from radio to infrared frequencies.

The target for CORE in the separation of diffuse polarised foreground emission corresponds to $E_{\rm for} \simeq 0.01$, i.e., to $\simeq 1$\,\% precision at the map level for angular scales larger than about $1^\circ$
(i.e., up to multipoles $\ell \lsim 200$), where the main information on primordial $B$-modes is contained, while at larger multipoles the main limitation comes from lensing subtraction and characterization
and secondarily through control of extragalactic source contributions. We note also that comparing CMB anisotropy maps available from the PLA at $N_{\rm side} = 2048$ derived with four different component-separation methods and degraded
to various resolutions, shows that the rms of the six difference maps does not scale strongly with the adopted pixel size, at least if we exclude regions close to the Galactic plane.
For example, outside the {\it Planck\/} common mask 76, if we pass from $N_{\rm side} = 2048$ to $N_{\rm side} = 256$ or $64$, i.e., increasing the pixel linear size by a factor of 8 or 32 (with the exception of the comparison of {\tt SEVEM} versus {\tt SMICA}),
the rms values of the cross-comparisons range from about 8--9\,$\mu$K to about 3--5\,$\mu$K, i.e., a decreases by a factor of only about 2.5. This suggests that, at least for temperature analyses,
the angular scale adopted to set $E_{\rm for}$ is not so critical.

Data calibration represents one of the most delicate aspects of CMB experiments. The quality of CMB anisotropy maps does not rely on absolute calibration of the signal (as it would, for example, in experiments
dedicated to absolute measurements of the CMB temperature, i.e., in the direct determination of the CMB spectrum).  However, the achievement of very high accuracy in the relative calibration
of the maps (sometimes referred to as absolute calibration of the anisotropy maps), as well as the inter-frequency calibration of the maps taken in different bands, is crucial for enabling the scientific goals of CMB projects.
Although this calibration step could in principle benefit from the availability of precise instrumental reference calibrators (implemented for example in FIRAS \cite{1999ApJ...512..511M} and foreseen in
PIXIE \cite{2011JCAP...07..025K},
or -- but with much less accurate requirements -- in {\it Planck}-LFI \cite{2009JInst...4T2006V}), this is not necessary for anisotropy experiments, as shown for example by WMAP and {\it Planck}-HFI.
This represents a huge simplification in the design of anisotropy experiments with respect to absolute temperature ones. {\it Planck\/} demonstrated the possibility to achieve relatively calibration of anisotropy data at a level of accuracy of about 0.1\,\% up to about 300\,GHz, while recent analyses of planet flux density measurements and modelling \cite{2016arXiv161207151P} indicate the possibility to achieve a calibration accuracy of $\simeq 1$\,\% even above 300\,GHz, with only moderate improvements over what is currently realised.

The goal of CORE is to achieve a calibration accuracy level around 0.01\,\%, while the requirement of $0.1$\,\% is clearly feasible on the basis of current experiments, with some possible relaxation at
high frequencies.
Methods for improving calibration are fundamental in astrophysical and cosmological surveys, and clearly critical in CMB experiments. In principle, improvements in various directions can be pursued:
from a better characterization of all instrument components to cross-correlation between different CMB surveys; from the implementation of external precise artificial calibration sources
to the search for a better characterization (and increasing number) of astronomical calibration sources; and, in general, with the improvement of data analysis methods.

To parametrically model potential residuals due to imperfect calibration we follow an approach similar to that described above for foreground contamination.
We note that calibration uncertainty implies an error proportional to the global effective (anisotropy in our case) signal. We therefore produce templates as described above, but do so by adding the foreground, ZLE,
and CMB anisotropy maps, keeping their signs and maintaining all the $a_{lm}$ modes contained in the maps. The absolute values of these templates are then
multiplied by a tunable parameter, $E_{\rm cal}$ (possibly dependent on frequency), which globally characterizes the amplitude of potential residuals arising from imperfect calibration.
These are then used to define the pixel-by-pixel rms amplitudes, which are adopted to construct maps, $T_{\rm res,cal}$, of Gaussian random fields at each CORE frequency.

In fact, we might also expect calibration errors to affect the level of foreground residuals.
Hence, as a final step, we include in the model a certain coupling between the
two types of residuals.
At each frequency,
we multiply the above simulated maps of foreground residuals by $(1+T_{\rm res,cal}/T_{\rm amp,for})$.

\section{The CMB dipole: forecasts for CORE including potential residuals}
\label{sec:dipsyst}

We now extend the analysis presented in Sect.~\ref{sec:dipideal} by including two sources of systematic effects, namely calibration errors and sky foreground residuals. We consider two pairs of calibration uncertainty and sky residuals (parameterised by $E_{\rm for} = 0.04$ and $E_{\rm cal} = 0.004$, and by $E_{\rm for} = 0.64$ and $E_{\rm cal} = 0.064$) at $N_{\rm side} = 1024$ in order to explore different resolutions through pixel degradation. Rescaled to $N_{\rm side} = 64$, the two cases correspond to a set-up respectively better and worse by a factor of 4 with respect to the case $E_{\rm cal} = 10^{-3}$ and $E_{\rm for} = 10^{-2}$.

\begin{figure}[ht!]
\centering
  \includegraphics[angle=90,width=0.45\textwidth]{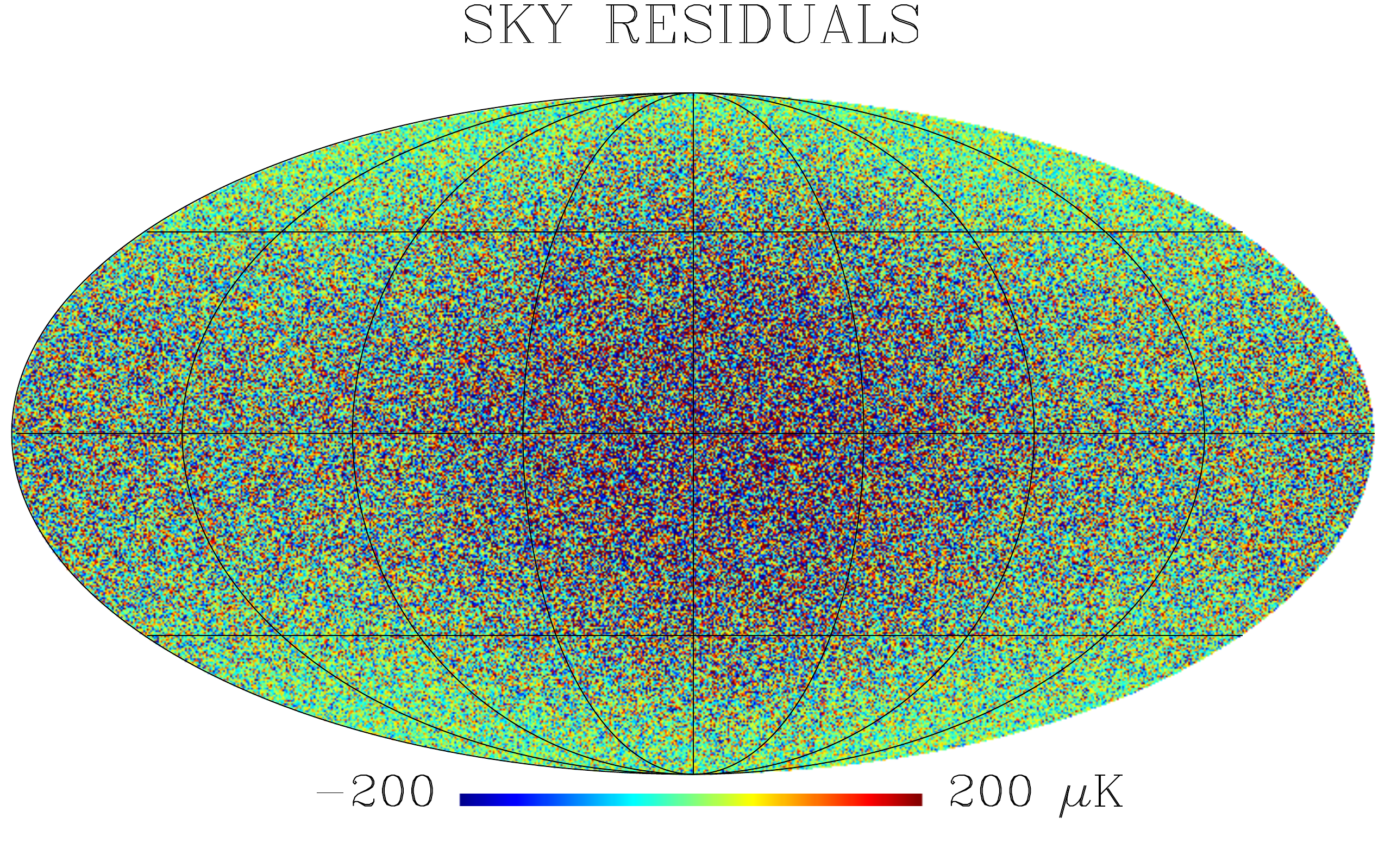}
  \includegraphics[angle=90,width=0.45\textwidth]{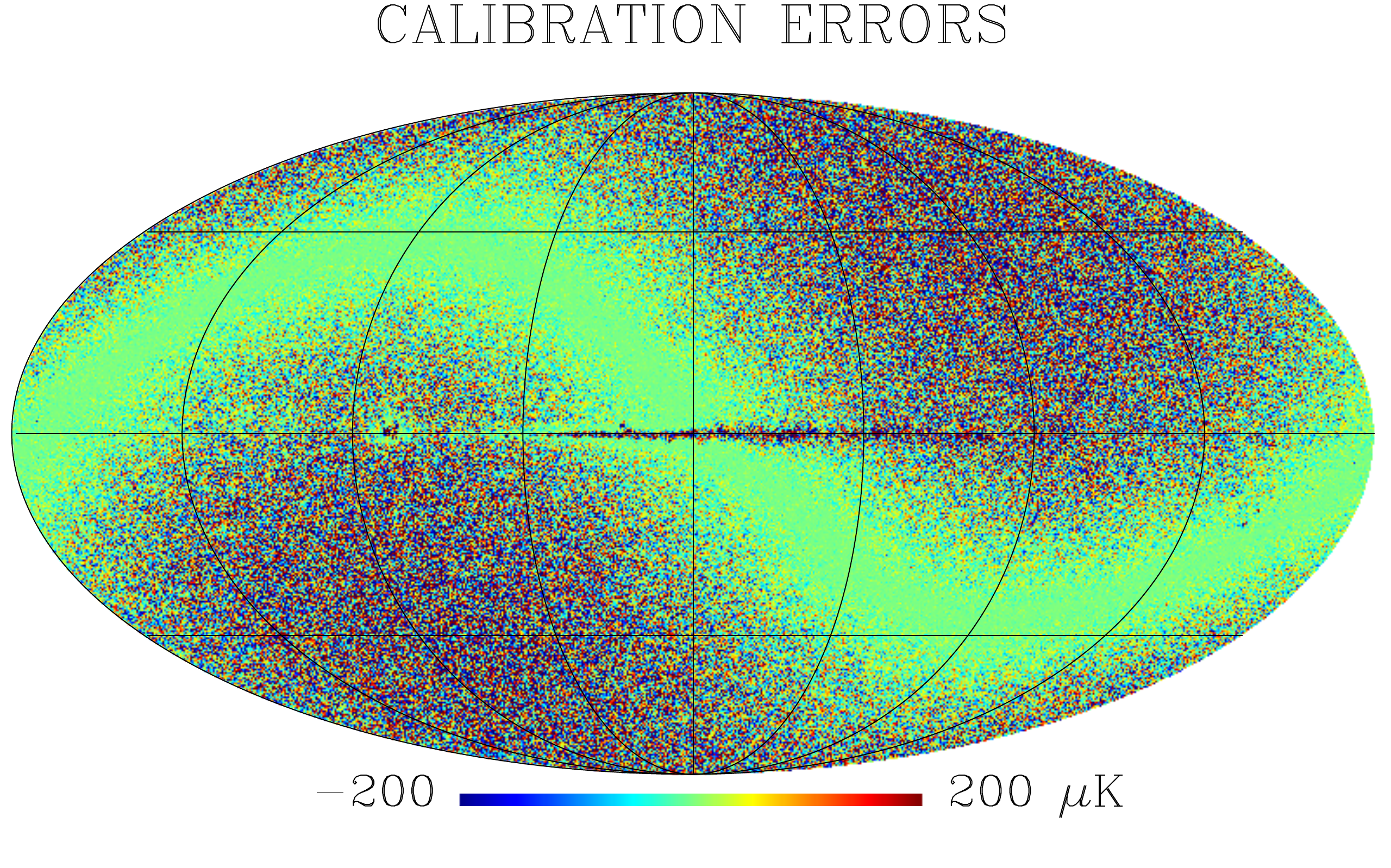}
  \caption{Sky residual and calibration error maps (in Galactic coordinates)
  in the 60-GHz band employed in the simulations.
  Their amplitudes correspond to the pessimistic case,
  $E_{\rm for} = 0.64$ and $E_{\rm cal} = 0.064$, for maps at resolution {\tt HEALPix} $N_{\rm side}=1024$.}
\label{fig:dipsystematics}
\end{figure}

\begin{figure}[ht!]
\centering
 \includegraphics[angle=90,scale=0.32]{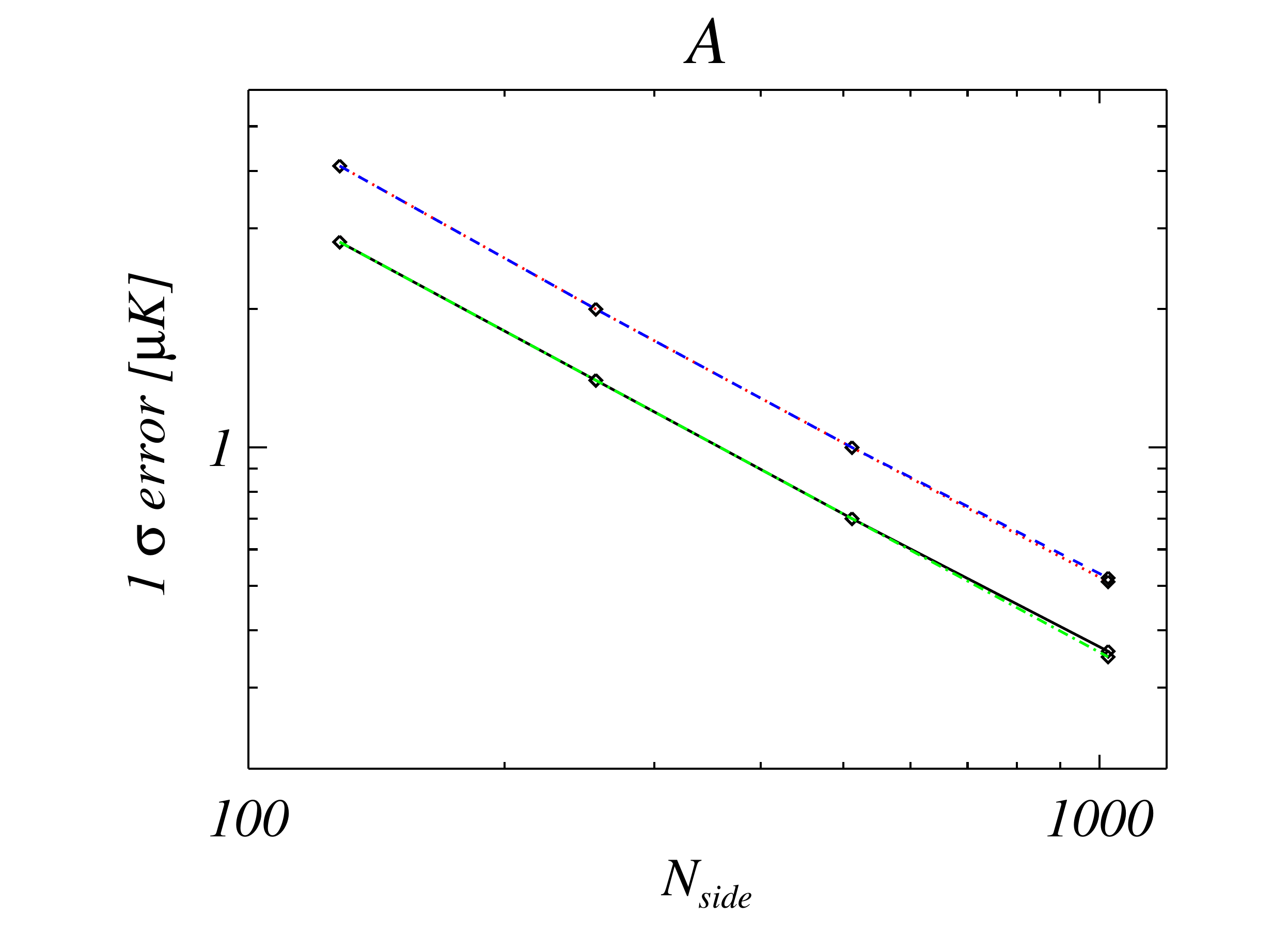}%
 \includegraphics[angle=90,scale=0.32]{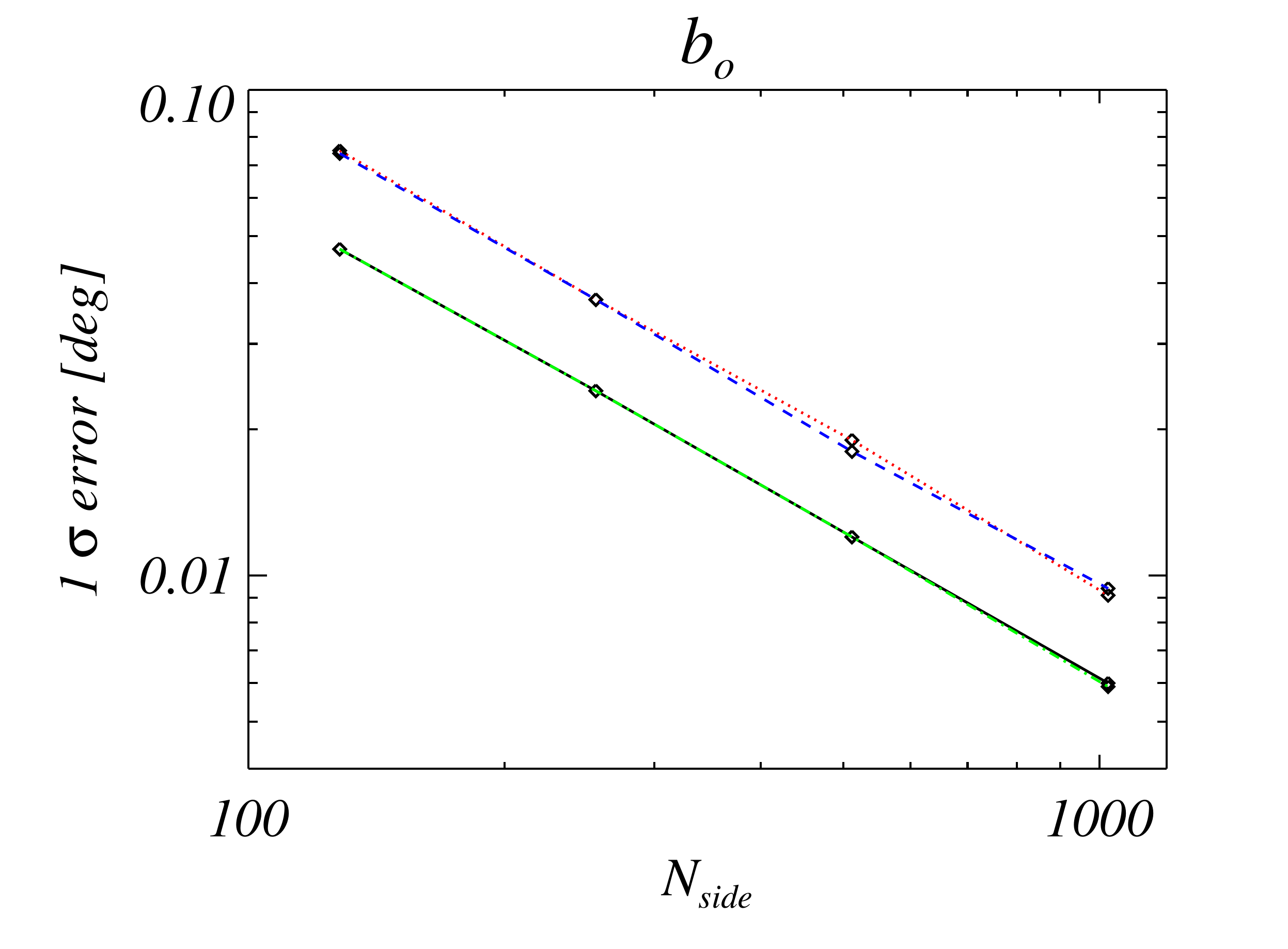}
 \includegraphics[angle=90,scale=0.32]{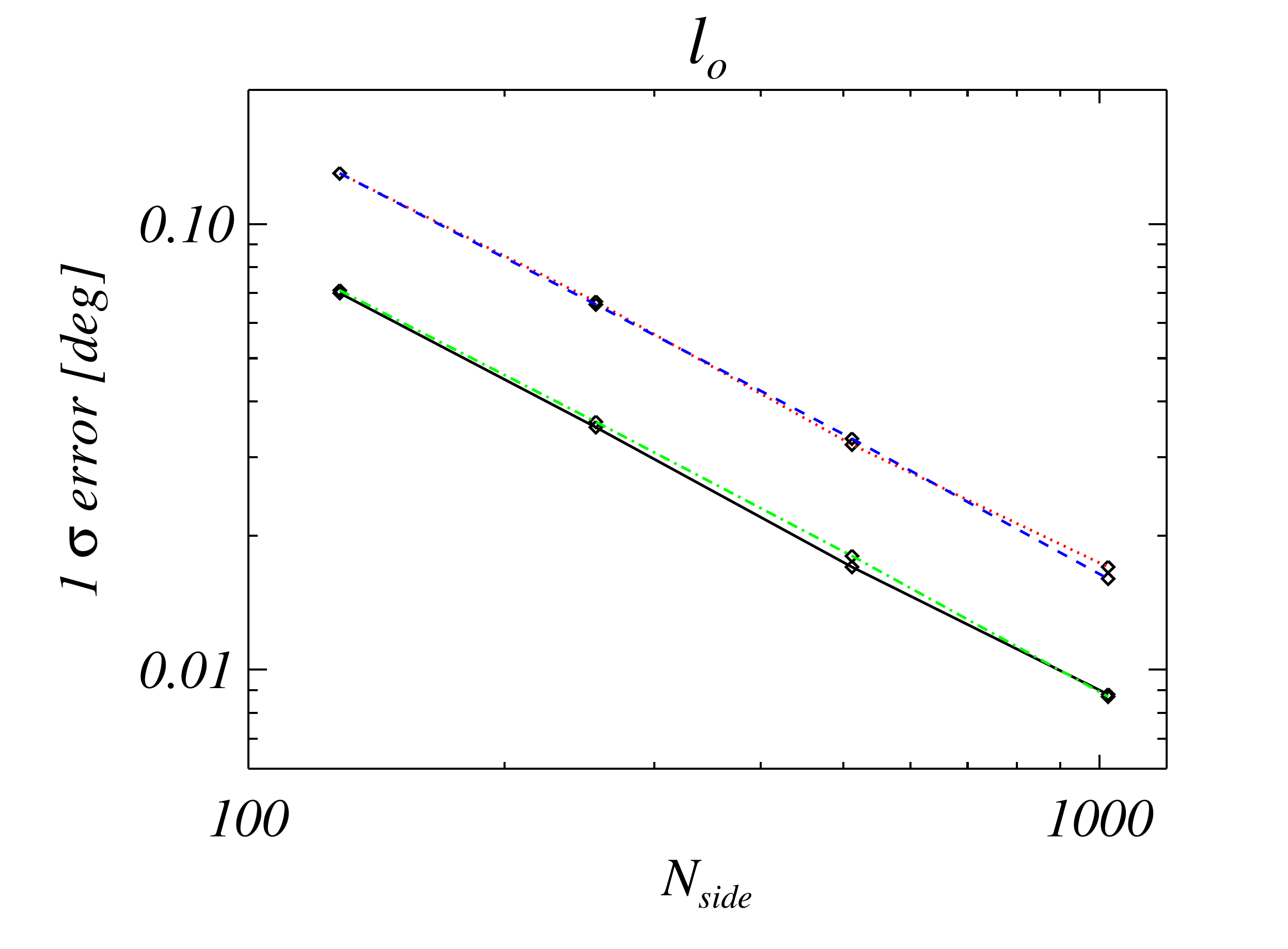}%
 \includegraphics[angle=90,scale=0.32]{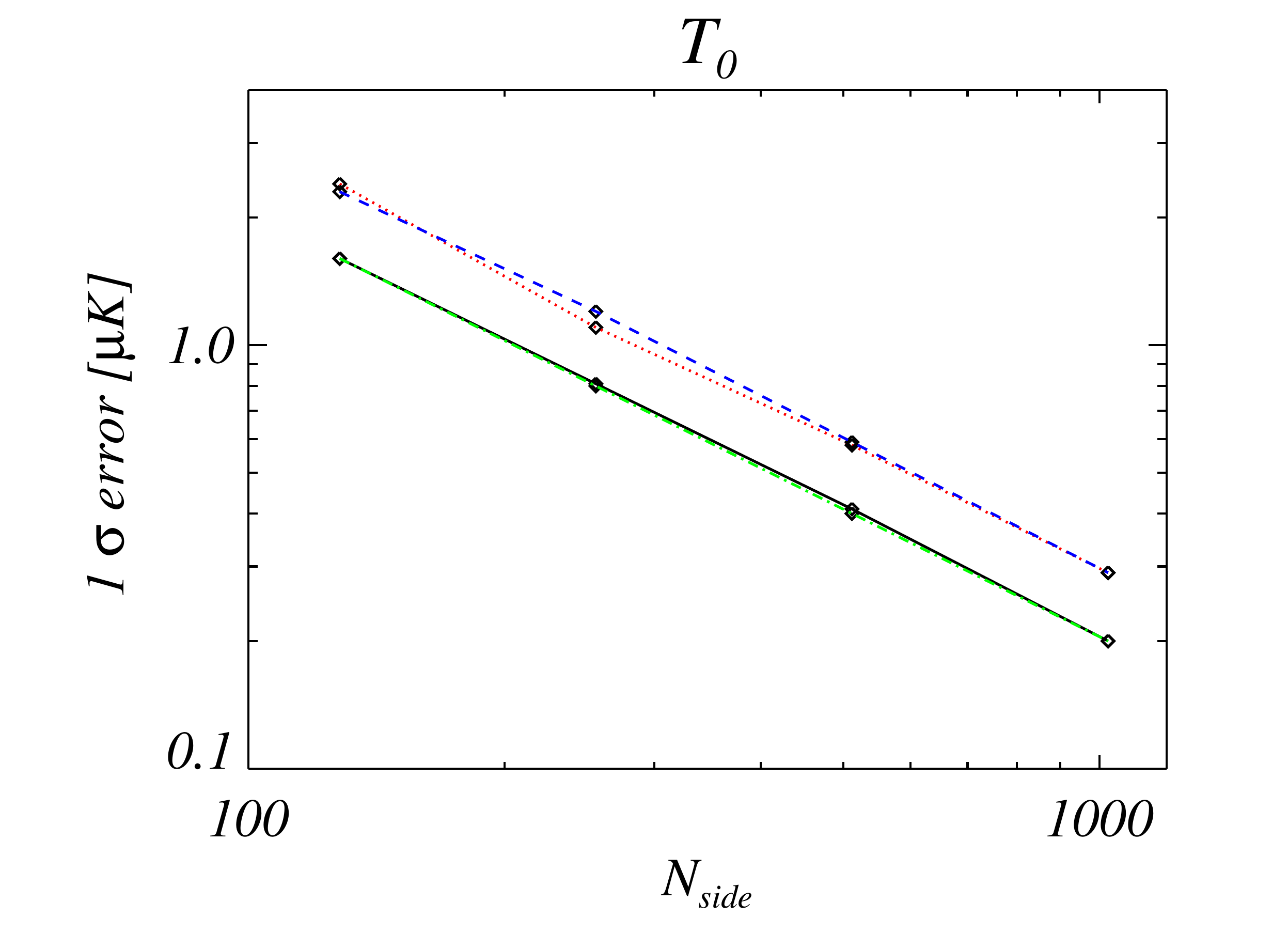}
 \caption{1\,$\sigma$ errors as function of {\tt HEALPix} $N_{\rm side}$ values
for the parameters $A$, $b_0$, $l_0$, and $T_{0}$: dipole-only (solid
black line); dipole+noise (green dot-dashed line); dipole+noise+mask (red dotted line); and dipole+noise+mask+systematics (blue dashed line).
The chosen frequency channel is 60\,GHz and the noise map corresponds to 7.5\,$\mu$K.arcmin. The adopted mask is the {\it Planck\/} Galactic
mask extended to cut out $\pm30^\circ$ of the Galactic plane. The systematics correspond to the pessimistic expectation of calibration errors and
sky (foreground, etc.) residuals. Notice that the pixelization error, due to the finite map resolution, is dominant over the noise for any $N_{\rm side}$. While
the impact of noise and systematics is negligible, we find that the effect of reducing the effective sky fraction is important.}%
\label{fig:diperrors}
\end{figure}

In Fig.~\ref{fig:dipsystematics} we display the maps used in the simulations (for the 60-GHz band). The amplitudes correspond to the worse expected case; the most optimistic case is not shown, since the amplitude is just rescaled by a factor 16. The corresponding likelihood plots and 68\% confidence levels are collected in Appendix~\ref{sec:Likelihoods}.

We find that the impact of systematic effects on the parameter errors is negligible. In fact, as shown in Fig.~\ref{fig:diperrors}, calibration errors and sky residuals do not noticeably worsen the 1\,$\sigma$ uncertainty at any sampling resolution. Furthermore, the frequency analysis confirms that the impact of systematic effects is not relevant in any of the bands under consideration (from 60 to 220\,GHz).

While the effect of the systematics studied here on the \emph{precision} of the parameter reconstruction is negligible, we find instead that they may have a moderate impact on the \emph{accuracy}, introducing a bias in the central values of the  estimates. Nonetheless, the bias is usually buried within the 1\,$\sigma$ error, with the marginal exception of the estimate of $l_0$ for the 220-GHz band (in the case of pessimistic systematics).

In conclusion, our results show that the dipole recovery (in
both amplitude $A$ and direction angles $b_0$ and $l_0$) is
completely dominated by the sky sampling resolution. We find that: the
noise impact is negligible; the reduction of the sky fraction due to the
presence of the Galactic mask impacts on the parameter error amplitude
by increasing the 1\,$\sigma$ errors on $A$, $b_0$ and $l_0$  by a factor
of about 1.5, 1.6, and 1.9, respectively; and the effect of systematics slightly
worsens the accuracy of the MCMC chain without affecting the error
estimate.

The main point of our analysis is that, in order to achieve an increasing
precision in the dipole reconstruction, high resolution measurements are
required, in particular when a sky mask has to be applied.
This is especially relevant for dipole spectral distortion analyses,
based on the high-precision, multi-frequency observations that are
necessary to study the tiny signals expected.

\section{Measuring Doppler and aberration effects in different maps}\label{sect:aberration}

\subsection{Boosting effects on the CMB fields}

As discussed in the previous sections, a relative velocity between an observer and the CMB rest frame induces a dipole in the CMB temperature through the Doppler effect.
The CMB dipole, however, is completely degenerate with an {\it intrinsic\/} dipole,
which could be produced by the Sachs-Wolfe effect at the last-scattering
surface due to a large-scale dipolar Newtonian potential~\cite{Roldan:2016ayx}.
For $\Lambda$CDM such a dipole should be of order of the Sachs-Wolfe plateau
amplitude
(i.e., $10^{-5}$) \cite{2008PhRvD..78h3012E,2008PhRvD..78l3529Z}, nevertheless
the dipole could be larger in the case of more exotic models.
In addition to the dipole, a moving observer will also see velocity imprints at
$\ell>1$ in the CMB due to Doppler and aberration
effects~\cite{Challinor:2002zh,Burles:2006xf}. Such effects can be measured as
correlations among different $\ell$s, as has been proposed
in Refs.~\cite{Kosowsky:2010jm,Amendola:2010ty,Notari:2011sb} and subsequently
demonstrated in Ref.~\cite{2014A&A...571A..27P}.

The aberration effect changes the arrival direction of photons from
$\hat{n}^{\prime}$ to $\hat{n}$, which, at linear order in $\beta$, is
completely degenerate with a lensing dipole.
The Doppler effect modulates the CMB (an effect that is partly
degenerate with an intrinsic CMB dipole\footnote{It has been shown in~\cite{Roldan:2016ayx} that, in the Gaussian case, an intrinsic large scale dipolar potential exactly mimics on large scales a Doppler modulation.}) changing the specific intensity $I'$
in the CMB rest frame to the intensity $I$ in the observer's frame\footnote{In
this section we will use primes for the CMB frame and non-primes for the
observer frame, following Ref.~\cite{2014A&A...571A..27P}.} by a multiplicative,
direction-dependent factor as~\cite{1986rpa..book.....R,Challinor:2002zh}
\begin{eqnarray}
    I'(\nu',\,\hat{n}^{\prime})=I(\nu,\,\hat{n}) \left(\frac{\nu'}{\nu} \right)^3,
    \label{eq:i_boost}
\end{eqnarray}
where
\begin{align}
    \nu \,=\, \nu^{\prime}\,\gamma\,\big(1+ \vec{\beta}\cdot\hat{n}^{\prime}\big)\,,\qquad
    \hat{n} \,=\,  \frac{\hat{n}^{\prime} +  \left[\gamma\,\beta + (\gamma - 1) \big(\hat{n}^{\prime}\cdot\hat{\beta} \big)\right]\hat{\beta}}{\gamma (1+ \vec{\beta} \cdot\hat{n}^{\prime}) } \,,
    \label{eq:nu_boost}
\end{align}
with $\gamma \equiv (1-\vec{\beta}^2)^{-1/2}$.
The temperature and polarization fields $X(\hat{n})$ in the CMB rest frame (where $X$ stands for $T,\,E$ or $B$) are similarly transformed as
\begin{align}
    X^\prime ( \hat{n}^{\prime} ) \,=\, X(\hat{n}) \gamma \big( 1 - \vec{\beta}\cdot\hat{n} \big) \,.
    \label{eq:Xfields}
\end{align}

Decomposing Eq.~\eqref{eq:Xfields} into spherical harmonics leads to an effect in
the multipole $\ell$ of order $\beta^\ell.$ Although this effect is dominant in
the dipole, it also introduces a small, non-negligible correction to the
quadrupole, with a different frequency dependence, due to the conversion of
intensity to temperature~\cite{2003PhRvD..67f3001K, 2004A&A...424..389C,
Notari:2015kla, Quartin:2015kaa}. In addition, both aberration and Doppler
effects couple multipoles $\ell$ to $\ell\pm n$~\cite{Chluba:2011zh,
Notari:2011sb}. This coupling is largest in the correlation between $\ell$ and
$\ell\pm 1$~\cite{Challinor:2002zh,Kosowsky:2010jm,Amendola:2010ty}, which was
measured by {\it Planck\/} at 2.8 and 4.0\,$\sigma$ significance for the
aberration and Doppler effects,
respectively~\cite{2014A&A...571A..27P}. These ${\cal O}(\beta)$ couplings are
present on all scales and the measurability of aberration is mostly limited by cosmic
variance, which constrains our ability to assume fully uncorrelated modes for $\ell
\neq \ell^\prime$. Hence, in order to improve their measurement, it is
important to have as many modes as possible, which drives us to
cosmic-variance-limited measurements of temperature and polarization up to
very high $\ell_{\rm max}$ and coverage of a large fraction of the sky $f_{\rm
sky}$. CORE probes a larger $\ell_{\rm max}$ and covers a larger effective $f_{\rm sky}$
than {\it Planck} (as the extra frequency channels and the better sensitivity allow for an improved capability in doing component separation), hence it should achieve a detection of almost $13\,\sigma$
even with a 1.2-m telescope, as shown below.

As discussed in Ref.~\cite{Challinor:2002zh}, upon a boost of a CMB map $X$,
the \alm~coefficients of the spherical harmonic decomposition transform as
\begin{equation}\label{eq:aberrated-alm}
    a^{X}_{\ell m} \;=\; \sum_{\ell'=0}^\infty {}_s K_{\ell' \ell m} \, a'^{X}_{\ell' m}\,,
\end{equation}
where $s$ indicates the spin of the quantity $X$. For scalars (such as the temperature), $s=0$, while for spin-2 quantities (such as the polarization), $s=2.$

The kernels ${}_s K_{\ell' \, \ell\, m}$ in general cannot be computed analytically and their numerical computation is not trivial, since this involves highly oscillatory integrals~\cite{Chluba:2011zh}.
However, efficient methods using an operator approach in harmonic space have been developed \cite{2014PhRvD..89l3504D}, although for our estimates more approximate methods will suffice. It was shown in Ref.~\cite{Notari:2011sb, 2014PhRvD..89l3504D} that the kernels can be well approximated by Bessel functions as follows:
\begin{equation}\label{eq:non-linear-fit-general}
\begin{aligned}
    K_{(\ell-n) \ell m}^X   &\;\simeq\; J_n\!\left(\!-2\, \beta \left[\prod_{k=0}^{n-1} \big[(\ell-k) \;{}_sG_{(\ell-k) m} \big]\right]^{1/n} \right); \\
    K_{(\ell+n) \ell m}^X   &\;\simeq\; J_n\!\left(\,2\, \beta \left[\prod_{k=1}^n \big[(\ell+k) \;{}_sG_{(\ell+k) m} \big]\right]^{1/n} \right).
\end{aligned}
\end{equation}
Here
\begin{equation}\label{eq:Glm}
    {}_{s}G_{\ell m} \equiv \sqrt{\frac{\ell^2-m^2}{4\ell^2-1} \left[1-\frac{s^2}{\ell^2}\right]}\,,
\end{equation}
and $n \ge1$ (where $n$ is the difference in multipole between a pair of coupled multipoles, namely $\ell$ and $\ell \pm n$ ). It is also assumed that $\beta \ll 1$, although the formula above can be generalised to large $\beta$~\cite{Notari:2011sb, 2014PhRvD..89l3504D}. These kernels couple different multipoles so that, by Taylor expanding, we find $\big<a_{\ell m}~a_{(\ell+n)m}^{\ast}\big> = {\cal O}(\beta \ell)^n$. For $\ell \ll 1/\beta,$ the most important couplings are between neighbouring multipoles, $\ell$ and $\ell \pm 1$ (e.g.~\cite{Challinor:2002zh}). One may wonder about the importance of the couplings between non-neighbouring multipoles, i.e., $\ell$ and $ \ell \pm n$, for $\ell \gtrsim 1/\beta$. However, quite surprisingly, for $\ell \gg 1/\beta$ we find that: (1) in the $(\ell, \ell \pm 1)$ correlations, terms that are higher order in $\beta \ell$ are negligible~\cite{Chluba:2011zh, Notari:2011sb}; and (2) most of the correlation seems to remain in the $(\ell, \ell \pm 1)$ coupling. For these reasons, from here onwards, we will ignore terms that are higher order in $\beta$ and couplings between non-neighbouring multipoles (i.e., $n>1).$

In order to measure deviations from isotropy due to the proper motion of the observer, we therefore compute the off-diagonal correlations
$\big<a_{\ell m}^{X}~a_{(\ell+1)m}^{X\ast}\big>.$
Assuming that in the rest frame the Universe is statistically isotropic and that parity is conserved, then in the boosted frame,
for $\ell^{\prime}=\ell+1,$ we find that (see Refs.~\cite{Challinor:2002zh,Kosowsky:2010jm,Amendola:2010ty})
\begin{equation}
    a_{\ell \, m}^{X} \, \simeq \, c_{\ell m}^{-} a^{' \, X}_{(\ell-1) m} + c_{\ell m}^{+} a^{' \, X}_{(\ell+1) m} \, ,
\end{equation}
where
\begin{equation}
    c_{\ell m}^{+} = \beta(\ell+2 -d) {}_sG_{(\ell+1) m}\,,\qquad
    c_{\ell m}^{-} = -\beta(\ell-1+ d ){}_sG_{\ell m}\,,\label{eq:alm-coef}
\end{equation}
and $d$ parametrizes the Doppler effect of dipolar modulation. It
then follows that
\begin{equation}\label{eq:almcorr}
    \left<a_{\ell m}^{X}~a_{(\ell+1)m}^{Y\ast}\right>
    = \beta
    \left[ (\ell+2 -d) \,{}_{s_X}\! G_{(\ell+1)m} C_{\ell+1}^{XY} - (\ell+d)\,{}_{s_Y}\! G_{(\ell+1)m} C_{\ell}^{XY}
    \right] + O(\beta^2)\,.
\end{equation}
For $\ell \gtrsim 20$, we have ${}_{2}G_{\ell m} \simeq {}_{0}G_{\ell m}$. As will be shown, large scales are not important for measuring the boost, and thus it is not important to keep the indication of the spin. Thus from here onwards, we will drop $s$. The above equation reduces to
\begin{equation}
    \left<a_{\ell m}^{X}~a_{(\ell+1)m}^{Y\ast}\right>
    = \beta G_{(\ell+1)m}
    \left[ (\ell+2 -d) C_{\ell+1}^{XY} - (\ell+d) C_{\ell}^{XY}
    \right] + O(\beta^2)\,,
    \label{eq:aberration_covariance}
\end{equation}
where the angular power spectra $C^{XY}_\ell$ are measured in the CMB
rest frame. For the CMB temperature and polarization, $d=1$, as observed from
Eqs.~\eqref{eq:i_boost}--\eqref{eq:nu_boost}.
In this case, no mixing of $E$- and $B$-polarization modes occurs, not even in
higher orders in $\beta$ \cite{Notari:2011sb,2014PhRvD..89l3504D}. However, for
$d \neq 1$, the coupling is non-zero already at first order in $\beta$
\cite{Challinor:2002zh,2014PhRvD..89l3504D}. Maps estimated from spectra that
are not blackbody have different Doppler coefficients,\footnote{Note that the
kernel defined as in Eq.~\eqref{eq:aberrated-alm} for $d\neq 1$ can be obtained
from $_{s}K_{\ell' \ell m}$ using recursions~\cite{2014PhRvD..89l3504D}.} as we
discuss in the next subsection.

Note that in practice one never measures temperature and polarization
anisotropies directly, instead one measures anisotropies in \emph{intensity}
and then converts this to temperature and polarization. This distinction
(though perhaps seeming trivial)
is relevant for the Doppler effect, which induces a dipolar
modulation of the CMB anisotropies, appearing with frequency-dependent
factors~\cite{2014A&A...571A..27P,2016PhRvD..94d3006N}.
In particular such factors were shown to be proportional to
a Compton $y$-type spectrum (exactly like the quadrupole
correction~\cite{2003PhRvD..67f3001K, 2004A&A...424..389C, Notari:2015kla,
Quartin:2015kaa} and therefore degenerate with the tSZ effect); they are
measurable in the {\it Planck\/} maps at about 12$\,\sigma$ and in the CORE maps
even at 25--60\,$\sigma$~\cite{2016PhRvD..94d3006N}, depending on the template
that is used for contamination due to the tSZ effect. Such S/N ratios are much larger than those that can be obtained in temperature
and polarization and so, at first sight, they may appear to represent a better
way to measure the boosting effects.
However, the peculiar frequency
dependence is strictly a consequence of the intensity-to-temperature (or
intensity-to-polarization) conversion and thus agnostic to the source of the
dipole~\cite{2014A&A...571A..27P, 2016PhRvD..94d3006N} (i.e., whether
it is from our peculiar velocity or is an intrinsic CMB dipole).
For this reason we
focus on the frequency-\emph{independent} part of the dipolar modulation signal
in Eq.~\eqref{eq:aberration_covariance} (with $d=1$), which is unlikely to be
caused by an intrinsically large CMB dipole (see Ref.~\cite{Roldan:2016ayx}
for details), in our forecast.

\subsection{Going beyond the CMB maps}

Since CORE will also measure the thermal Sunyaev-Zeldovich effect, the CIB, and the weak lensing signal over a wide multipole range, it is interesting to examine if these maps could also be used to measure the aberration and Doppler couplings.

The intensity of a tSZ Compton-$y$ map is given by
\begin{equation}\label{eq:TSZ-spectrum}
  I'_{tSZ}(\nu')  \,=\, y \cdot g\left(\frac{h\nu'}{k_{\rm B}T_{0}}\right) \,K(\nu') \, ,
\end{equation}
where $g(x')=x' \coth(x'/2)-4$, $K(\nu')$ is the conversion factor that derives from setting $T=T_{0}+\delta T$ in the
Planck distribution and expanding to first order in $\delta T$,
and $x' \equiv h\nu'/k_{\rm B} T_{0}$ ($T_{0}$ being the present temperature of the CMB).
Explicitly $K(\nu')$ is given by
\begin{equation}
    K(\nu') \,=\, \frac{2\,h \nu'^3}{c^2} \frac{x' \exp (x')}{(\exp (x') -1)^2} \,.
\end{equation}
A boosted observer will see an intensity as defined in Eq.~\eqref{eq:i_boost}.
Such intensity, expanded at first order in $\beta$, will contain Doppler
couplings with a non-trivial frequency dependence, similarly to what happens in
the case of CMB fluctuations, where frequency-dependent boost factors are
generated, as discussed in the previous subsection. For simplicity we only
analyse the couplings that retain the same frequency dependence of the
original tSZ signal, which come from aberration,\footnote{Also, sub-leading
contributions, namely the kinetic Sunyaev-Zeldovich effect
\cite{1980MNRAS.190..413S} and changes in the tSZ signal induced by the
observer motion relative to the CMB rest frame \cite{2005A&A...434..811C}, as
well as relativistic corrections
\cite{1979ApJ...232..348W,1981Ap&SS..77..529F}, are specific to each particular
cluster. Their inclusion could be considered in more detailed predictions in future, but represent higher-order corrections for the present study.}
and so we here set $d=0$ in Eq.~\eqref{eq:aberration_covariance}.

For the intensity of the CIB map (see Sect.~\ref{sect:DipCIB} for further
details), we assume the template obtained by Ref.~\cite{Fixsen:1998kq},
\begin{equation}\label{eq:CIB-spectrum}
       I'_{\rm CIB} \,\propto\, \nu'^{0.64}  \frac{\nu'^3}{\exp\!\left[\frac{h \nu'}{k_{\rm B}\, 18.5{\rm K}}\right]-1} \, .
\end{equation}
At low frequencies, the intensity scales as
\begin{equation}\label{eq:CIB-spectrum-RJ}
    I'_{\rm CIB} = A_{\rm CIB} \, \nu'^{2.64} \, ,
\end{equation}
where $A_{\rm CIB}$ is a constant related to the amplitude. In the boosted frame and to lowest order in $\beta,$ we find that
\begin{equation}
    I_{\rm CIB}(\nu) = \left(\frac{\nu}{\nu'} \right)^3 A'_{\rm CIB} \, \nu'^{2.64}  \simeq
    A'_{\rm CIB} \big[ \gamma (1- \vec{\beta} \cdot {\hat{n}}) \big]^{-0.36}   \nu^{2.64} \, .
\end{equation}
Therefore, the boosted amplitude is
$A_{\rm CIB} \equiv A'_{\rm CIB} / \big[\gamma (1- \vec{\beta}\cdot\hat{n})\big]^{0.36}, $
which implies $d=0.36$.
Note that in this case, since we work in a low-frequency approximation (relative to the peak of the CIB at around 3000\,GHz), we do
not have any frequency-dependent boost factors.

The CMB weak lensing maps can also be used to measure the boost. However, since the estimation of the weak lensing potential involves 4-point correlation functions of the CMB fields, the boost effect is more complex to estimate; hence we leave this analysis for a future study.

\subsection{Estimates of the Doppler and aberration effect}

For full-sky experiments, it has been shown in Ref.~\cite{Challinor:2002zh}
that, under a boost, the corrections to the power spectra are ${\cal
O}(\beta^2),$ whereas for experiments with partial sky coverage there can be an
${\cal O}(\beta)$ correction~\cite{Pereira:2010dn,Jeong:2013sxy,Louis:2016ahn}.
Nevertheless, even for the partial-sky case, this correction to $C_{\ell}^{XY}$
would only propagate at ${\cal O}(\beta^2)$ in the correlations above. In what
follows, we will neglect the effect of the sky coverage in the boost
corrections. Also, since we will be restricting ourselves to ${\cal O}(\beta)$
effects, from here onwards we will drop ${\cal O}(\beta^2)$ from the equations.

For the CMB fields, as it was shown in Refs.~\cite{Amendola:2010ty,Notari:2011sb}, that the fractional uncertainty in the estimator of $ \left<a_{\ell m}^{X}~a_{(\ell+1)m}^{Y\ast}\right>$ is given by
\begin{equation}\label{eq:delta-beta}
    \left. \frac{\delta \beta}{\beta}\right|_{XY} \simeq \left[\sum_{\ell}\sum_{m=-\ell}^\ell \frac{\left< a_{\ell m}^{X}~a_{(\ell+1)m}^{Y\ast}\right>^{2}} {{\mathfrak C}^{XX}_{\ell} {\mathfrak C}^{YY}_{\ell+1}} \right]^{-1/2}
\end{equation}
(see also Ref.~\cite{2016A&A...594A..16P}).
Here, ${\mathfrak C}^{XX}_{\ell} \equiv (C^{XX}_{\ell} + N^{XX}_{\ell,\text{total}}) / \sqrt{f_{\rm sky}},$ where $f_{\rm sky}$ is the fraction of the sky covered by the experiment and $\,N^{XX}_{\ell,\text{total}}\,$ is the effective noise level on the map $X.$ Thus ${\mathfrak C}^{XX}_\ell$ represents the sum of instrumental noise and cosmic variance. The effective noise is obtained by taking the inverse of the sum over the different channels $i$ of the inverse of the individual $N_{\ell,i}^2$~\cite{Notari:2011sb},
\ba
    N_{\ell,\text{total}}=
    \left[\sum_{i}^{\text{nchannel}} {1\over N_{\ell,i}^2}\right]^{-1/2}\,.
\ea
The noise in each channel is given by a constant times a Gaussian beam characterised by the beam width $\theta_{\text{FWHM}}$:
\ba \label{eq:Nell-formula}
    N_{\ell,i}^{X} \,=\, \big(\sigma^{X}\big)^2 \exp\left[\frac{\ell(\ell+1)\theta_{\text{FWHM}}^2 }{8\ln 2}\right]\,,
\ea
where $\sigma^{X}$ is the noise in $\mu$K.arcmin for the map $X.$

\begin{figure}[ht!]
    \centering
    \hspace{-0.5cm}
    \includegraphics[width=0.521\textwidth]{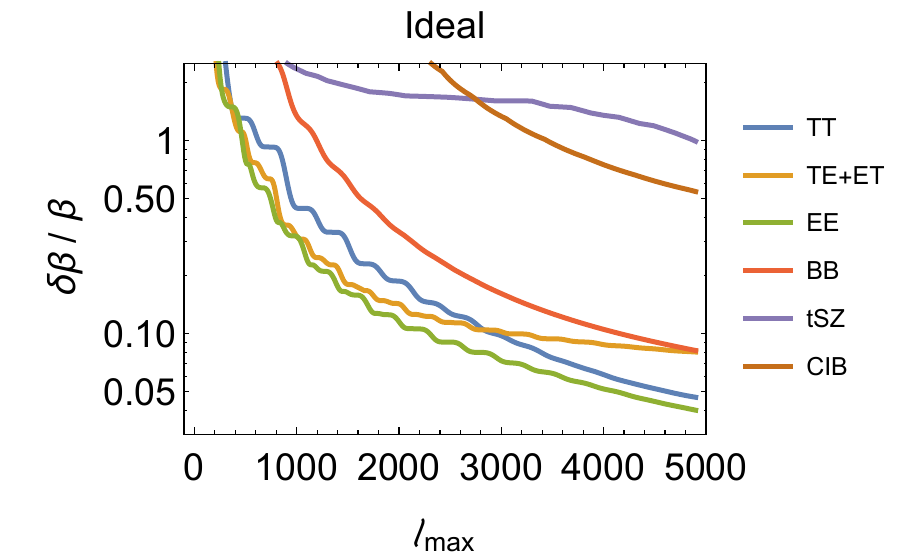}\hspace{-0.4cm}
    \includegraphics[width=0.521\textwidth]{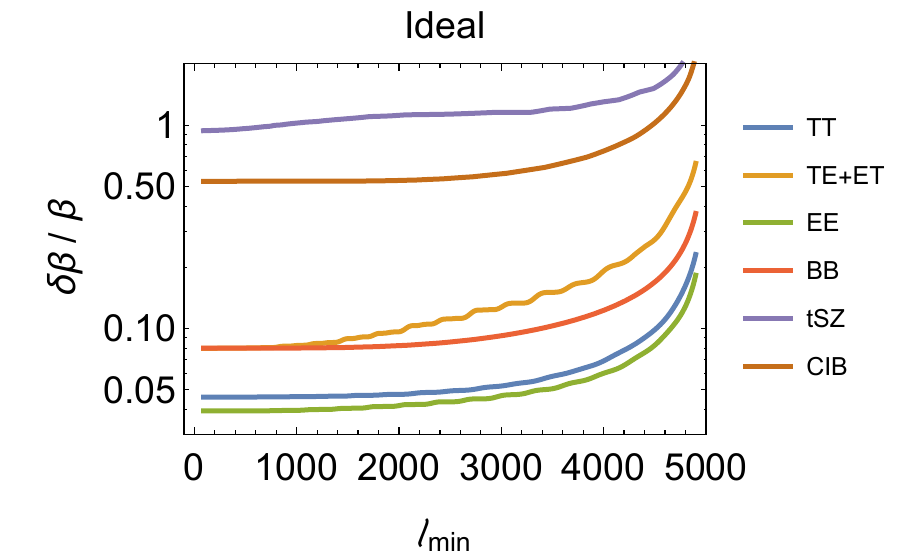}
    \includegraphics[width=0.521\textwidth]{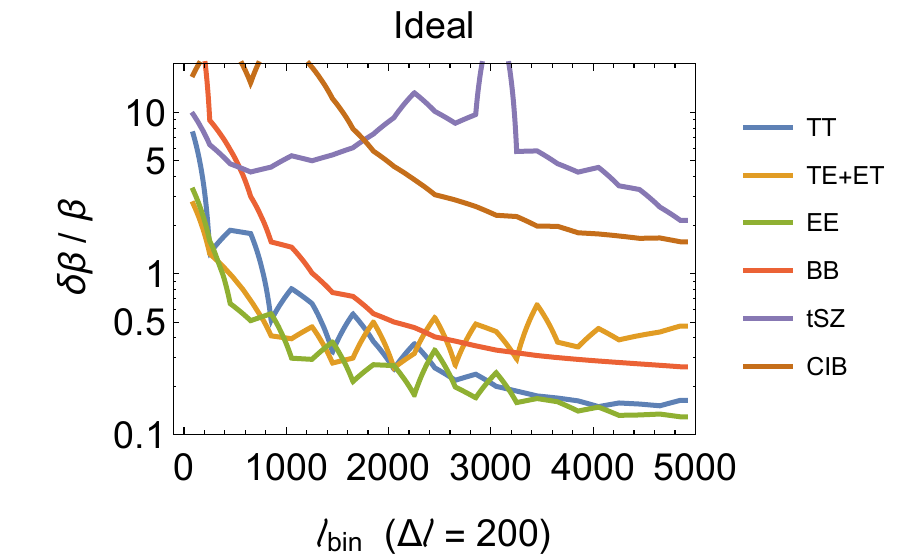}
    \caption{Achievable precision in estimating the velocity through aberration
    and Doppler effects in an ideal experiment (with $f_{\rm sky} = 1$ and
    limited by cosmic variance only) for different maps. \emph{Left:} as a
    function of $\ell_{\rm max}$. \emph{Right:} as a function of $\ell_{\rm
    min}$ (with fixed $\ell_{\rm max}=5000$). \emph{Bottom:} for individual
    bins with $\Delta \ell = 200$. We see that: (i) the first hundred $\ell$s
    are not important for achieving a high S/N; and (ii) the non-CMB diffuse
    maps exhibit low precision and are not very useful for measuring $\beta$.
    Note that for simplicity we have assumed no primordial $B$-modes (our
    constraints are very weakly sensitive to this choice).
    \label{fig:SN-Ideal}}
\end{figure}

For the tSZ signal, we assume as a fiducial spectrum the one obtained in ref.~\cite{Aghanim:2015eva} (slightly extrapolated to higher $\ell$s).
For the forecast noise spectrum we use the estimates obtained in ref.~\cite{2017arXiv170310456M} using the NILC component separation technique (see figure 14 therein),
where it was shown that residual foreground contamination is a large fraction of the total noise.
For the CIB signal, we use the spectra obtained in Ref.~\cite{Cai2013}; for the noise, we rely on the simulations carried out in
Ref.~\cite{2016arXiv160907263D}.
We also make the conservative assumption that the different channels of the CIB
are 100\,\% correlated. Since different channels pick up different redshifts,
effectively the correlation is not going to be total and some extra signal can be obtained from multiple channels;
however, since this makes the analysis much more complex (due to the need to
have all
the covariance matrices) and since the CIB turns out not to be promising
for measuring aberration (see Fig.~\ref{fig:SN-Ideal}), we neglect these
corrections.

We computed Eq.~\eqref{eq:delta-beta} for the different maps of different
experiments. We compared the detection potentials of CORE (see
Table~\ref{tab:CORE-bands}) with those expected from both {\it Planck\/} and
LiteBIRD \cite{2016SPIE.9904E..0XI}. For the {\it Planck\/} specifications, we
use the values of the 2015 release, while the LiteBIRD specifications used in
this analysis are listed in Table~\ref{tab:LiteBIRD-bands}.

In Fig.~\ref{fig:SN-Ideal} we show the precision that could be reached by an
ideal experiment with $f_{\rm sky} = 1$ and limited by cosmic variance only. We
show the results for: the range $\ell \in [2, \ell_{\rm max}]$; the range $\ell
\in [\ell_{\rm min}, 5000]$; and for individual $\ell$ bins of width $\Delta\ell =
200.$ The signal-to-noise ratios in the tSZ and CIB maps are considerably
lower than in the CMB maps, which is due to the fact that the spectra are
smoother, as explained later. For instance, for $\ell_{\rm max}=4000,$ in the
$TT$ and $EE$ maps separately we have ${\rm S/N} > 16,$ whereas in tSZ and in CIB we
have ${\rm S/N} \simeq 1.$

\begin{table}[ht!]
\begin{center}
\scalebox{0.93}{\begin{tabular}{|c|c|c|c|c|c|c|c|c|c|c|c|}
    \hline
    Channel & Beam  &   $\Delta T$  &  $\Delta P$ \\ \relax
    [GHz]    &  [arcmin]  &   [$\mu$K.arcmin]  &   [$\mu$K.arcmin]  \\
    \hline
    \hline
     40 & 108 & 42.5 & 60.1 \\
     50 & 86 & 26 & 36.8 \\
     60 & 72 & 20 & 28.3 \\
     68.4 & 63 & 15.5 & 21.9 \\
     78 & 55 & 12.5 & 17.7 \\
     88.5 & 49 & 10 & 14.1 \\
     100 & 43 & 12 & 17. \\
     118.9 & 36 & 9.5 & 13.4 \\
     140 & 31 & 7.5 & 10.6 \\
     166 & 26 & 7 & 9.9 \\
     195 & 22 & 5 & 7.1 \\
     234.9 & 18 & 6.5 & 9.2 \\
     280 & 37 & 10 & 14.1 \\
     337.4 & 31 & 10 & 14.1 \\
     402.1 & 26 & 19 & 26.9 \\
    \hline
\end{tabular}}
\end{center}
\caption{\small LiteBIRD specifications used in this analysis.
}
\label{tab:LiteBIRD-bands}
\end{table}

\begin{figure}[ht!]
    \centering
    \hspace{-0.5cm}\includegraphics[width=0.51\textwidth]{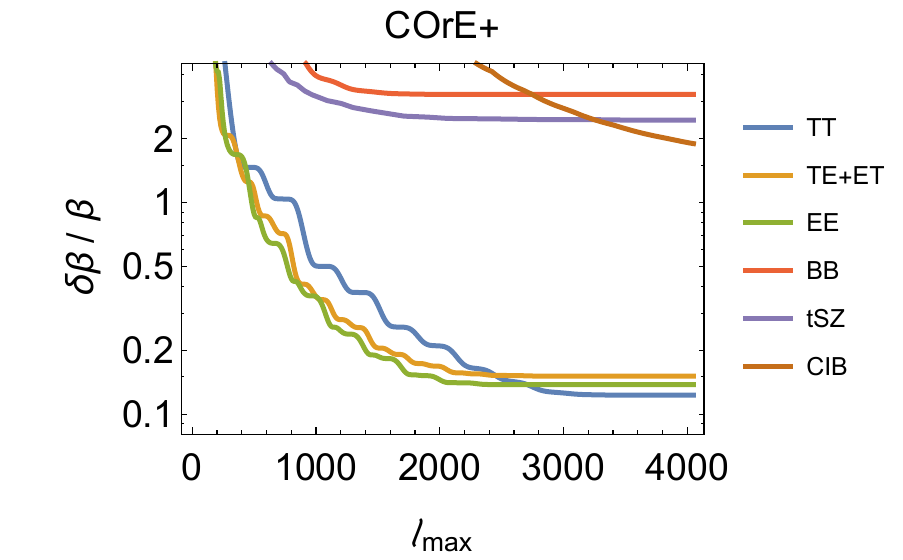}
    \hspace{-0.4cm}\includegraphics[width=0.51\textwidth]{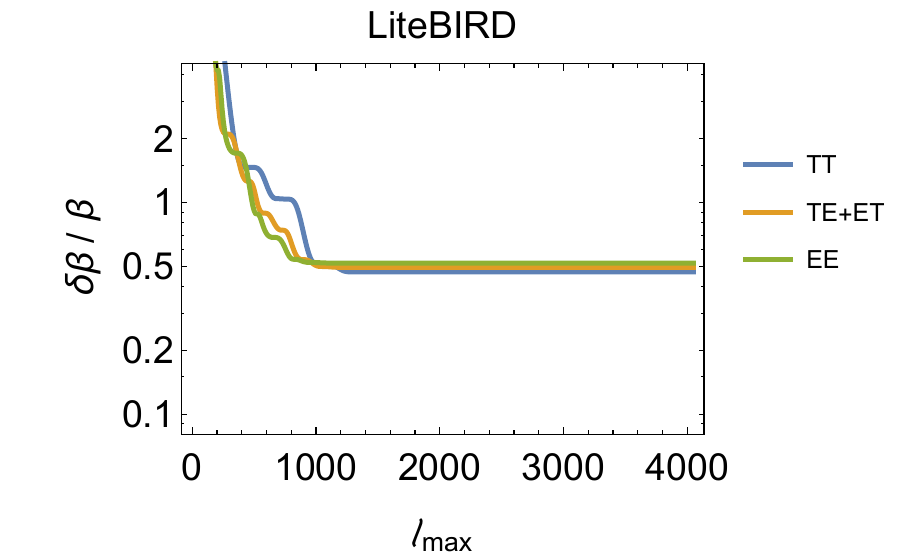}\\
    \hspace{-0.5cm}\includegraphics[width=0.51\textwidth]{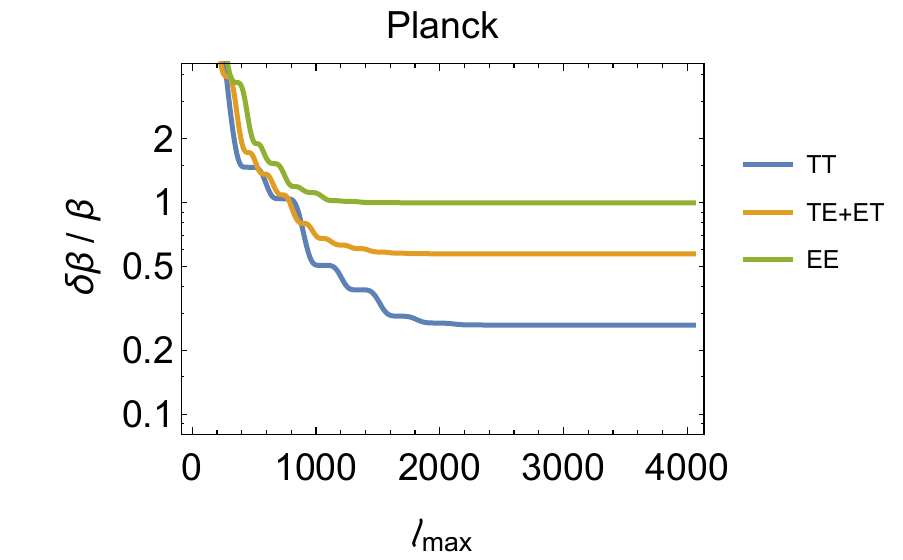}
    \hspace{-0.4cm}
    \includegraphics[width=0.51\textwidth]{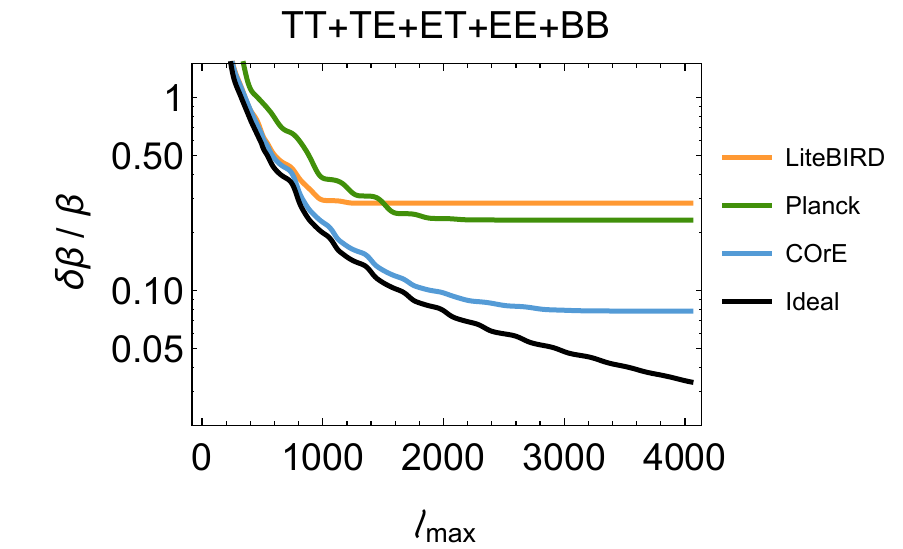}
    \caption{Similar to the left panel of Fig.~\ref{fig:SN-Ideal} but for realistic experiments (described in detail in Table~\ref{tab:CORE-bands}) and assuming $f_{\rm sky} = 0.8.$ In the bottom right panel we compare the total precision after combining all temperature and polarization maps, including also the case of an ideal experiment (no instrumental noise and $f_{\rm sky} = 1$). \label{fig:SN-Core}}
\end{figure}

In Fig.~\ref{fig:SN-Core} and Table~\ref{tab:ston-boost} we summarise our
forecasts for CORE and compare them with both {\it Planck\/} and LiteBIRD
forecasts.
These results differ from the ideal case due to the inclusion of instrumental
noise, foreground contamination (in the case of tSZ) and $f_{\rm sky} \neq 1.$
In the last panel we also show the total precision by combining all temperature
and polarization channels assuming a negligible correlation among them (which
was shown in Ref.~\cite{Amendola:2010ty} to be a good approximation).
Note also that the $TE$ and $ET$ correlation functions were shown to be
independent in Ref.~\cite{Amendola:2010ty} and both carry the same S/N. So we
usually present the combined S/N for $TE+ET$, which is $\sqrt{2}$ times their
individual S/N values.

\begin{table}[ht!]
{\small
\centering
\begin{tabular}{|c|ccc|cccc|}
\hline
\multirow{2}{*}{Experiment} & Channel & $\theta_{\rm{FWHM}}$ & $\sigma^{T}$ & S/N  & S/N & S/N & S/N \\
& [GHz] & [arcmin] & [$\mu$K.arcmin] & $TT$ & $TE+ET$ & $EE$ & Total \\
\hline\hline
{\it Planck}  & (all) & $\simeq 5.5$ & $\simeq 13$ & 3.8 & 1.7 & 1.0 & 4.3 \\
\hline
LiteBIRD & (all) & $\simeq 19$ & $\simeq 1.7$  & 2.0 & 1.8 & 1.8 & 3.3\\
\hline
\multirow{20}{*}{CORE}
   & 60 & 17.87 & 7.5 & 2.1 & 1.9 & 1.8 & 3.4 \\
   & 70 & 15.39 & 7.1 & 2.5 & 2.4 & 2.2 & 4.1 \\
   & 80 & 13.52 & 6.8 & 2.8 & 2.8 & 2.6 & 4.8 \\
   & 90 & 12.08 & 5.1 & 3.5 & 3.4 & 3.3 & 5.9 \\
   & 100 & 10.92 & 5 & 3.9 & 3.7 & 3.7 & 6.5 \\
   & 115 & 9.56 & 5 & 4.3 & 4.2 & 4.2 & 7.3 \\
   & 130 & 8.51 & 3.9 & 5.1 & 4.9 & 5. & 8.6 \\
   & 145 & 7.68 & 3.6 & 5.7 & 5.3 & 5.5 & 9.5 \\
   & 160 & 7.01 & 3.7 & 6.1 & 5.6 & 5.8 & 10.1 \\
   & 175 & 6.45 & 3.6 & 6.5 & 5.8 & 6.1 & 10.7 \\
   & 195 & 5.84 & 3.5 & 7.1 & 6.1 & 6.5 & 11.4 \\
   & 220 & 5.23 & 3.8 & 7.5 & 6.3 & 6.7 & 11.9 \\
   & 255 & 4.57 & 5.6 & 7.5 & 5.9 & 6.2 & 11.4 \\
   & 295 & 3.99 & 7.4 & 7.5 & 5.7 & 5.8 & 11. \\
   & 340 & 3.49 & 11.1 & 7. & 5.1 & 4.9 & 9.9 \\
   & 390 & 3.06 & 22 & 5.8 & 3.8 & 3.1 & 7.6 \\
   & 450 & 2.65 & 45.9 & 4.5 & 2.3 & 1.4 & 5.3 \\
   & 520 & 2.29 & 116.6 & 2.9 & 1. & 0.3 & 3.1 \\
   & 600 & 1.98 & 358.3 & 1.4 & 0.3 & 0. & 1.4 \\
\cline{2-8}
& (all) & $\simeq 4.5$ & $\simeq 1.4$ & 8.2 & 6.6 & 7.3 & 12.8 \\
\hline
Ideal  ($\ell_{\rm max}$ = 2000) & (all) & 0 & 0 & 5.3 & 7.1 & 8.7 & 12.7 \\
\hline
Ideal  ($\ell_{\rm max}$ = 3000) & (all) & 0 & 0 & 10 & 9.8 & 14 & 21 \\
\hline
Ideal  ($\ell_{\rm max}$ = 4000) & (all) & 0 & 0 & 16 & 11.4 & 19 & 29 \\
\hline
Ideal  ($\ell_{\rm max}$ = 5000) & (all) & 0 & 0 & 22 & 12.6 & 26 & 38 \\
\hline
\end{tabular}
\caption{\baselineskip=0.4cm{\bf Aberration and Doppler effects with CORE.} We assume $f_{\rm sky} = 0.8$ for all experiments (and $f_{\rm sky} = 1$ in the ideal cases) in order to make comparisons simpler. For CORE we assume the 1.2-m telescope configuration, but with extended mission time to match the 1.5-m noise in $\mu$K.arcmin. For CORE and LiteBIRD we assume $\sigma^{P}$ = $\sqrt{2}\sigma^{T}$, while for {\it Planck\/} we use the 2015 values. The combined channel estimates are effective values that best approximate Eq.~\eqref{eq:Nell-formula} in the $\ell$ range of interest. Note that CORE will have ${\rm S/N} \ge 5$ in 14 different frequency bands. Also, by combining all frequencies, CORE will have similar S/N in $TT$, $TE+ET$ and $EE$.
\label{tab:ston-boost}}
}
\end{table}

As a side note, since the estimators for $\left<a_{\ell m}^{X}~a_{(\ell+1)m}^{Y\ast}\right>$ involve a sum over all $\ell$s and $m$s and since $m$ enters through $G_{\ell m}$ only, it is useful to use the following approximations, which are valid to very good accuracy for $\ell \gtrsim 20$~\cite{Notari:2011sb,2016PhRvD..94d3006N}:
\begin{equation}\label{eq:Glm-simple}
    \sum_m G_{\ell, m} = 0.39 (2\ell +1)\,; \qquad   \sum_m \big[G_{\ell, m}\big]^2 = 0.408^2 (2\ell +1)\,.
\end{equation}
Although we did not use these approximations in our results, they yield up to 1\,\%-level accuracy and by allowing the sum over $m$s to be removed, they significantly simplify the calculation of the estimators.

The achievable precision in $\beta$ through this method depends strongly on the
shape of the power spectrum -- strongly varying spectra give much lower
uncertainties compared to smooth spectra. For instance, for the tSZ and CIB
maps, many modes are in the cosmic-variance-limited regime, thus one might
think that they would yield a good measurement of $\beta$. However, since their
$C_\ell$s are smooth functions of $\ell,$ they do not carry much information
on the boost. To understand this and gain some insight, we rewrite
Eq.~\eqref{eq:almcorr} by approximating $C_{\ell+1}$ as $C_{\ell} + \dd
C_{\ell}/\dd \ell$ and adding the approximation that $\dd C_{\ell}/\dd \ell \ll
C_{\ell}$ (note, however, that $ \ell \dd C_{\ell}/\dd \ell$ could be comparable
to $C_{\ell}$ at small scales). We thus find that
\begin{equation}\label{eq:almcorr5}
    \sum_m \left<a_{\ell m}^{X}~a_{(\ell+1)m}^{Y\ast}\right>
    = 0.39 (2\ell +1) \beta
    \left[ (2-2d) C_{\ell}^{XY} -(\ell+d) \frac{\dd C_{\ell}^{XY}}{\dd \ell}
    \right]\,.
\end{equation}
Assuming the cosmic-variance dominated regime (i.e., ${\mathfrak C}^{XX}_{\ell} \simeq C^{XX}_{\ell}$) for $\ell \gtrsim 20$ and putting $X=Y$, we find that
\begin{equation}\label{eq:delta-beta-simp3}
    \left. \frac{\delta \beta}{\beta}\right|_{XX} \,\simeq  \, \frac{1}{0.408\beta}\left[\sum_{\ell}  (2\ell+1)  \left[(2-2d) - \ell \left(1 - \frac{C_{\ell+1}^{XX}}{C_{\ell}^{XX}} \right) \right]^2
    \right]^{-\frac{1}{2}}.
\end{equation}
For the $TE$ case, the formula is less useful. For the CMB temperature and polarization ($d=1$), only the derivative term survives:
\begin{equation}\label{eq:delta-beta-simp2}
   \left. \frac{\delta \beta}{\beta}\right|_{XX=TT,EE,BB} \,\simeq  \, \frac{1}{0.408\beta}\left[\sum_{\ell}  (2\ell+1)  \left[\frac{\dd \ln C_{\ell}^{XX}}{\dd \ln \ell}\right]^2
    \right]^{-\frac{1}{2}}.
\end{equation}

Note that for the CIB the precision is smaller than for the CMB temperature and polarization, not only because the spectra are smoother, but also because there is a partial cancellation between the two terms in the
summand of Eq.~\eqref{eq:delta-beta-simp3}.

In this analysis we relied only on the diffuse background components of the
measured maps. Aberration and Doppler effects can in principle also be detected
using point sources, since the boosting effects will change both their number
counts, angular distribution, and redshift. For the upcoming CMB experiments,
however, the number density of point sources is probably insufficient for a
significant signal, since one needs more than about $10^6$ objects to have a
detection at greater than $1\,\sigma$ \cite{Yoon:2015lta}.

\section{Differential approach to CMB spectral distortions and the CIB}\label{sect:DiffCMB}

Using the complete description of the Compton-Getting effect \cite{1970PSS1825F} we compute full-sky maps of the expected effect at desired frequency.
We start discussing the frequency dependence of the dipole spectrum \cite{DaneseDeZotti1981, Balashev2015} and then extend the analysis beyond the dipole.

\subsection{The CMB dipole}\label{sect:DipCMB}

The dipole amplitude is directly proportional to the first derivative of the photon occupation number, $\eta(\nu)$, which is related to the thermodynamic temperature, $T_{\rm therm}(\nu)$, i.e., to the temperature of the blackbody having the same $\eta(\nu)$ at the frequency $\nu$, by
\begin{equation}
    T_{\rm therm}={h\nu\over k_{\rm B}\ln(1+1/\eta(\nu))}.
    \label{eq:t_therm}
\end{equation}
The difference in $T_{\rm therm}$ measured in the direction of motion and in the perpendicular direction is given by \cite{DaneseDeZotti1981}:
\begin{equation}\label{eq:DeltaTtherm}
    \Delta T_{\rm therm}={h\nu\over k}\left\{{1\over
    \ln\left[1+1/\eta(\nu)\right]}-{1\over \ln\left[1+1/\eta(\nu(1+\beta))\right]}   \right\} \, ,
\end{equation}
\noindent
which, to first order, can be approximated by:
\begin{equation}\label{eq:DeltaTtherm_firstord}
    \Delta T_{\rm therm}
    \simeq -{x \beta T_0\over (1+\eta)\ln^2(1+1/\eta)}{d\ln\eta\over d\ln x},
\end{equation}
\noindent
where
$x\equiv h\nu/kT_0$
is the dimensionless frequency.

\begin{figure}[ht!]
\vskip -2.cm
\hskip -0.7cm
    \includegraphics[width=0.62\textwidth]{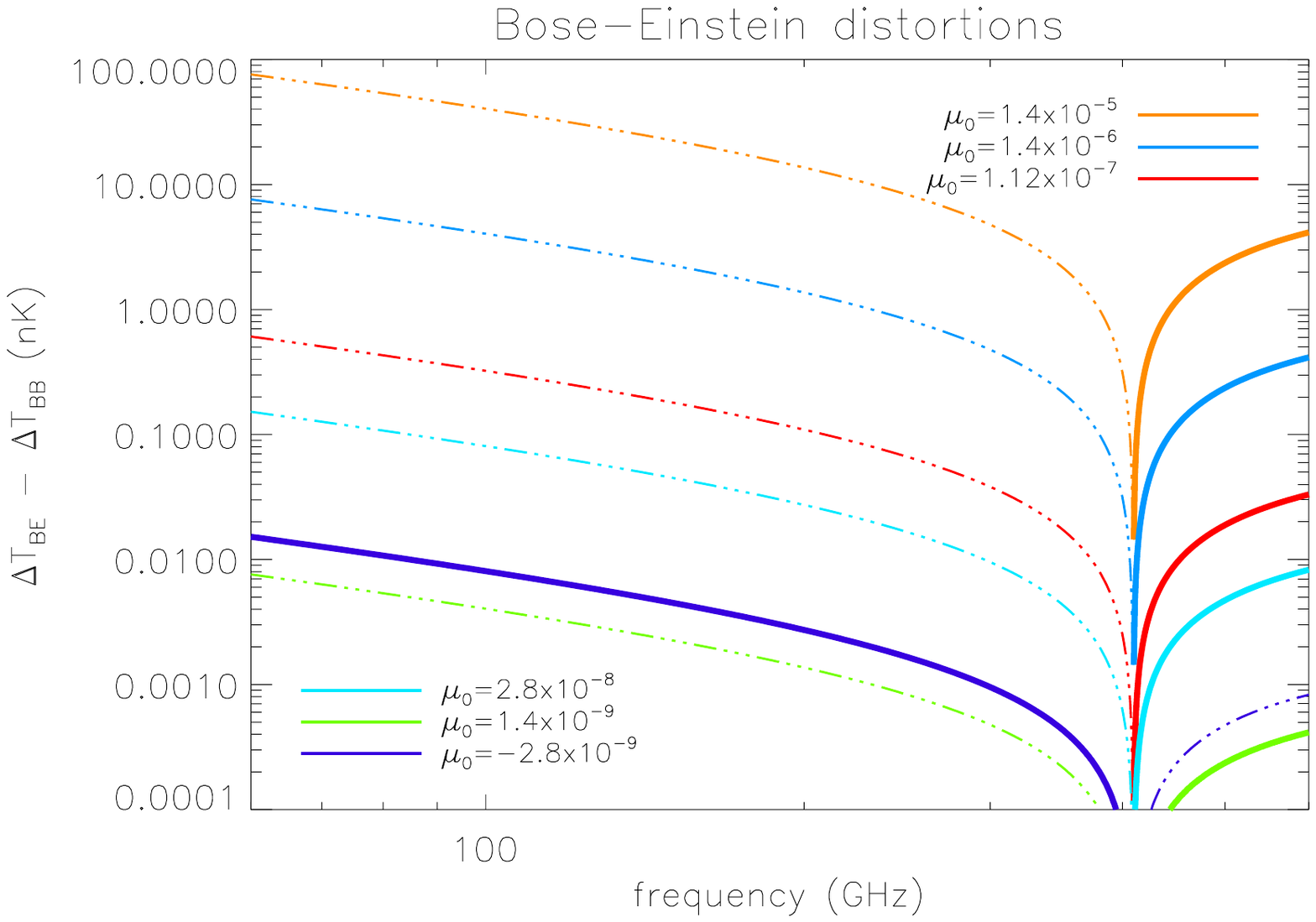}
\hskip -1.8cm
    \includegraphics[width=0.62\textwidth]{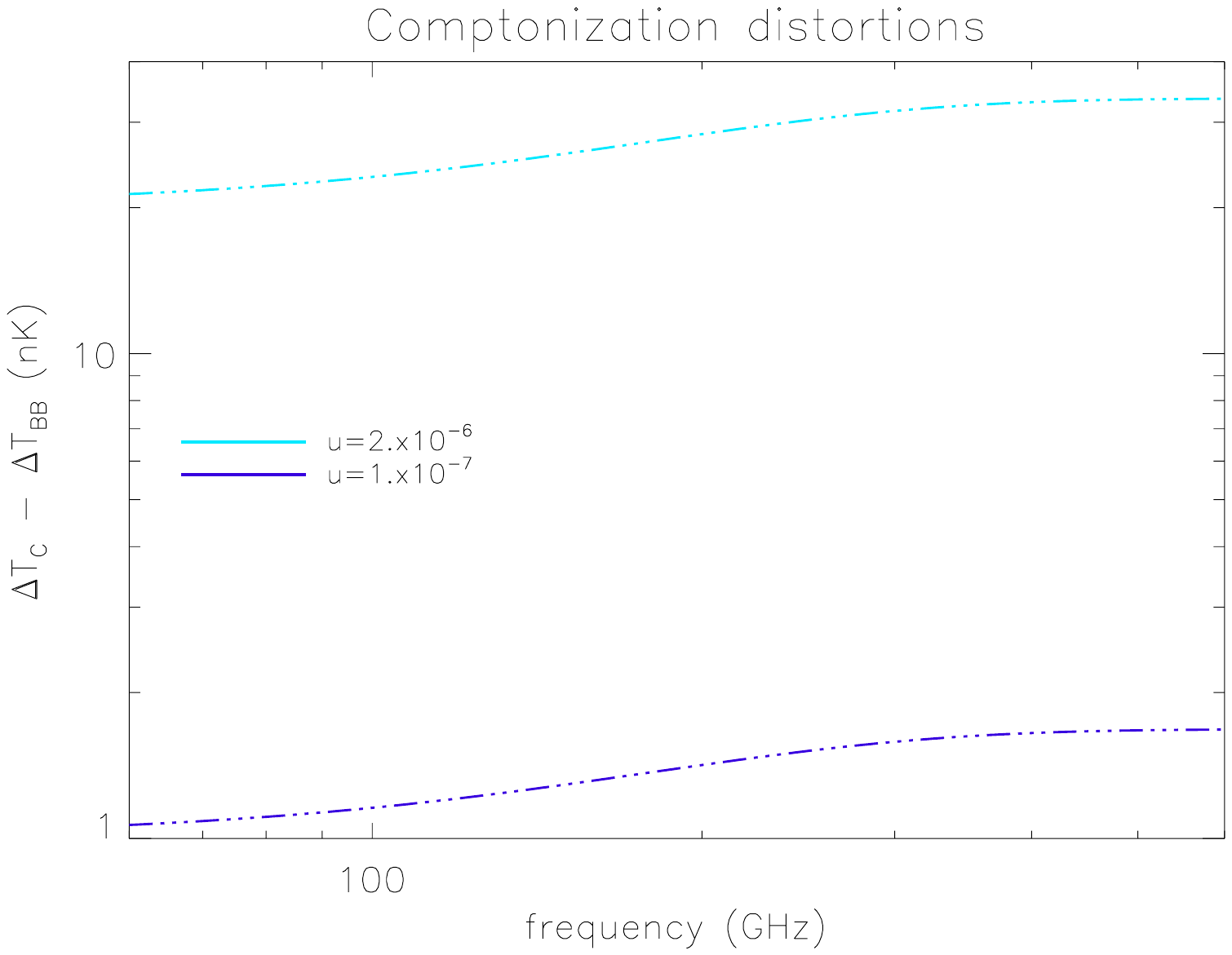}
\vskip -6.5cm
    \caption{Spectrum of dipole (in equivalent thermodynamic, or CMB,
    temperature) expressed as the difference between that produced by a
    distorted
    spectrum and that corresponding to the blackbody at the current temperature
    $T_0$. Thick solid lines (or thin three dots-dashes) correspond to positive (or
    negative) values. {\it Left}: the case of BE distortions for $\mu_0=
    -2.8\times10^{-9}$ (representative of adiabatic cooling; green dots, note
    the opposite signs with respect to the cases with positive $\mu_0$),
    $\mu_0= 1.4\times10^{-5}$, $1.4\times10^{-6}$ (representative of
    improvements
    with respect to FIRAS upper limits), $\mu_0= 1.12\times10^{-7}$,
    $2.8\times10^{-8}$, and $1.4\times10^{-9}$
    (representative of primordial adiabatic perturbation dissipation).
    {\it Right}: the case of Comptonization distortions for $u=2\times10^{-6}$
    (upper curves) and $u=10^{-7}$ (lower curves), representative of imprints
    by astrophysical or minimal reionization models, respectively.
}
    \label{fig:Dip_C_BE}
\end{figure}

In Fig.~\ref{fig:Dip_C_BE} we show the dipole spectrum derived for two well-defined deviations from the Planck distribution, namely the BE and Comptonization distortions
induced by unavoidable energy injections in the radiation field occurring at different cosmic times, early and late, respectively. We briefly discuss below their basic properties and the
signal levels expected from different processes.

A BE-like distorted spectrum is produced by two distinct processes.
Firstly there is the dissipation of primordial perturbations at small scales~\cite{1994ApJ...430L...5H, chlubasunyaev2012},
which generates a positive chemical potential.
Secondly we have Bose condensation of CMB photons by colder electrons, as a consequence of the faster decrease of the matter temperature relative to the radiation temperature in an expanding Universe,
which generates a negative chemical potential \cite{2012MNRAS.419.1294C,sunyaevkhatri2013}.

The photon occupation number of the BE spectrum is given by \cite{1970Ap&SS...7...20S}
\begin{equation}\label{eq:etaBE}
    \eta_{\rm BE}= {1\over e^{x_{\rm e}+\mu} -1},
\end{equation}
where $\mu$ is the chemical potential that quantifies the fractional energy,
$\Delta \epsilon/ \varepsilon_{\rm i}$, exchanged in the plasma during the
interaction,\footnote{Here, the subscript ${\rm i}$ denotes the initial time of the
dissipation process.} $x_{\rm e} = x / \phi (z)$, $\phi (z) = T_{\rm e}(z)/T_{\rm
CMB}(z)$, with $T_{\rm e}(z)$ being the electron temperature. For a BE spectrum,
$\phi = \phi_{\rm BE}(\mu)$. The dimensionless frequency $x$ is redshift
invariant, since in an expanding Universe both $T_{\rm CMB}$ and the physical
frequency $\nu$ scale as $(1+z)$. For small distortions, $\mu \simeq 1.4 \Delta
\epsilon/ \varepsilon_{\rm i}$ and $\phi_{\rm BE} \simeq (1-1.11\mu)^{-1/4}$. The
current FIRAS 95\,\% CL upper limit is $|\mu_0|<9 \times 10^{-5}$
\cite{1996ApJ...473..576F}, where $\mu_0$ is the value of $\mu$ at the redshift
$z_1$ corresponding to the end of the kinetic equilibrium era. At earlier times
$\mu$ can be significantly higher, and the ultimate limits on $\Delta \epsilon/
\varepsilon_{\rm i}$ before the thermalization redshift (when any distortion can be
erased) comes from cosmological nucleosynthesis.

These two kinds of distortions are characterised by a $|\mu_0|$ value in the
range, respectively, $\sim 10^{-9}$--$10^{-7}$ (and in particular $\simeq 2.52 \times 10^{-8}$ for a primordial
scalar perturbation spectral index $n_{\rm s}=0.96$, without running),
and $\simeq 3 \times 10^{-9}$. Since very small scales that are not explored by current CMB anisotropy data are relevant in this context, a broad set of primordial spectral indices needs to be explored.
A wider range of chemical potentials is found by \cite{chlubaal12}, allowing also for variations in the amplitude of primordial perturbations at very small scales, as motivated by some inflation models.

Cosmological reionization associated with the early stages of structure and star formation is an additional source of photon and energy production. This mechanism induces electron heating that is responsible for Comptonization distortions \cite{1972JETP...35..643Z}. The characteristic parameter for describing this effect is
\begin{equation}\label{eq:uComp}
u(t)=\int_{t_{\rm i}}^{t} [(\phi-\phi_{\rm i})/\phi] (k_{\rm B}T_{\rm e}/m_{\rm e}c^2) n_{\rm e} \sigma_{\rm T} c dt \, .
\end{equation}

\noindent In the case of small energy injections and integrating over the relevant epochs then
$u \simeq (1/4) \Delta\varepsilon/\varepsilon_{\rm i}$.
In Eq.~\eqref{eq:uComp},
$\phi_{\rm i} = \phi(z_{\rm i}) = (1+ \Delta \epsilon/ \varepsilon_{\rm i})^{-1/4} \simeq 1-u$ is
the ratio between the equilibrium matter temperature and the radiation temperature evaluated at the beginning of the heating process (i.e., at $z_{\rm i}$).
The distorted spectrum is then
\begin{equation}\label{eq:etaC}
\eta_{\rm C} \simeq \eta_{\rm i} + u {x / \phi_{\rm i} {\rm exp}(x/\phi_{\rm i}) \over [{\rm exp}(x/\phi_{\rm i}) - 1]^{2}} \left ( {x/\phi_{\rm i} \over {\rm tanh}(x/2\phi_{\rm i}) - 4} \right ),
\end{equation}

\noindent where $\eta_{\rm i}$ is the initial photon occupation number (before the energy injection).\footnote{Here
and in Eq.~\eqref{eq:etaBE} we neglect the effect of photon emission/absorption processes,
which is instead remarkable at low frequencies (see \cite{1980A&A....84..364D} and \cite{1995A&A...303..323B}).}

Typically, reionization induces Comptonization distortions with {\it minimal} values $u \simeq 10^{-7}$
\cite{buriganaetal08}.
In addition to this, the variety of energy injections expected in astrophysical
reionization models, including: energy produced by nuclear reactions in stars
and/or
by nuclear activity that mechanically heats the intergalactic medium (IGM); super-winds from supernova explosions
and active galactic nuclei; IGM heating by quasar radiative energy; and shocks associated with structure formation.  Together these induce much larger values of
$u$ ($\simeq \hbox{several}\times 10^{-6}$) \cite{2000PhRvD..61l3001R, 2015PhRvL.115z1301H}, i.e., not much below the current FIRAS 95\,\% CL upper limit of
$|u|<1.5\times 10^{-5}$ \cite{1996ApJ...473..576F}. Free-free distortions associated with reionization \cite{2014MNRAS.437.2507T}
are instead more relevant at the lowest frequencies (below $10\,$GHz), and thus we do not consider them in this paper.

We could also consider the possible presence of unconventional heating sources.
Decaying and annihilating particles during the pre-recombination epoch
may affect the CMB spectrum, with the exact distorted shape
depending on the process timescale and, in some cases, being different
from the one produced by energy release.  This is especially interesting
for decaying particles with lifetimes $t_{X} \simeq $\,few$\times 10^{8}$--$10^{11}$\,sec \cite{1993PhRvL..70.2661H, daneseburigana94, 2013MNRAS.436.2232C}.
Superconducting cosmic strings would also produce copious
electromagnetic radiation, creating CMB spectral distortion shapes \cite{ostrikerthompson87}
that would be distinguishable with high accuracy measurements.
Evaporating primordial black holes provide another possible source of energy injection, with the shape of the resulting distortion depending on the black hole mass function \cite{carretal2010}.
CMB spectral distortion measurements could also be used to constrain the spin of non-evaporating black holes \cite{paniloeb2013}.
The CMB spectrum could additionally set constraints on the power spectrum of small-scale magnetic fields \cite{jedamziketal2000},
the decay of vacuum energy density \cite{BartlettSilk1990}, axions \cite{ejllidolgov2014}, and other new physical processes.

\subsection{The CIB dipole}\label{sect:DipCIB}

Multi-frequency measurements of the dipole spectrum will allow us to constrain
the CIB intensity spectrum \cite{DaneseDeZotti1981,Balashev2015}.
The spectral shape of the CIB is hard to determine directly because it requires
absolute intensity measurements, which are also compromised by Galactic and
other foregrounds.  Although the dipole amplitude is about $10^{-3}$ of the
monopole, its spatial form is already known and hence this indirect route may
provide the most robust measurements of the CIB in the future.

\begin{figure}[ht!]
\begin{center}
\vskip -1.5cm
\includegraphics[width=0.6\columnwidth]{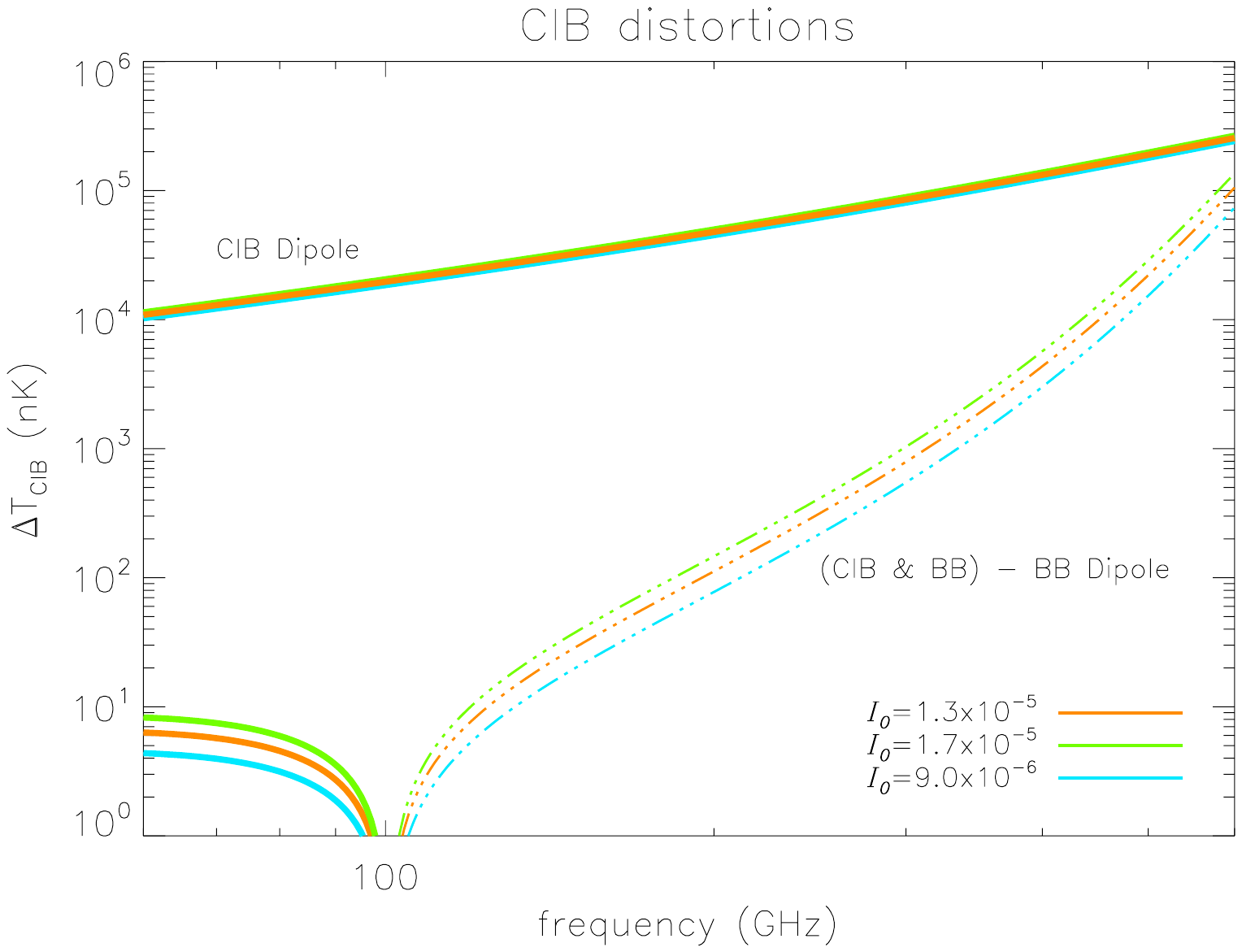}
\vskip -6.cm
\caption{Expected behaviour of the dipole spectrum.  The upper lines show the
spectrum of the (pure) CIB dipole, while the lower lines show the spectrum
coming from the dipole pattern computed from the CIB distribution function
added to the blackbody (at temperature $T_0$) distribution function, minus
the dipole pattern computed by the blackbody distribution function.
Thick solid lines (or thin three dots-dashes) correspond to positive (or negative) values.
The analytic representation of the CIB spectrum by \cite{Fixsen:1998kq} is
adopted here, considering the best-fit amplitude and the range
of $\pm 1\,\sigma$.}
\label{fig:dipole}
\end{center}
\end{figure}

Fig.~\ref{fig:dipole} shows the CIB dipole spectrum computed according to
Eq.~\eqref{eq:DeltaTtherm}, using the analytic representation of the CIB
spectrum (observed at present time) given in Ref.~\cite{Fixsen:1998kq}:
\begin{equation}\label{eq:eta_CIB}
\eta_{\rm CIB}={c^2\over 2 h \nu^3} I_{\rm CIB}(\nu)
= I_0 \left({k_{\rm B}T_{\rm CIB} \over h \nu_0}\right)^{k_F} {x^{k_{\rm F}}_{\rm CIB}\over \exp(x_{\rm CIB})-1}\, ,
\end{equation}
where $T_{\rm CIB}=(18.5\pm1.2)\,$K, $x_{\rm CIB}=h\nu/k_{\rm B}T_{\rm CIB}=
7.78(\nu/\nu_0)$, $\nu_0\simeq 3\times 10^{12}\,$Hz and $k_{\rm F}= 0.64 \pm
0.12$. Here $I_0$ sets the CIB spectrum amplitude, its best-fit value being
$1.3 \times 10^{-5}$ \citep{Fixsen:1998kq}. On the other hand, the uncertainty
of the CIB amplitude is currently quite high, with $I_0$ 
only known to a 1\,$\sigma$ accuracy of about 30\,\%.

The CIB dipole amplitude, in terms of thermodynamic temperature, increases
rapidly with frequency, reaching $257\,\mu$K (or
$652\,\hbox{Jy}\,\hbox{sr}^{-1}$) at 600\,GHz and $420\,\mu$K (or
$1306\,\hbox{Jy}\,\hbox{sr}^{-1}$) at 800\,GHz.  The measurement of the
CIB dipole amplitude will be dependent on systematic effects from the
foreground Galaxy subtraction, which has a similar spectrum to the CIB
\cite{2011ApJ...734...61F}. Although the calibration of the dipole signal at
different frequencies is not trivial (since the orbital part of the dipole
will be used for calibration), the {\it Planck\/} experience is that with
sufficient care the limitation is removal of the Galactic signals, not
calibration uncertainty.  Hence the CIB dipole should be clearly
detectable by \mission in its highest frequency bands. Such a detection will
provide important constraints on the CIB intensity;
its amplitude uncertainty constitutes a major current limitation in our
understanding of the dust-obscured star-formation phase of galaxy evolution.

\begin{figure}
\minipage{0.32\textwidth}
\includegraphics[width=\linewidth]{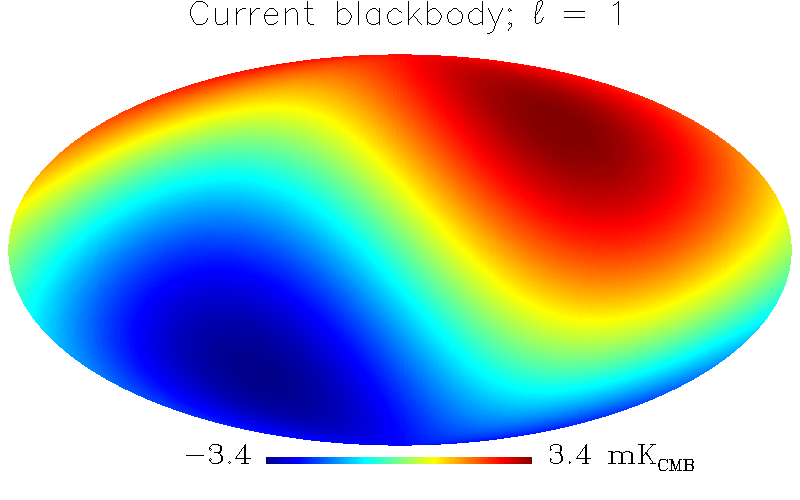}
\endminipage\hfill
\minipage{0.32\textwidth}
\includegraphics[width=\linewidth]{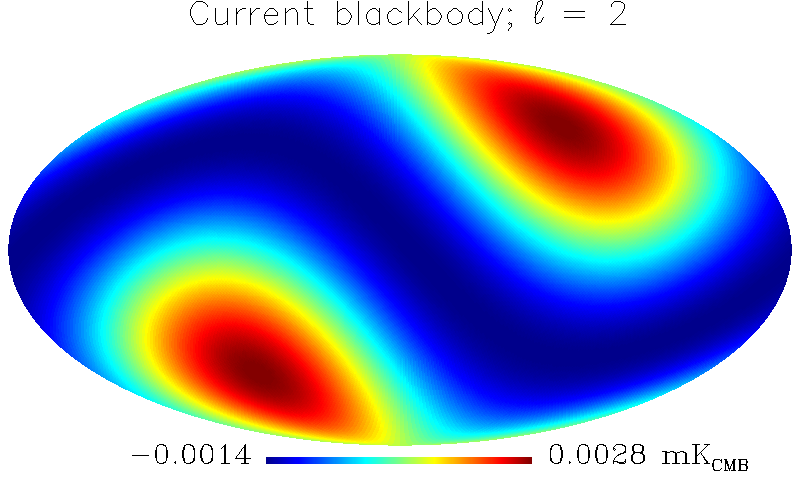}
\endminipage\hfill
\minipage{0.32\textwidth}
\includegraphics[width=\linewidth]{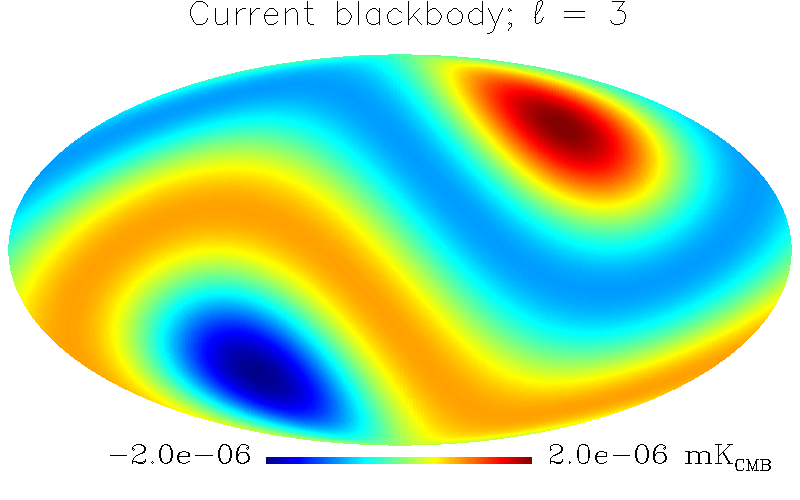}
\endminipage\hfill
\minipage{0.32\textwidth}
\includegraphics[width=\linewidth]{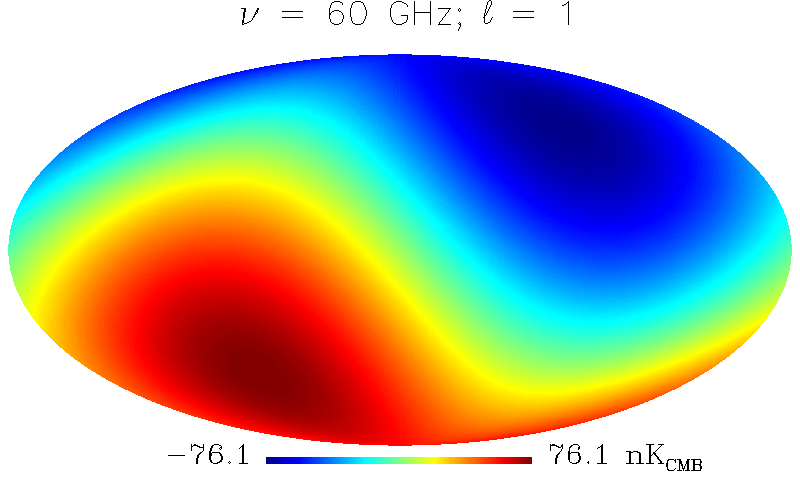}
\endminipage\hfill
\minipage{0.32\textwidth}
\includegraphics[width=\linewidth]{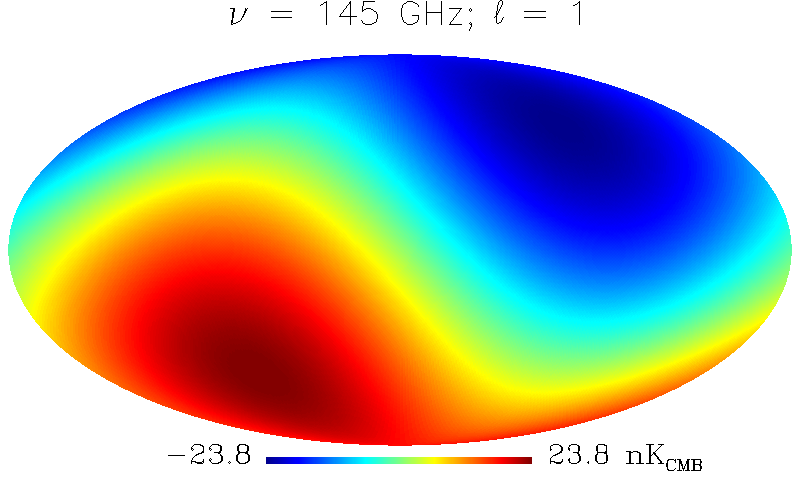}
\endminipage\hfill
\minipage{0.32\textwidth}
\includegraphics[width=\linewidth]{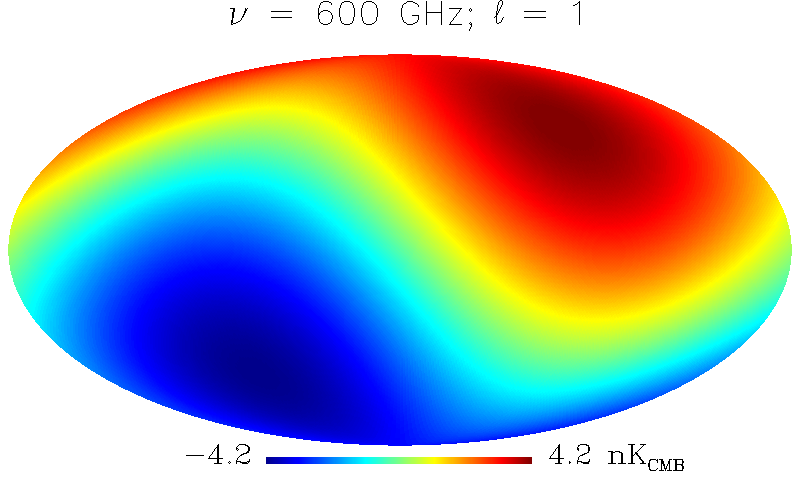}
\endminipage\hfill
\minipage{0.32\textwidth}
\includegraphics[width=\linewidth]{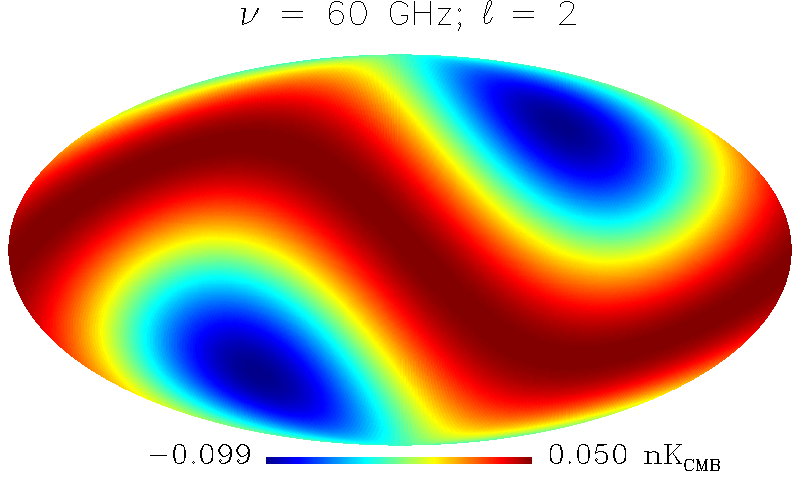}
\endminipage\hfill
\minipage{0.32\textwidth}
\includegraphics[width=\linewidth]{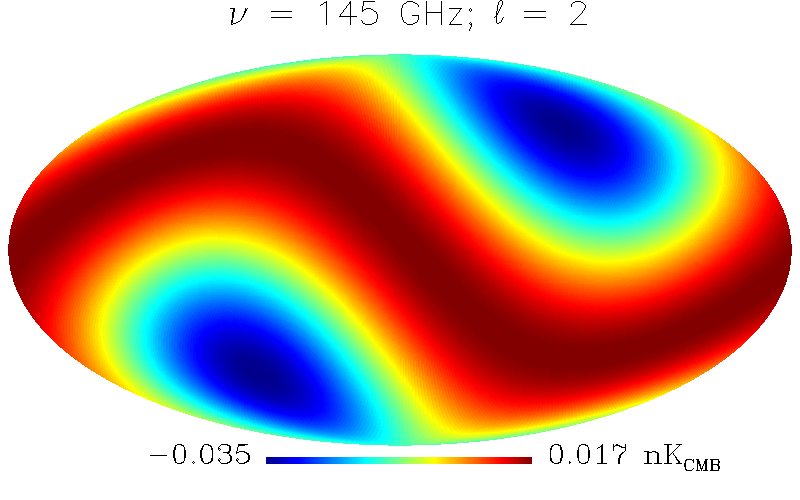}
\endminipage\hfill
\minipage{0.32\textwidth}
\includegraphics[width=\linewidth]{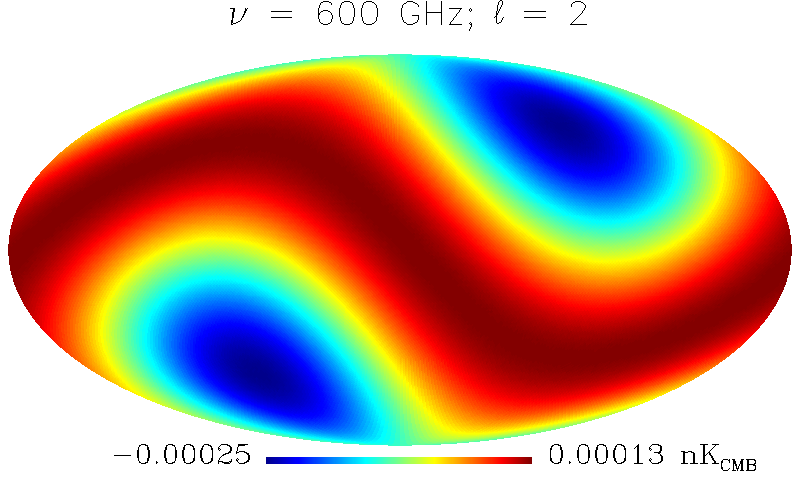}
\endminipage\hfill
\minipage{0.32\textwidth}
\includegraphics[width=\linewidth]{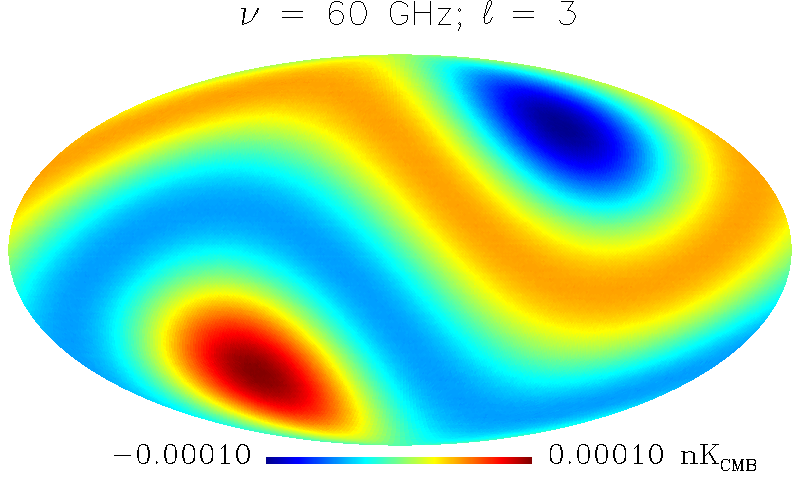}
\endminipage\hfill
\minipage{0.32\textwidth}
\includegraphics[width=\linewidth]{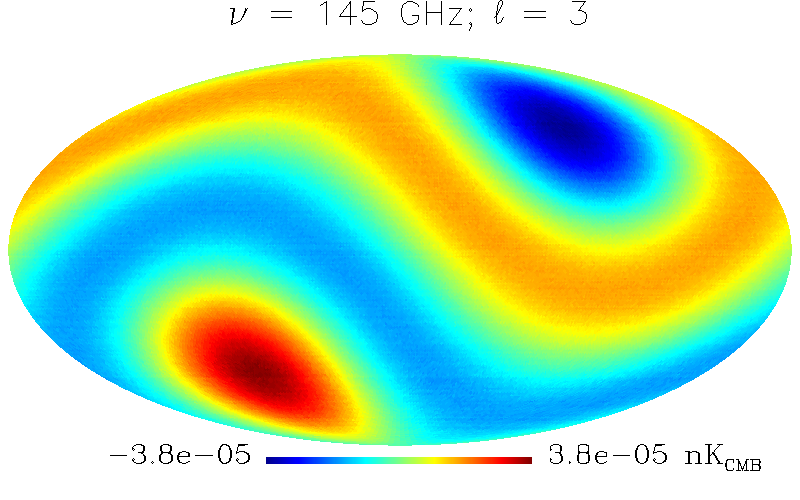}
\endminipage\hfill
\minipage{0.32\textwidth}
\includegraphics[width=\linewidth]{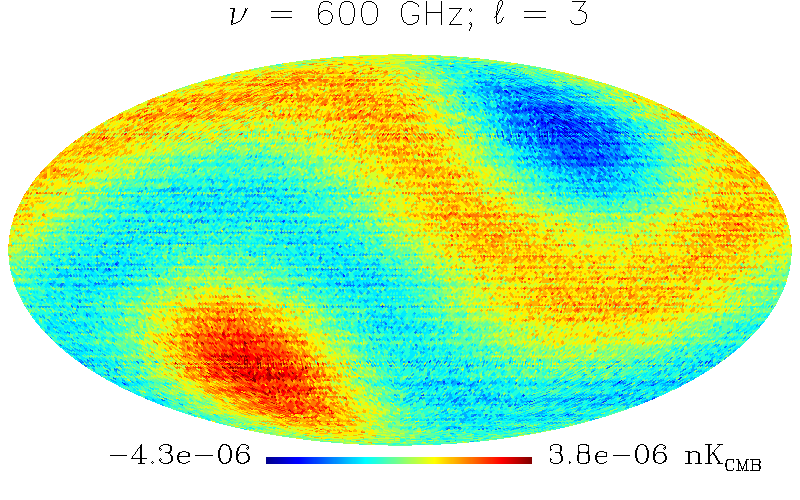}
\endminipage\hfill
\caption{{\it Top row}: maps of the dipole, quadrupole and octupole computed
assuming a CMB blackbody spectrum at the current temperature $T_0$, for
reference.  In all other cases we show the maps of the dipole ({\it second row}), quadrupole ({\it third row}), and octupole ({\it bottom row}) at three different frequencies
(namely 60, 145, and 600\,GHz, from left to right), in terms of the difference
between the pattern computed for a BE distortion with
$\mu_0 = 1.5 \times 10^{-5}$ and that computed for a blackbody at the
present-day temperature $T_0$.}
\label{fig:map_BE}
\end{figure}

\begin{figure}
\minipage{0.32\textwidth}
\includegraphics[width=\linewidth]{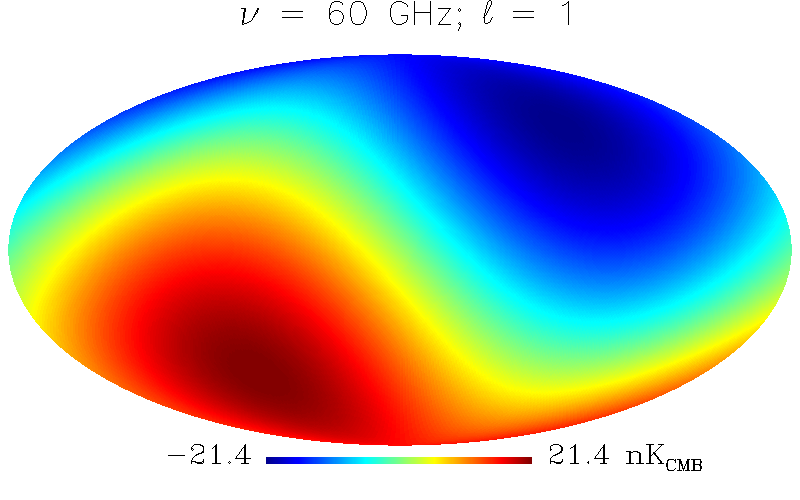}
\endminipage\hfill
\minipage{0.32\textwidth}
\includegraphics[width=\linewidth]{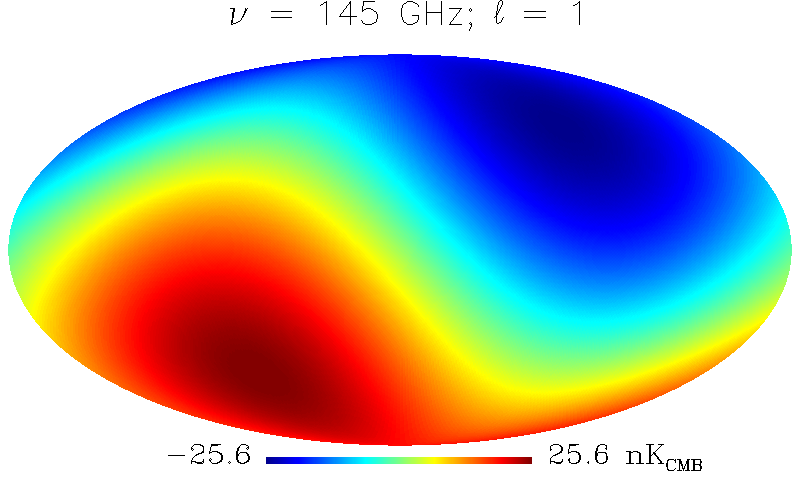}
\endminipage\hfill
\minipage{0.32\textwidth}
\includegraphics[width=\linewidth]{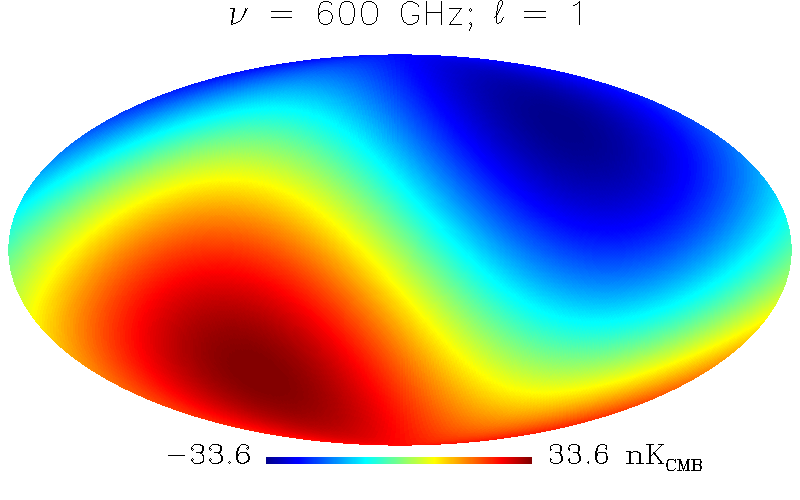}
\endminipage\hfill
\minipage{0.32\textwidth}
\includegraphics[width=\linewidth]{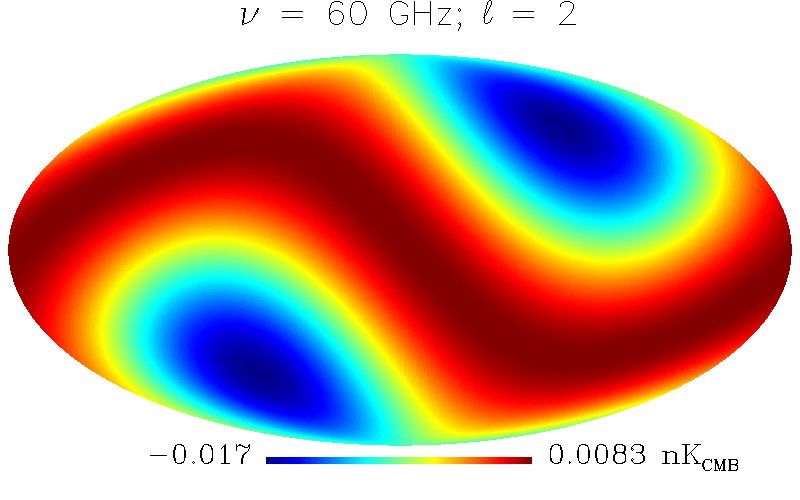}
\endminipage\hfill
\minipage{0.32\textwidth}
\includegraphics[width=\linewidth]{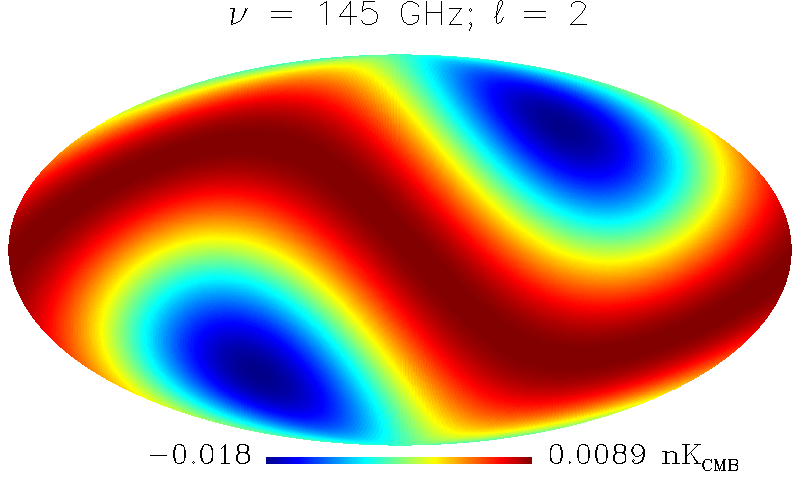}
\endminipage\hfill
\minipage{0.32\textwidth}
\includegraphics[width=\linewidth]{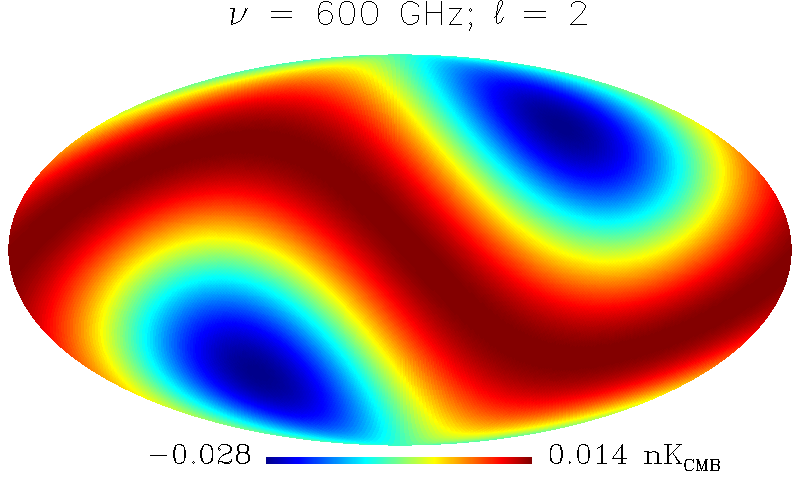}
\endminipage\hfill
\minipage{0.32\textwidth}
\includegraphics[width=\linewidth]{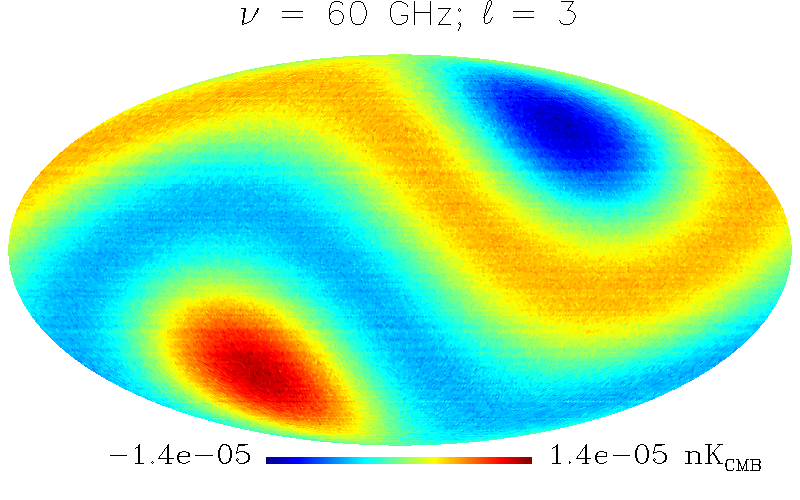}
\endminipage\hfill
\minipage{0.32\textwidth}
\includegraphics[width=\linewidth]{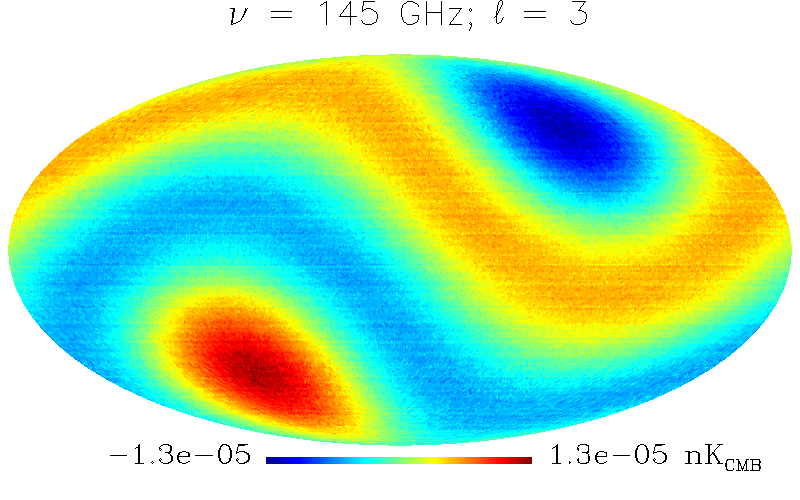}
\endminipage\hfill
\minipage{0.32\textwidth}
\includegraphics[width=\linewidth]{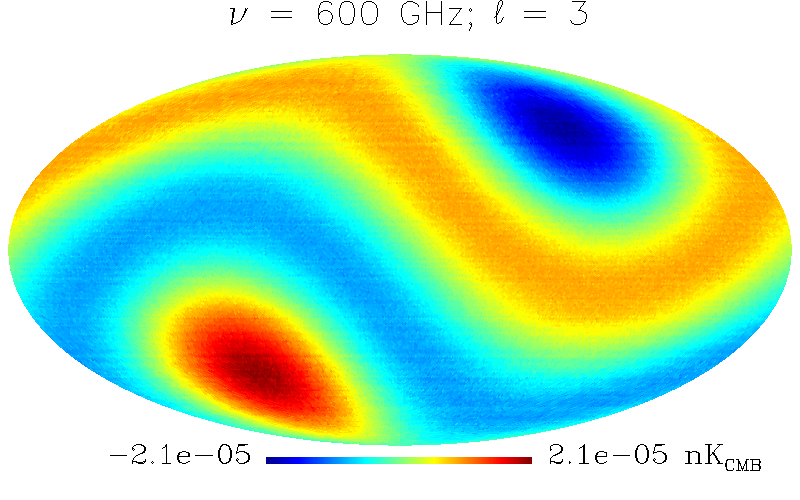}
\endminipage\hfill
\caption{The same as in Fig.~\ref{fig:map_BE}, but for the case of a
Comptonization distortion with $ u = 2 \times 10^{-6}$.}
\label{fig:map_C}
\end{figure}

\begin{figure}
\minipage{0.32\textwidth}
\includegraphics[width=\linewidth]{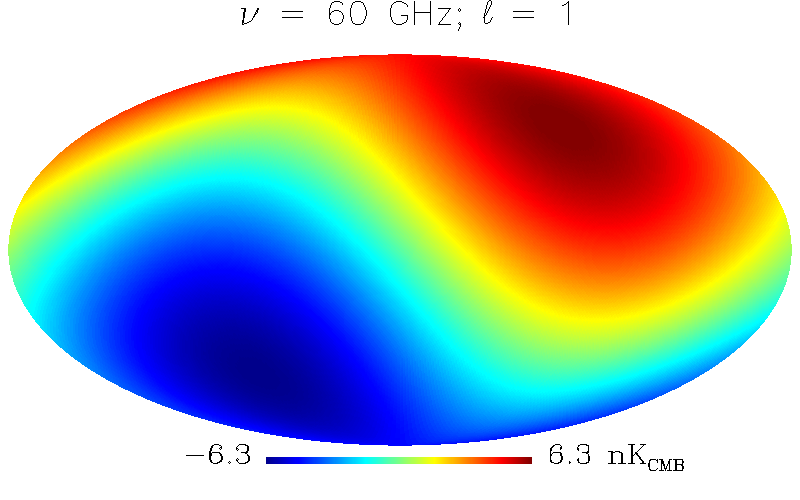}
\endminipage\hfill
\minipage{0.32\textwidth}
\includegraphics[width=\linewidth]{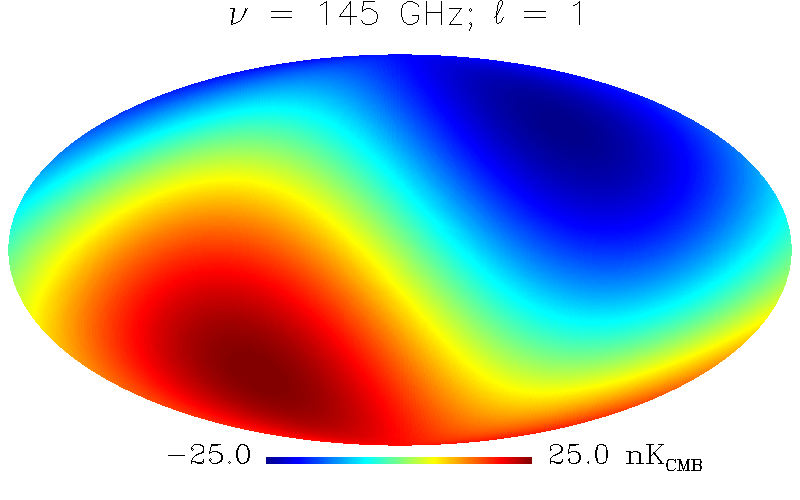}
\endminipage\hfill
\minipage{0.32\textwidth}
\includegraphics[width=\linewidth]{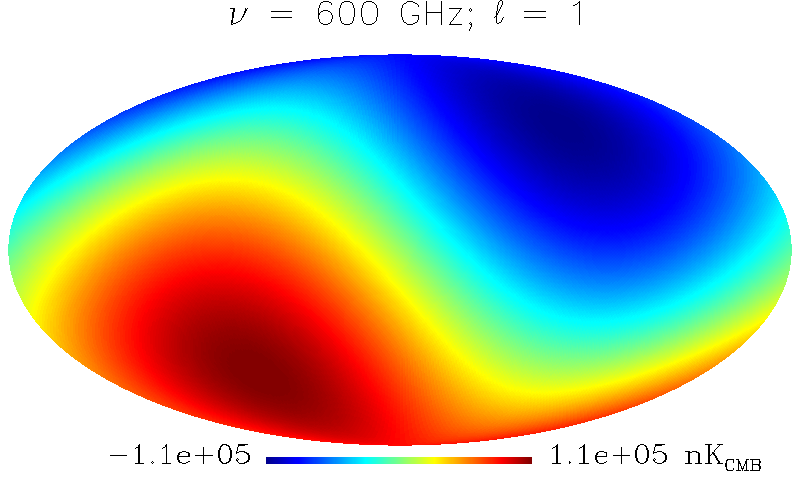}
\endminipage\hfill
\minipage{0.32\textwidth}
\includegraphics[width=\linewidth]{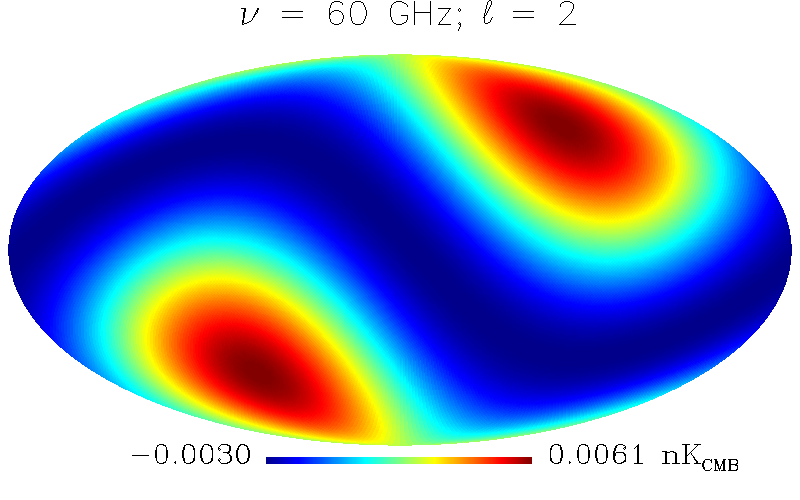}
\endminipage\hfill
\minipage{0.32\textwidth}
\includegraphics[width=\linewidth]{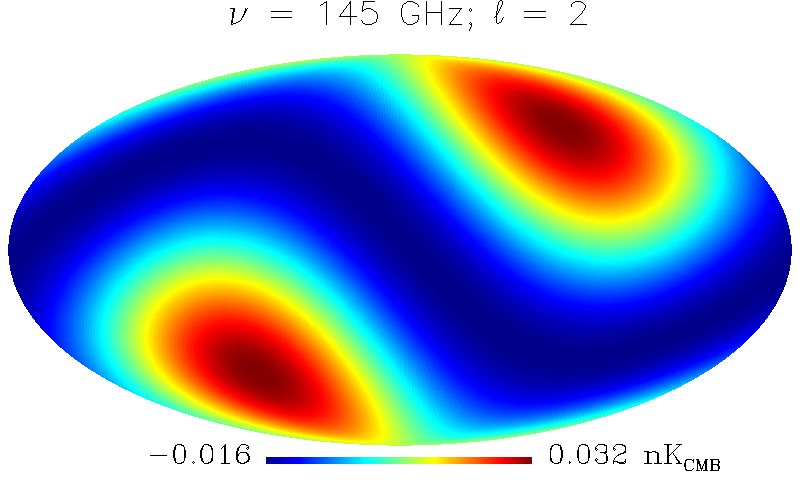}
\endminipage\hfill
\minipage{0.32\textwidth}
\includegraphics[width=\linewidth]{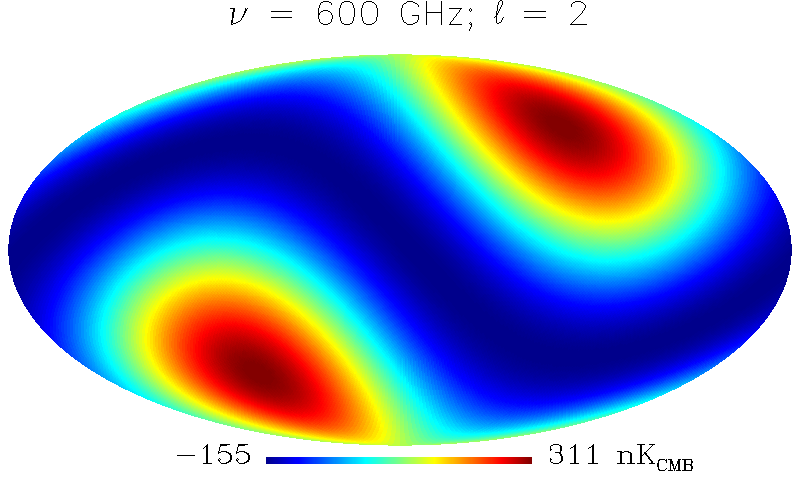}
\endminipage\hfill
\minipage{0.32\textwidth}
\includegraphics[width=\linewidth]{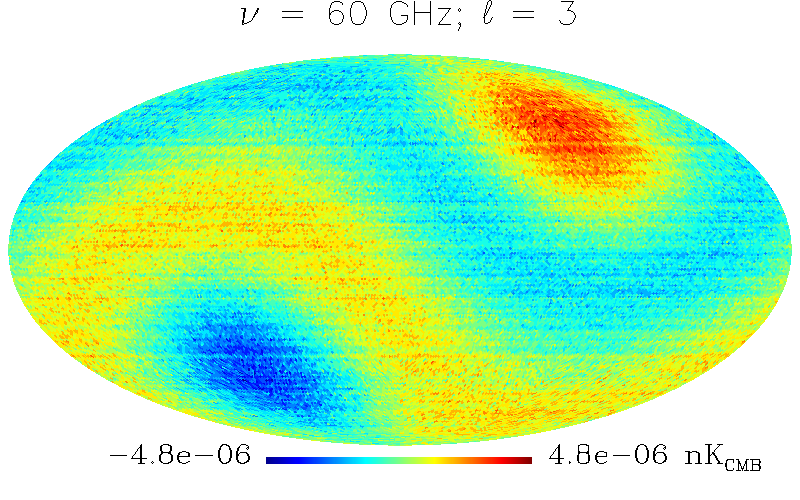}
\endminipage\hfill
\minipage{0.32\textwidth}
\includegraphics[width=\linewidth]{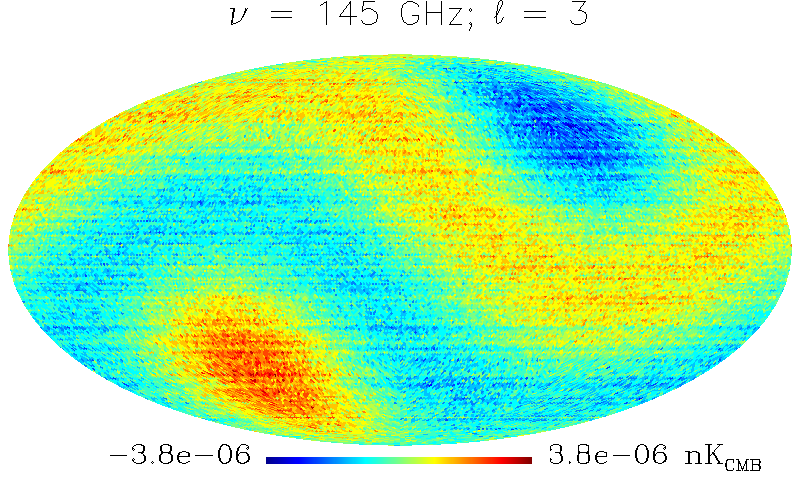}
\endminipage\hfill
\minipage{0.32\textwidth}
\includegraphics[width=\linewidth]{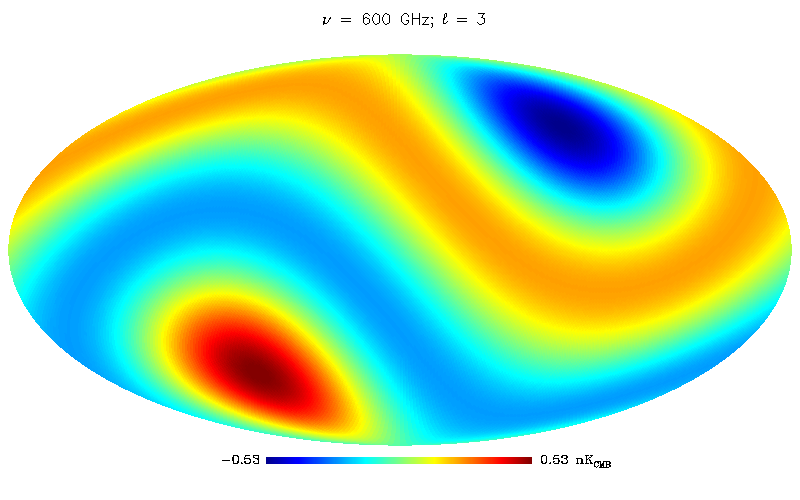}
\endminipage\hfill
\caption{The same as in Fig.~\ref{fig:map_BE}, but for the case of the CIB with amplitude set at the best-fit value found by FIRAS.
More precisely, we display the temperature pattern of the CIB distribution function added to the blackbody one, minus the temperature pattern coming from the blackbody.}
\label{fig:map_CIB}
\end{figure}

\subsection{Beyond the dipole}
\label{sect:LowEll}

A generalization of the considerations of the
previous section allows us to evaluate the effect of
peculiar velocity on the whole sky. To achieve this, we generate maps and,
using the Lorentz-invariance of the distribution function, we can include all
orders of the effect, coupling them with the geometrical properties induced at
low multipoles. To compute the maps at each multipole\footnote{For the sake of
generality and for the purpose of cross-checking, we also include the monopole
term, which can be
easily subtracted afterwards.} $\ell \ge 1$, we first derive the maps at all
angular scales, both for the distorted spectra and for the blackbody at the
current temperature $T_0$. 
From the dipole direction found in the {\it Planck} (HFI+LFI combined) 2015 release
and defining the motion vector of the observer, we produce the maps in a pixelization scheme at a given observational frequency $\nu$ by computing the photon distribution function, $\eta^{\rm BBdist}$, 
for each considered type of spectrum at a frequency given by the observational frequency $\nu$ but multiplied by the product $(1 - \hat{n} \cdot \vec{\beta})/(1 - \vec{\beta}^2)^{1/2}$ 
to account for all the possible sky directions with respect to the observer peculiar velocity. Here the notation `BBdist' stands for BB, CIB, BE, or Comptonization (C).
Hence, the map of the observed signal in terms of thermodynamic temperature is given by generalising Eq.~\eqref{eq:t_therm}:

\begin{equation}
T_{\rm therm}^{\rm BB/dist} (\nu, {\hat{n}}, \vec{\beta}) =
\frac{xT_{0}} {{\rm{log}}(1 / (\eta(\nu, {\hat{n}}, \vec{\beta}))^{\rm BB/dist}  + 1) } \, ,
\label{eq:eta_boost}
\end{equation}

\noindent where $\eta(\nu, {\hat{n}}, \vec{\beta}) = \eta(\nu')$ with
$\nu' = \nu (1 - {\hat{n}} \cdot \vec{\beta})/(1 - \vec{\beta}^2)^{1/2}$.

We adopt the {\tt HEALPix} pixelization scheme to discretise the sky at the desired
resolution. We decompose the maps into spherical harmonics and
then regenerate them considering the $a_{lm}$ only up to a desired multipole $\ell_{\rm{max}}$.
We start setting $\ell_{\rm{max}}= 5$ and then iterate the process with a decreasing $\ell_{\rm{max}}$.
We produce maps containing the power at a single multipole by taking the difference of the map at
$\ell_{\rm{max}}$ from the map at $\ell_{\rm{max}-1}$. We then compute the
difference of maps having specific spectral distortions from the purely
blackbody
maps. As seen in Figs.~\ref{fig:map_BE}--\ref{fig:map_CIB}, the expected
signal is important for the dipole, can be considerable for the quadrupole and,
depending on the distortion parameters, still not negligible for the octupole
(although this will depend on the amplitude relative to experimental noise
levels, as we discuss below).  For higher-order multipoles, the signal
is essentially negligible.

Note that the maps present a clear and obvious symmetry with respect to the
axis of the observer's peculiar velocity.\footnote{For real experiments, these
patterns are weakly modulated (and their perfect symmetry broken) by the
second-order (`orbital dipole') effect coming from the Earth's motion around
the Sun and (for spacecraft moving around the Earth-Sun L2 point), by the
further contribution from motion in the
Lissajous orbit.} This is simply due to the angular dependence in
Eq.~\eqref{eq:eta_boost}. For coordinates in which the positive $z$-axis is
aligned with the dipole, the only angular dependence comes from
$\hat{n}\cdot\vec{\beta} \equiv \beta \cos{\theta_{\rm d}}$.
In terms of the spherical harmonic expansion, this implies that higher-order
multipoles will appear as polynomial functions of $\cos{\theta_{\rm d}}$, with
different frequency-dependent factors depending on the specific type of
spectral distortion being considered.

In the above considerations we assumed that each multipole pattern can be isolated from that of the other multipoles. In reality, a certain leakage is expected (particularly between adjacent multipoles), especially as a result of masking for foregrounds.  The sources of astrophysical emission are highly complex, and their geometrical properties mix with their frequency behaviour. Furthermore, in real data analysis, there is an interplay between the determination of the calibration and zero levels of the maps, and this issue is even more critical when data in different frequency domains are used to improve the component-separation process. The analysis of these aspects is outside the scope of the present paper, but deserves further investigation.

\begin{figure}[ht!]
    \includegraphics[width=0.49\textwidth]{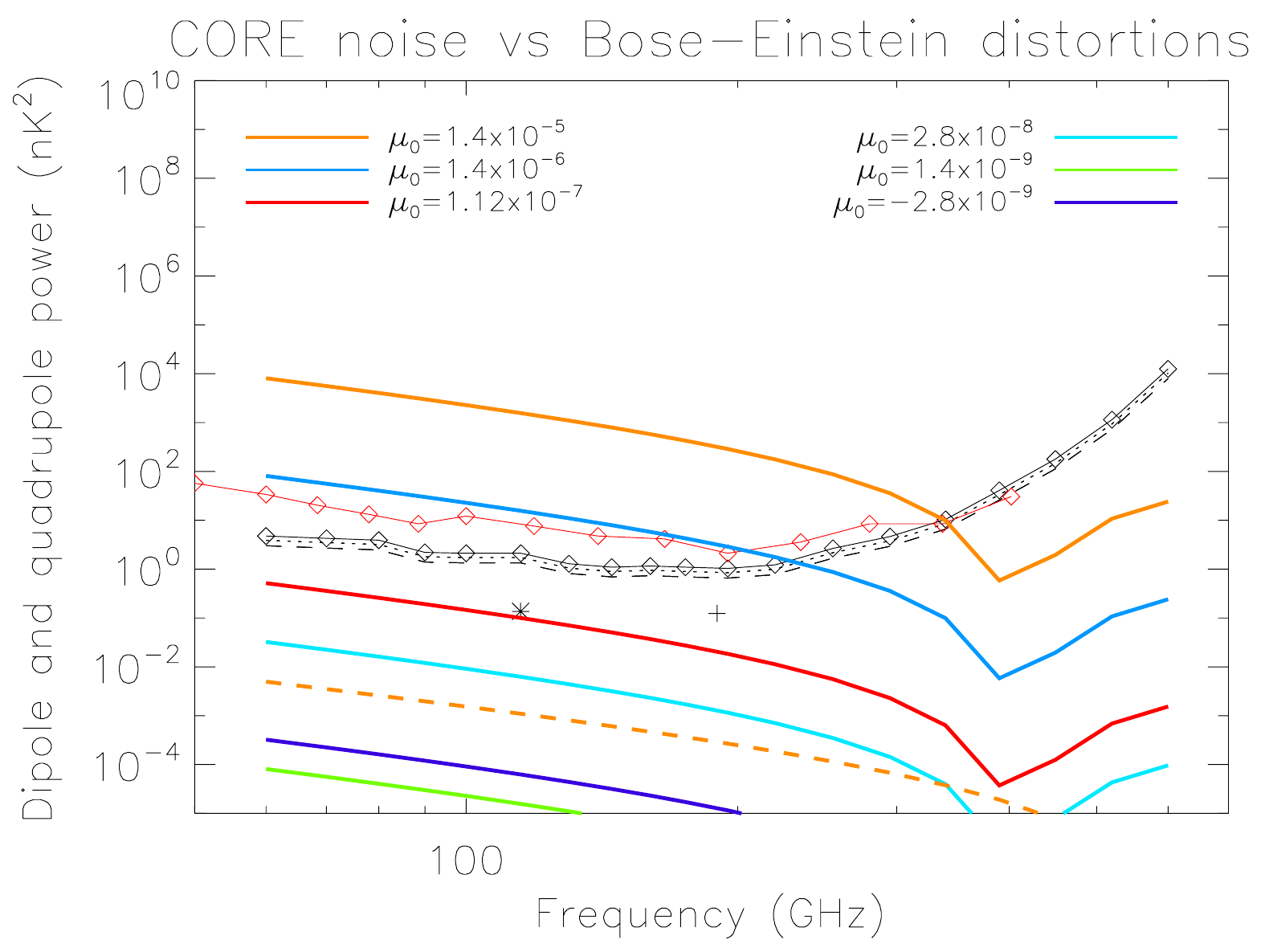}
    \includegraphics[width=0.496\textwidth]{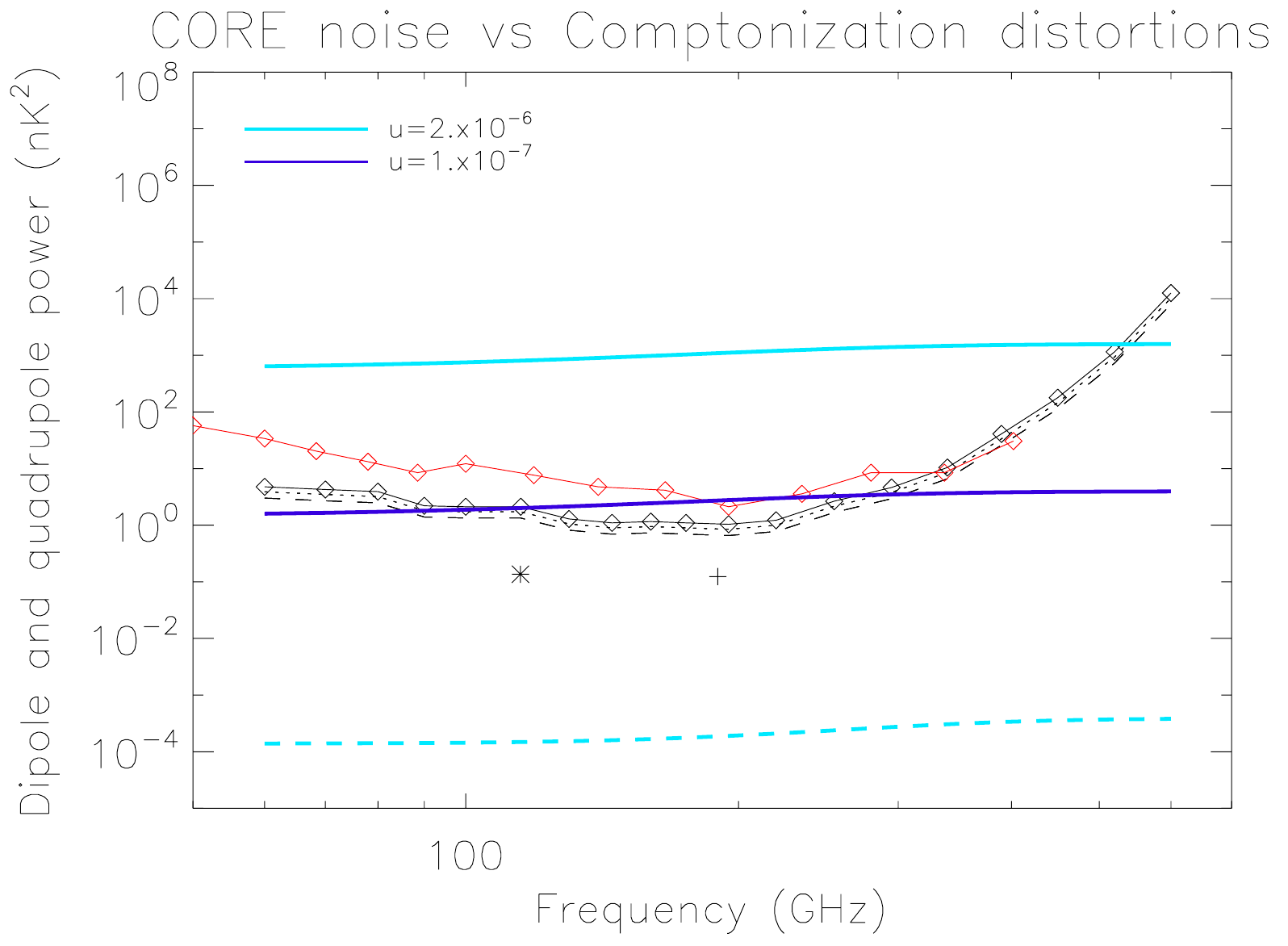}
    \caption{Angular power spectrum of the dipole map, derived from the
    difference between distorted spectra and the current blackbody
    spectrum versus CORE sensitivity.  The CORE white noise power spectrum
    (independent of multipole, shown as the black upper solid curve and with
    diamonds for different frequency channels) and its rms uncertainty (for
    $\ell = 1$, using dots, and for $\ell = 2$, using dashes) are plotted in black. The cross (asterisk) displays
    aggregated CORE noise from all channels (up to 220\,GHz). 
    We shown also for comparison the LiteBIRD white noise power spectrum (red solid curve and 
    diamonds for different frequency channels). {\it Left}:
    BE distortions for $\mu_0= -2.8\times10^{-9}$ (representative of adiabatic
    cooling), $\mu_0= 1.4\times10^{-5}$, $1.4\times10^{-6}$
    (representative of improvements with respect to FIRAS upper limits),
    $\mu_0= 1.12\times10^{-7}$, $2.8\times10^{-8}$, and $1.4\times10^{-9}$
    (representative of primordial adiabatic perturbation dissipation). 
    For $\mu_0= 1.4\times10^{-5}$ we show also the angular power spectrum of the quadrupole map.
    {\it Right}: Comptonization distortions for $u=2\times10^{-6}$ (upper solid
    curve for the dipole map and bottom dashed curve for the quadrupole map) and $u=10^{-7}$ (lower solid curve for the dipole map), 
    representative of imprints by astrophysical and minimal reionization models, respectively.
    }
    \label{fig:APS_C_BE_sens}
\end{figure}

\begin{figure}[ht!]
\begin{minipage}[t]{0.49\linewidth}
    \centering
    \includegraphics[width=1.01\textwidth]{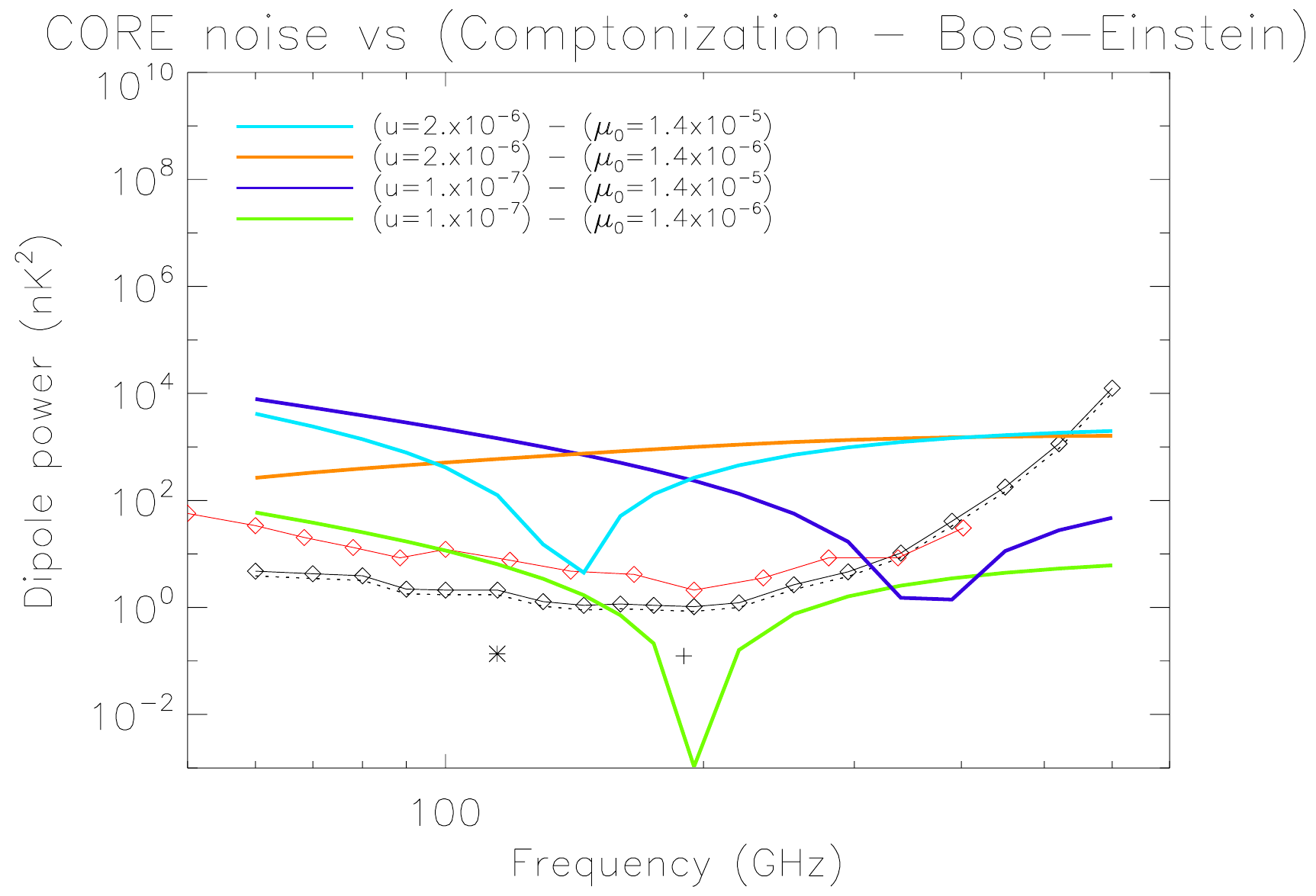}
    \caption{The same as in Fig.~\ref{fig:APS_C_BE_sens}, but comparing the two above cases of Comptonization distortions with the two above cases of BE distortions with the largest values of $\mu_0$.}
    \label{fig:APS_CvsBE_sens}
\end{minipage}
\hspace{0.1cm}
\begin{minipage}[t]{0.49\linewidth}
    \centering
    \includegraphics[width=0.92\textwidth]{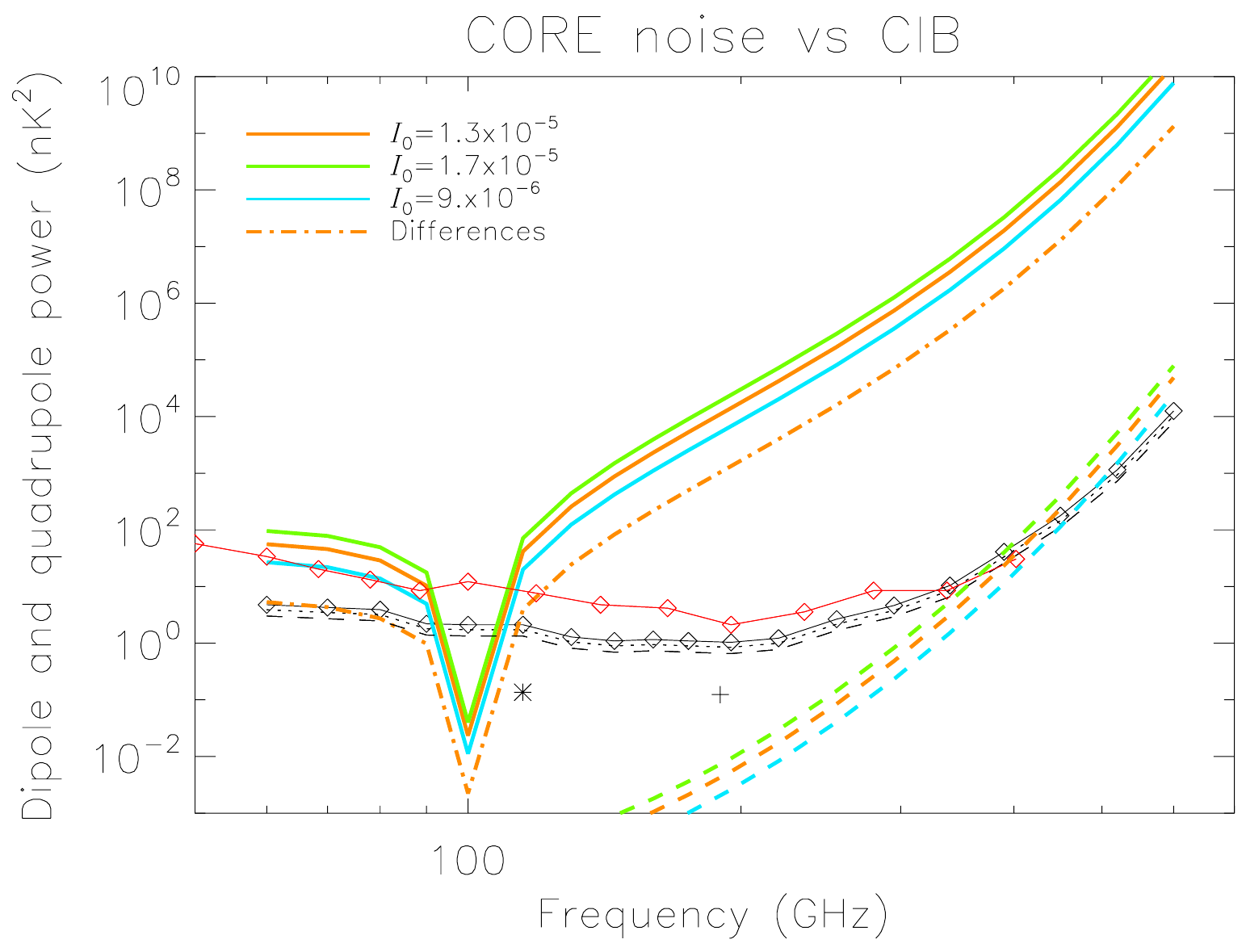}
    \caption{The same as in Fig.~\ref{fig:APS_C_BE_sens}, but for the CIB (assuming the
    model from \cite{Fixsen:1998kq}). Also shown is the quadrupole signal
    (dashes). The different values of $I_0$ in Eq.~\eqref{eq:eta_boost} are the
    best-fit value and deviations by $\pm1\,\sigma$.}
    \label{fig:APS_CIB_sens}
\end{minipage}
\end{figure}

\subsection{Detectability}
\label{sect:DiffCMB_SN}

Here we discuss the detectability of the dipolar and quadrupolar signals
introduced in Sects.~\ref{sect:DipCMB}--\ref{sect:LowEll}. To this end we
compare the dipole signal with the noise dipole as a function of frequency. Note
that since the prediction includes the specific angular dependence of the
dipole, there is no cosmic-variance related component in the noise. 
The noise for each frequency is determined by Table~\ref{tab:CORE-bands},
assuming full-sky coverage for simplicity.\footnote{Sampling variance, as
specified by the adopted masks, will be taken into account in the next section.}

In Fig.~\ref{fig:APS_C_BE_sens} we show the dipole signal for BE and 
Comptonization distortions (left and right, respectively), defined as the temperature dipole
coming from Eq.~\eqref{eq:eta_boost} subtracted from the CMB dipole (shown as
coloured lines). In black we show the CORE noise as a function of frequency.
For BE distortions, the signal is clearly above the CORE noise up to about
200\,GHz for $\mu_{0} \gsim 10^{-6}$ and comparable or slightly above the
aggregated noise below about 100\,GHz for $\mu_{0} \gsim 10^{-7}$, while for
Comptonization distortions, the signal is clearly above the noise up to around
500\,GHz for $u \gsim 2 \times 10^{-6}$ and comparable to or above the noise
between approximately 100\,GHz and 300\,GHz for $u \gsim 10^{-7}$. The analogous analysis for the quadrupole 
(shown for simplicity only for the largest values of $\mu_0$ and $u$) shows
that, for CORE sensitivity, noise dominates at any frequency for CMB spectral distortion parameters 
compatible with FIRAS limits, thus experiments beyond CORE are needed to use the quadrupole pattern to infer constraints on 
CMB spectral distortions. In Fig.~\ref{fig:APS_CvsBE_sens} we show the dipole signal of the difference
between Comptonization and BE distortion maps. 
In Fig.~\ref{fig:APS_CIB_sens} we show the size
of the dipole signal (the quadrupole is shown as dashed curves) for the CIB
(where we have removed the CMB dipole) compared to noise. The signal is always
above the noise except at about 100\,GHz. Due to the large uncertainty in the
amplitude of the CIB spectra ($I_0$), we show also deviations of $\pm 1\,\sigma$ from
the best-fit value of $1.3\times10^{-5}$ (as well as their difference from the best fit). The signal is orders of magnitude
above the noise at high frequencies and moreover, the quadrupole is above the
noise at frequencies greater than about 400\,GHz, although it is always much
smaller than the dipole (since it is suppressed by an extra factor of $\beta$).

Comparing Fig.~\ref{fig:APS_CIB_sens} with Fig.~\ref{fig:APS_C_BE_sens}, it is
evident that the dipole power expected from the CIB is above those predicted for
CMB spectral distortions at $\nu \gsim 200$\,GHz for the classes of processes
and parameter values discussed here. Since the dependence of the
quoted power on the CMB spectral distortion parameter is quadratic, the above 
statement does not hold for larger CMB
distortions, even just below the FIRAS limits.
Although they are not predicted by standard scenarios, they may be
generated by unconventional dissipation processes, such those discussed at the end of Sect. \ref{sect:DipCMB},
according to their characteristic parameters. 

We computed for comparison the power spectrum sensitivity of LiteBIRD (see Table~\ref{tab:LiteBIRD-bands}):
it is similar to that of CORE around 300\,GHz and significantly worse
at $\nu \lsim 150$\,GHz, a range suitable in particular for BE distortions. 
As discussed in Sects.~\ref{sec:dipideal} and~\ref{sec:dipsyst}, resolution is
important to achieve the sky sampling necessary for ultra-accurate dipole
analysis, thus adopting a resolution changing from a range of
$\simeq 2$--18 arcmin to a range of $\simeq 0.5^{\circ}$--$1.5^{\circ}$
is certainly critical. Furthermore, the number of frequency channels
is relevant, in particular (see next section) when one compares between pairs
of frequencies, the number of which scales approximately as the square of the
number of frequency channels.  In addition, a large number of frequency
channels and especially the joint analysis of frequencies around
300\,GHz and above 400\,GHz (not foreseen in LiteBIRD) is crucial for
separating the various types of signals, and, in particular, to accurately
control the contamination by Galactic dust emission.

The analysis carried out here will be extended to include all frequency
information in the following section. This will also include a discussion of
the impact of residual foregrounds.

\section{Simulation results for CMB spectral distortions and CIB intensity}
\label{sec:dip_spec_sim}

In order to quantify the ideal CORE sensitivity to measure spectral distortion
parameters and the CIB amplitude, we carried out some detailed simulations.
The idea here
is to simulate the sky signal assuming a certain model and to quantify the
accuracy level at which (in the presence of noise and of potential residuals)
we can recover the key input parameters. We consider twelve reference cases, physically
or observationally motivated, based on considerations and works quoted in Sect. \ref{sect:DipCIB} (cases 2--4) and 
Sect. \ref{sect:DipCMB} (cases 8--12), namely:

\vskip 0.5cm

\noindent
(1) a (reference) blackbody spectrum defined by $T_0$; \\
(2) a CIB spectrum at the FIRAS best-fit amplitude; \\
(3) a CIB spectrum at the FIRAS best-fit amplitude plus $1\,\sigma$ error; \\
(4) a CIB spectrum at the FIRAS best-fit amplitude minus $1\,\sigma$ error; \\
(5) a BE spectrum with $\mu_0 = 1.12 \times 10^{-7}$, representative of a
distortion induced by damping of primordial adiabatic perturbations in the
case of relatively high power at small scales; \\
(6) a BE spectrum with $\mu_0 = 1.4 \times 10^{-5}$, a value 6.4 times smaller
than the FIRAS $95\,\%$ upper limits; \\
(7) a BE spectrum with $\mu_0 = 1.4 \times 10^{-6}$, a value 64 times smaller
than the FIRAS $95\,\%$ upper limits; \\
(8) a BE spectrum with $\mu_0 = 1.4 \times 10^{-9}$, representative of the
typical minimal distortion induced by the damping of primordial adiabatic
perturbations; \\
(9) a BE spectrum with $\mu_0 = 2.8 \times 10^{-8}$, representative of the
typical distortion induced by damping of primordial adiabatic perturbations; \\
(10) a BE spectrum with $\mu_0 = -2.8 \times 10^{-9}$, representative of the
typical distortion induced by BE condensation (adiabatic cooling); \\
(11) a Comptonised spectrum with $u = 10^{-7}$, representative of minimal
reionization models; \\
(12) a Comptonised spectrum with $u = 2 \times 10^{-6}$, representative of
typical astrophysical reionization models. \\

For each model listed we generate both an ideal sky (the ``prediction'') and a
sky with noise realizations (``simulated data'') according to the sensitivity
of CORE (see Table~\ref{tab:CORE-bands}), at each of its 19 frequency channels.
For a suitable number of cases we repeated the analysis working with maps simply containing only the dipole term
and verified that the major contribution to the significance comes from the
dipole, i.e., the quadrupole (the only other possibly
relevant term) contributes almost negligibly,\footnote{In some cases we found it affects only
the last digit (reported in the tables) of the estimated $\sqrt{|\Delta
\chi^2|} \; {\rm sign}(\Delta \chi^2)$.} in agreement with
Sect.~\ref{sect:DiffCMB_SN}. For the sake of simplicity our noise realizations
assume Gaussian white noise. Our simulation set consists of 10 realizations for
each of the 19 CORE frequencies (giving 190 independent noise 
realizations). These are generated at $N_{\rm side} = 64$ (roughly $1^\circ$ resolution).
We will also consider
the inclusion of certain systematics in the following subsections. We then
compare each theoretical prediction with all maps of our simulated data. We
calculate $\Delta \chi^2$ for each combination, summarised in a $12\times12$
matrix, quantifying the significance level at which each model can be
potentially detected or ruled out. We report our results in terms of
$\sqrt{|\Delta \chi^2|} \; {\rm sign}(\Delta \chi^2)$, which directly gives
the significance in terms of $\sigma$ levels, since we only consider a single
parameter at a time.\footnote{The adopted number of realizations
allows to provide an estimate the rms of the quoted significance values suitable to check 
(particularly for some results based, for simplicity, on a single realization) they are not in the tail of distribution,
to quantitatively compare pros and cons of the three adopted approaches, and to spot in the results the effects
of coupling between signal and noise/residuals realizations.
With much more realizations it is obviously possible to refine these estimates, 
but it is not relevant in this work that deals with wide ranges of residual parameters.}

We perform the $\sqrt{\Delta \chi^2}$ analysis for three different approaches:
\begin{enumerate}[label=(\alph*)]
  \item using each of the 19 frequency channels, assuming they are independent; 
  \item using the 171 ($19\times18/2$) combinations coming from the differences of the maps from pairs of frequency
 bands;
  \item combining cases (a) and (b) together.
\end{enumerate}
When differences of maps from pairs of frequency
 bands are included in the analysis, in the corresponding contributions to the $\chi^2$ the variance comes from the
sum of the variances at the two considered frequencies. 

Approach (a) essentially compares the amplitude of dipole of a distorted spectrum with that of the blackbody,
being so sensitive to the overall difference between the two cases, while approach (b) compares the dipole signal at different
frequencies for each type of spectrum, being so sensitive to its slope.

\subsection{Ideal case: perfect calibration and foreground subtraction}
\label{sec:MC_ideal}

\begin{table}[ht!]
\centering
\begin{adjustbox}{width=1\textwidth}
\begin{tabular}{|c|c|ccc|cccccc|cc|}
\hline
{\bf $\sigma$ level} & \multicolumn{1}{c|}{Current}
& \multicolumn{3}{c|}{FIRAS CIB amplitude}
& \multicolumn{6}{c|}{$(\Delta \varepsilon / \varepsilon_{\rm i})_{z_1}$}
& \multicolumn{2}{c|}{$(\Delta \varepsilon / \varepsilon_{\rm i})_{\rm late}$}\\[\defaultaddspace]
{\bf significance} & blackbody  &  &  (units $10^{-5}$) & & $8 \times 10^{-8}$ & $10^{-5}$ & $10^{-6}$ & $10^{-9}$ & $2 \times 10^{-8}$ & $-2 \times 10^{-9}$ & $4 \times 10^{-7}$ & $8 \times 10^{-6}$ \\
\hline\hline
& \multicolumn{1}{c|}{} &
$I_0^{\rm bf}$ & $+ 1\,\sigma$ & $ - 1\,\sigma$
& \multicolumn{6}{c|}{$\mu_0$}
& \multicolumn{2}{c|}{$u$}\\[\defaultaddspace]
&  & $1.3$ & $1.7$ & $0.9$ & $1.12 \times 10^{-7}$ & $1.4 \times 10^{-5}$ & $1.4 \times 10^{-6}$ & $1.4 \times 10^{-9}$ & $2.8 \times 10^{-8}$ & $-2.8 \times 10^{-9}$ & $10^{-7}$ & $2 \times 10^{-6}$ \\
\hline\hline
Case & (1) & (2) & (3) & (4) & (5) & (6) & (7) & (8) & (9) & (10) & (11) & (12) \\
\hline
$(1)$ & $0$ & $3650$ & $4740$ & $2540$ & $1.63$ & $175.$ & $17.9$ & $0.0998$ & $0.532$ & $-0.137$ & $7.95$ & $154.$ \\[\defaultaddspace]
$(2)$ & $ 3650$ & $0$ & $1100$ & $1110$ & $3650$ & $3640$ & $3650$ & $3650$ & $3650$ & $3650$ & $3650$ & $3620$ \\[\defaultaddspace]
$(3)$ & $ 4740$ & $1100$ & $0$ & $2210$ & $4740$ & $4740$ & $4740$ & $4740$ & $4740$ & $4740$ & $4740$ & $4710$ \\[\defaultaddspace]
$(4)$ & $ 2540$ & $1110$ & $2200$ & $0$ & $2540$ & $2540$ & $2540$ & $2540$ & $2540$ & $2540$ & $2540$ & $2510$ \\[\defaultaddspace]
$(5)$ & $0.403$ & $3650$ & $4740$ & $2540$ & $0$ & $174.$ & $16.5$ & $0.384$ & $-0.111$ & $0.445$ & $6.83$ & $153.$ \\[\defaultaddspace]
$(6)$ & $174.$ & $3640$ & $4740$ & $2540$ & $173.$ & $0$ & $157.$ & $174.$ & $174.$ & $174.$ & $168.$ & $108.$ \\[\defaultaddspace]
$(7)$ & $17.0$ & $3650$ & $4740$ & $2540$ & $15.6$ & $158.$ & $0$ & $17.0$ & $16.6$ & $17.0$ & $11.8$ & $140.$ \\[\defaultaddspace]
$(8)$ & $-0.0975$ & $3650$ & $4740$ & $2540$ & $1.62$ & $175.$ & $17.9$ & $0$ & $0.514$ & $-0.165$ & $7.93$ & $154.$ \\[\defaultaddspace]
$(9)$ & $-0.346$ & $3650$ & $4740$ & $2540$ & $1.28$ & $175.$ & $17.6$ & $-0.342$ & $0$ & $-0.352$ & $7.66$ & $153.$ \\[\defaultaddspace]
$(10)$ & $0.142$ & $3650$ & $4740$ & $2540$ & $1.67$ & $175.$ & $17.9$ & $0.176$ & $0.567$ & $0$ & $7.98$ & $154.$ \\[\defaultaddspace]
$(11)$ & $ 7.25$ & $3650$ & $4740$ & $2540$ & $6.22$ & $169.$ & $12.8$ & $7.23$ & $6.98$ & $7.27$ & $0$ & $146.$ \\[\defaultaddspace]
$(12)$ & $ 153.$ & $3620$ & $4710$ & $2510$ & $152.$ & $108.$ & $140.$ & $153.$ & $153.$ & $153.$ & $145.$ & $0$ \\[\defaultaddspace]
\hline
\end{tabular}
\end{adjustbox}
\caption{Average values of $\sqrt{|\Delta \chi^2|} \; {\rm sign}(\Delta \chi^2)$ from a Monte Carlo simulation at $N_{\rm side} = 64$, full-sky coverage,
adopting perfect foreground subtraction and calibration, and considering each of the 19 frequency channels.}
\label{MC_corr_avg_ideal}
\end{table}

\begin{table}
\centering
\begin{adjustbox}{width=1\textwidth}
\begin{tabular}{|c|c|ccc|cccccc|cc|}
\hline
{\bf $\sigma$ level} & \multicolumn{1}{c|}{Current}
& \multicolumn{3}{c|}{FIRAS CIB amplitude}
& \multicolumn{6}{c|}{$(\Delta \varepsilon / \varepsilon_{\rm i})_{z_1}$}
& \multicolumn{2}{c|}{$(\Delta \varepsilon / \varepsilon_{\rm i})_{\rm late}$}\\[\defaultaddspace]
{\bf significance} & blackbody  &  &  (units $10^{-5}$) & & $8 \times 10^{-8}$ & $10^{-5}$ & $10^{-6}$ & $10^{-9}$ & $2 \times 10^{-8}$ & $-2 \times 10^{-9}$ & $4 \times 10^{-7}$ & $8 \times 10^{-6}$ \\
\hline\hline
& \multicolumn{1}{c|}{} &
$I_0^{\rm bf}$ & $+ 1\,\sigma$ & $ - 1\,\sigma$
& \multicolumn{6}{c|}{$\mu_0$}
& \multicolumn{2}{c|}{$u$}\\[\defaultaddspace]
&  & $1.3$ & $1.7$ & $0.9$ & $1.12 \times 10^{-7}$ & $1.4 \times 10^{-5}$ & $1.4 \times 10^{-6}$ & $1.4 \times 10^{-9}$ & $2.8 \times 10^{-8}$ & $-2.8 \times 10^{-9}$ & $10^{-7}$ & $2 \times 10^{-6}$ \\
\hline\hline
Case & (1) & (2) & (3) & (4) & (5) & (6) & (7) & (8) & (9) & (10) & (11) & (12) \\
\hline
$(1)$ & $0$ & $14000$ & $18200$ & $9740$ & $1.96$ & $271.$ & $27.8$ & $0.124$ & $0.666$ & $-0.170$ & $-0.229$ & $43.7$\\[\defaultaddspace]
$(2)$ & $14000$ & $0$ & $4180$ & $4240$ & $14000$ & $14100$ & $14000$ & $14000$ & $14000$ & $14000$ & $14000$ & $14000$\\[\defaultaddspace]
$(3)$ & $18200$ & $4190$ & $0$ & $8430$ & $18200$ & $18200$ & $18200$ & $18200$ & $18200$ & $18200$ & $18200$ & $18200$\\[\defaultaddspace]
$(4)$ & $9740$ & $4240$ & $8420$ & $0$ & $9740$ & $9810$ & $9750$ & $9740$ & $9740$ & $9740$ & $9740$ & $9730$\\[\defaultaddspace]
$(5)$ & $0.0596$ & $14000$ & $18200$ & $9740$ & $0$ & $269.$ & $25.6$ & $0.0456$ & $-0.249$ & $0.0910$ & $1.44$ & $45.7$\\[\defaultaddspace]
$(6)$ & $ 269.$ & $14000$ & $18200$ & $9810$ & $267.$ & $0$ & $242.$ & $269.$ & $269.$ & $269.$ & $271.$ & $312.$\\[\defaultaddspace]
$(7)$ & $ 26.0$ & $14000$ & $18200$ & $9750$ & $23.8$ & $244.$ & $0$ & $26.0$ & $25.4$ & $26.1$ & $28.1$ & $69.6$\\[\defaultaddspace]
$(8)$ & $-0.122$ & $14000$ & $18200$ & $9740$ & $1.93$ & $271.$ & $27.8$ & $0$ & $0.644$ & $-0.205$ & $-0.209$ & $43.7$\\[\defaultaddspace]
$(9)$ & $-0.386$ & $14000$ & $18200$ & $9740$ & $1.51$ & $271.$ & $27.2$ & $-0.384$ & $0$ & $-0.391$ & $0.228$ & $44.2$\\[\defaultaddspace]
$(10)$ & $0.178$ & $14000$ & $18200$ & $9740$ & $2.00$ & $271.$ & $27.9$ & $0.220$ & $0.708$ & $0$ & $-0.267$ & $43.6$\\[\defaultaddspace]
$(11)$ & $ 2.66$ & $14000$ & $18200$ & $9740$ & $4.69$ & $273.$ & $29.9$ & $2.68$ & $3.11$ & $2.62$ & $0$ & $41.4$\\[\defaultaddspace]
$(12)$ & $ 46.4$ & $14000$ & $18200$ & $9730$ & $48.4$ & $314.$ & $72.0$ & $46.4$ & $46.9$ & $46.4$ & $44.1$ & $0$\\[\defaultaddspace]
\hline
\end{tabular}
\end{adjustbox}
\caption{The same as in Table~\ref{MC_corr_avg_ideal}, but considering all
171 independent combinations of pairs of different frequency channels.}
\label{MC_cross_avg_ideal}
\end{table}
\begin{table}
\centering
\begin{adjustbox}{width=1\textwidth}
\begin{tabular}{|c|c|ccc|cccccc|cc|}
\hline
{\bf $\sigma$ level} & \multicolumn{1}{c|}{Current}
& \multicolumn{3}{c|}{FIRAS CIB amplitude}
& \multicolumn{6}{c|}{$(\Delta \varepsilon / \varepsilon_{\rm i})_{z_1}$}
& \multicolumn{2}{c|}{$(\Delta \varepsilon / \varepsilon_{\rm i})_{\rm late}$}\\[\defaultaddspace]
{\bf significance} & blackbody  &  &  (units $10^{-5}$) & & $8 \times 10^{-8}$ & $10^{-5}$ & $10^{-6}$ & $10^{-9}$ & $2 \times 10^{-8}$ & $-2 \times 10^{-9}$ & $4 \times 10^{-7}$ & $8 \times 10^{-6}$ \\
\hline\hline
& \multicolumn{1}{c|}{} &
$I_0^{\rm bf}$ & $+ 1\,\sigma$ & $ - 1\,\sigma$
& \multicolumn{6}{c|}{$\mu_0$}
& \multicolumn{2}{c|}{$u$}\\[\defaultaddspace]
&  & $1.3$ & $1.7$ & $0.9$ & $1.12 \times 10^{-7}$ & $1.4 \times 10^{-5}$ & $1.4 \times 10^{-6}$ & $1.4 \times 10^{-9}$ & $2.8 \times 10^{-8}$ & $-2.8 \times 10^{-9}$ & $10^{-7}$ & $2 \times 10^{-6}$ \\
\hline\hline
Case & (1) & (2) & (3) & (4) & (5) & (6) & (7) & (8) & (9) & (10) & (11) & (12) \\
\hline
$(1)$ & $0$ & $14400$ & $18800$ & $10100$ & $2.67$ & $323.$ & $33.1$ & $0.157$ & $0.844$ & $-0.214$ & $7.83$ & $160.$\\[\defaultaddspace]
$(2)$ & $14500$ & $0$ & $4320$ & $4380$ & $14500$ & $14500$ & $14500$ & $14500$ & $14500$ & $14500$ & $14500$ & $14400$\\[\defaultaddspace]
$(3)$ & $18800$ & $4330$ & $0$ & $8710$ & $18800$ & $18800$ & $18800$ & $18800$ & $18800$ & $18800$ & $18800$ & $18800$\\[\defaultaddspace]
$(4)$ & $10100$ & $4380$ & $8710$ & $0$ & $10100$ & $10100$ & $10100$ & $10100$ & $10100$ & $10100$ & $10100$ & $10100$\\[\defaultaddspace]
$(5)$ & $0.369$ & $14400$ & $18800$ & $10100$ & $0$ & $320.$ & $30.5$ & $0.346$ & $-0.105$ & $0.413$ & $7.24$ & $159.$\\[\defaultaddspace]
$(6)$ & $ 321.$ & $14500$ & $18800$ & $10100$ & $318.$ & $0$ & $288.$ & $321.$ & $320.$ & $321.$ & $319.$ & $330.$\\[\defaultaddspace]
$(7)$ & $ 31.1$ & $14500$ & $18800$ & $10100$ & $28.5$ & $290.$ & $0$ & $31.0$ & $30.4$ & $31.1$ & $30.5$ & $157.$\\[\defaultaddspace]
$(8)$ & $-0.153$ & $14400$ & $18800$ & $10100$ & $2.64$ & $323.$ & $33.1$ & $0$ & $0.816$ & $-0.259$ & $7.82$ & $160.$\\[\defaultaddspace]
$(9)$ & $-0.537$ & $14400$ & $18800$ & $10100$ & $2.07$ & $322.$ & $32.4$ & $-0.532$ & $0$ & $-0.547$ & $7.61$ & $160.$\\[\defaultaddspace]
$(10)$ & $0.224$ & $14400$ & $18800$ & $10100$ & $2.73$ & $323.$ & $33.1$ & $0.277$ & $0.900$ & $0$ & $7.85$ & $160.$\\[\defaultaddspace]
$(11)$ & $ 7.92$ & $14400$ & $18800$ & $10100$ & $8.03$ & $321.$ & $32.5$ & $7.92$ & $7.87$ & $7.93$ & $0$ & $152.$\\[\defaultaddspace]
$(12)$ & $ 160.$ & $14400$ & $18800$ & $10000$ & $160.$ & $332.$ & $157.$ & $160.$ & $160.$ & $160.$ & $152.$ & $0$\\[\defaultaddspace]
\hline
\end{tabular}
\end{adjustbox}
\caption{The same as in Table~\ref{MC_corr_avg_ideal}, but considering each of
the 19 frequency channels independently and all 171 independent
combinations of pairs of different frequency channels.}
\label{MC_corr_cross_avg_ideal}
\end{table}

Tables~\ref{MC_corr_avg_ideal} and~\ref{MC_corr_rms_ideal} (and
Tables~\ref{MC_cross_avg_ideal} and~\ref{MC_cross_rms_ideal},
Tables~\ref{MC_corr_cross_avg_ideal} and~\ref{MC_corr_cross_rms_ideal},
respectively) report the results of approach a (approach b, and c, respectively)
in terms of average and rms of $\sqrt{|\Delta \chi^2|} \; {\rm sign}(\Delta
\chi^2)$ (see Appendix~\ref{app_rms_ideal}).

We find that, in general, the analysis of the difference of pairs of frequency
channels (approach b) tends to substantially increase the significance of the
recovery of the CIB amplitude, which is due to the very steep frequency shape
of the CIB dipole spectrum. For the opposite reason, the same does not occur in
general for CMB distortion parameters, and, in particular, approach (b) can
make the recovery of the Comptonization distortion more difficult. These
results are in agreement with the shapes displayed in
Figs.~\ref{fig:APS_C_BE_sens},~\ref{fig:APS_CvsBE_sens},
and~\ref{fig:APS_CIB_sens}. It is important to note that, in general, the rms
values found in approach (b) are larger than those found in approach
(a), seemingly relatively more stable. We interpret this as an effect of larger
susceptibility of approach (b) to realization combinations. On the other hand,
for the estimation of the CIB amplitude this rms amplification does not spoil
the improvement in significance. We find that combining the two approaches, as
in (c), typically results in an overall advantage, with an improvement in
significance larger than the possible increasing of the quoted rms. We
anticipate that these results will still be valid when including potential
residuals, as discussed below.

We remark that in the present analysis both pure theoretical maps and maps
polluted with noise are pixelised in the same way. So, the sampling problem
discussed in Sects.~\ref{sec:dipideal} and~\ref{sec:dipsyst} is automatically
by-passed. This is not a limitation for the present analysis, given the high
resolution achieved by CORE, and because it is clear that we could in principle
perform our simulations at the desired resolution. Working at roughly $1^\circ$
resolution makes our analysis feasible without supercomputing facilities, with
no significant loss of information. Nonetheless, we also report some results
carried out at higher resolution. In particular, in
Appendix~\ref{ideal_highres} we present results of the analysis repeated
at $N_{\rm side} = 512$ (i.e., at about 7 arcmin
resolution), for a single realization. 
The results are fully compatible, within the statistical variance, with those derived working at $N_{\rm side} = 64$.

The matrices reported in each of these tables perhaps require a little more
explanation.  Firstly, we should point out that the diagonals are zero by
construction. We found that the reduced $\chi^2$ ($\chi^2_{\rm r}
= \chi^2 / (n_{\rm d}-1)$, where $n_{\rm d}$ is the global number of
data being treated and we are considering the estimate of a single parameter,
namely CMB distortion or CIB
amplitude), is always extremely close to unity, which is an obvious validation
cross-check. Note that, in principle, when potential residuals are included,
one should specify the variance pixel-by-pixel in the estimation of
$\chi^2$.\footnote{This
holds also in the case that the instrument sensitivity varies across the
sky because of non-uniform sky coverage from the adopted scanning
strategy (an aspect that is not so crucial in the case of the relatively
uniform sky coverage expected for CORE \cite{CORE2016,2017MNRAS.466..425W})
is included in the analysis.
Note also that, in principle, pixel-to-pixel correlations introduced by noise
correlations and potential residual morphologies should be included in the
$\chi^2$. This aspect, although important in the analysis of real data, is
outside the scope of the present paper. Nonetheless, it does not affect the
main results of our forecasts.}
This requires a precise local characterization of residuals.
While this can easily be included by construction in our analyses, we
explicitly avoid implementing this in the $\chi^2$ analysis, but instead
perform our forecasts assuming knowledge of only the average level of the
residuals in the sky region being considered. Secondly, we note that
the matrices are not perfectly symmetric, due to the cross-terms in the
squares (from noise and signal) entering into the $\chi^2$. Thirdly, the
off-diagonal terms are sometimes negative, but with absolute values compatible
with the quoted rms. These second and third effects
are clearly statistical in nature.

The results found in this section (see also Appendix~\ref{ideal_highres})
identify the ideal sensitivity target for CMB spectral distortion parameters
and CIB amplitude that are achievable from the dipole frequency behaviour.

Elements\footnote{We adopt the convention (row index range, column index range).} (2:4, 2:4)
of the matrix quantify the sensitivity to the CIB amplitude. Comparison with
FIRAS in terms of the $\sigma$ level of significance can be extracted
directly from the tables; the
ideal improvement ranges from a factor of about 1000 to 4000.

The ideal improvement found for CMB spectral distortion parameters is also
impressive. Elements (1, 5:10) and (5:10, 1) and elements (1, 11:12) and
(11:12, 1) refer to comparisons between the blackbody and BE and Comptonization
distortions, respectively. The comparison with FIRAS is simply quoted by the
element of the matrix of the table multiplied by the ratio between the FIRAS
$1\,\sigma$ upper limit on $\mu_0$ or $u$ and the distortion parameter value
considered in the table. The sensitivity on $u$ is clearly enough to
disentangle between minimal models of reionization and a variety of
astrophysical models that predict larger amounts of energy injection by
various types of source. The ideal improvement with respect to FIRAS limits is
about 500--600. The level of (negative) BE distortions is much
lower, and the same holds also for BE distortions predicted for the damping of
primordial adiabatic perturbations. Only weak, tentative constraints on models
with high power at small scales could be set with this approach, for a mission
with the sensitivity of CORE. Nonetheless, the ideal improvement with respect
to FIRAS limits on BE distortions lies in the range 600--1000.

The other elements of the matrix refer to the comparison of distorted spectra;
note in particular the elements (6:7, 11:12) and (11:12, 6:7) that show how
Comptonization distortions can be distinguished from BE distortions, for the
two larger values considered for $\mu_0$, as suggested by
Fig.~\ref{fig:APS_CvsBE_sens}.

\subsection{Including potential foreground and calibration residuals}

We expect that potential residuals from imperfect foreground subtraction and
calibration may affect the results presented in the previous section,
depending on their level.  To assess this, we have carried
out a wide set of simulations in order to quantify the accuracy in
recovering the CMB distortion parameters and CIB amplitude under different
working assumptions.

We first perform simulations adopting $E_{\rm for} = 10^{-2}$ and
$E_{\rm cal} = 10^{-4}$ (defined by the parametric model introduced in
Sect.~\ref{sec:res_mod}) at $N_{\rm side} = 64$, and then add many tests
exploring combinations of possible improvements in foreground characterization
(assuming $E_{\rm for} = 10^{-3}$ or $E_{\rm for} = 10^{-2}$, but at larger
$N_{\rm side}$), as well as different levels of calibration accuracy (including
possible worsening at higher frequencies).

\begin{table}[ht!]
\centering
\begin{adjustbox}{width=1\textwidth}
\begin{tabular}{|c|c|ccc|cccccc|cc|}
\hline
{\bf $\sigma$ level} & \multicolumn{1}{c|}{Current}
& \multicolumn{3}{c|}{FIRAS CIB amplitude}
& \multicolumn{6}{c|}{$(\Delta \varepsilon / \varepsilon_{\rm i})_{z_1}$}
& \multicolumn{2}{c|}{$(\Delta \varepsilon / \varepsilon_{\rm i})_{\rm late}$}\\[\defaultaddspace]
{\bf significance} & blackbody  &  &  (units $10^{-5}$) & & $8 \times 10^{-8}$ & $10^{-5}$ & $10^{-6}$ & $10^{-9}$ & $2 \times 10^{-8}$ & $-2 \times 10^{-9}$ & $4 \times 10^{-7}$ & $8 \times 10^{-6}$ \\
\hline\hline
& \multicolumn{1}{c!}{} &
$I_0^{\rm bf}$ & $+ 1\,\sigma$ & $ - 1\,\sigma$
& \multicolumn{6}{c|}{$\mu_0$}
& \multicolumn{2}{c|}{$u$} \\[\defaultaddspace]
&  & $1.3$ & $1.7$ & $0.9$ & $1.12 \times 10^{-7}$ & $1.4 \times 10^{-5}$ & $1.4 \times 10^{-6}$ & $1.4 \times 10^{-9}$ & $2.8 \times 10^{-8}$ & $-2.8 \times 10^{-9}$ & $10^{-7}$ & $2 \times 10^{-6}$ \\
\hline\hline
Case & (1) & (2) & (3) & (4) & (5) & (6) & (7) & (8) & (9) & (10) & (11) & (12) \\
\hline
$(1)$ & $0$ & $13.7$ & $17.7$ & $9.61$ & $-0.00676$ & $10.4$ & $0.833$ & $-0.00264$ & $-0.00996$ & $ 0.00383$ & $0.166$ & $4.83$ \\[\defaultaddspace]
$(2)$ & $12.8$ & $0$ & $4.37$ & $3.48$ & $12.8$ & $16.6$ & $12.8$ & $12.8$ & $12.8$ & $12.8$ & $12.8$ & $13.1$ \\[\defaultaddspace]
$(3)$ & $16.8$ & $3.44$ & $0$ & $7.58$ & $16.8$ & $19.9$ & $16.9$ & $16.8$ & $16.8$ & $16.8$ & $16.8$ & $16.9$ \\[\defaultaddspace]
$(4)$ & $8.75$ & $4.39$ & $8.45$ & $0$ & $8.76$ & $13.7$ & $8.83$ & $8.75$ & $8.75$ & $8.75$ & $8.72$ & $9.47$ \\[\defaultaddspace]
$(5)$ & $0.0391$ & $13.7$ & $17.7$ & $9.61$ & $0$ & $10.3$ & $0.686$ & $0.0385$ & $0.0307$ & $0.0400$ & $0.133$ & $4.76$ \\[\defaultaddspace]
$(6)$ & $10.5$ & $17.3$ & $20.6$ & $14.3$ & $10.4$ & $0$ & $9.41$ & $10.5$ & $10.4$ & $10.5$ & $10.2$ & $6.51$ \\[\defaultaddspace]
$(7)$ & $0.821$ & $13.7$ & $17.7$ & $9.67$ & $0.739$ & $9.39$ & $0$ & $0.821$ & $0.799$ & $0.825$ & $0.611$ & $3.93$ \\[\defaultaddspace]
$(8)$ & $0.00270$ & $13.7$ & $17.7$ & $9.61$ & $-0.00701$ & $10.4$ & $0.830$ & $0$ & $-0.00967$ & $ 0.00473$ & $0.166$ & $4.83$ \\[\defaultaddspace]
$(9)$ & $0.0139$ & $13.7$ & $17.7$ & $9.61$ & $ -0.0100$ & $10.4$ & $0.788$ & $0.0135$ & $0$ & $0.0146$ & $0.163$ & $4.82$ \\[\defaultaddspace]
$(10)$ & $-0.00370$ & $13.7$ & $17.7$ & $9.61$ & $-0.00658$ & $10.4$ & $0.836$ & $-0.00451$ & $ -0.0102$ & $0$ & $0.166$ & $4.84$ \\[\defaultaddspace]
$(11)$ & $0.0337$ & $13.7$ & $17.6$ & $9.57$ & $ -0.0402$ & $10.2$ & $0.505$ & $0.0330$ & $0.0210$ & $0.0346$ & $0$ & $4.60$ \\[\defaultaddspace]
$(12)$ & $4.56$ & $13.9$ & $17.7$ & $10.1$ & $4.48$ & $6.28$ & $3.57$ & $4.56$ & $4.54$ & $4.56$ & $4.32$ & $0$ \\[\defaultaddspace]
\hline
\end{tabular}
\end{adjustbox}
\caption{Average values of $\sqrt{|\Delta \chi^2|} \; {\rm sign}(\Delta \chi^2)$ from a Monte Carlo simulation at $N_{\rm side} = 64$, with full-sky coverage,
adopting $E_{\rm for} = 10^{-2}$ and $E_{\rm cal} = 10^{-4}$, and considering each of the 19 frequency channels.}
\label{MC_corr_avg}
\end{table}

\begin{table}[ht!]
\centering
\begin{adjustbox}{width=1\textwidth}
\begin{tabular}{|c|c|ccc|cccccc|cc|}
\hline
{\bf $\sigma$ level} & \multicolumn{1}{c!}{Current}
& \multicolumn{3}{c!}{FIRAS CIB amplitude}
& \multicolumn{6}{c!}{$(\Delta \varepsilon / \varepsilon_{\rm i})_{z_1}$}
& \multicolumn{2}{c|}{$(\Delta \varepsilon / \varepsilon_{\rm i})_{\rm late}$}\\[\defaultaddspace]
{\bf significance} & blackbody  &  &  (units $10^{-5}$) & & $8 \times 10^{-8}$ & $10^{-5}$ & $10^{-6}$ & $10^{-9}$ & $2 \times 10^{-8}$ & $-2 \times 10^{-9}$ & $4 \times 10^{-7}$ & $8 \times 10^{-6}$ \\
\hline\hline
& \multicolumn{1}{c|}{} &
$I_0^{\rm bf}$ & $+ 1\,\sigma$ & $ - 1\,\sigma$
& \multicolumn{6}{c|}{$\mu_0$}
& \multicolumn{2}{c|}{$u$}\\[\defaultaddspace]
&  & $1.3$ & $1.7$ & $0.9$ & $1.12 \times 10^{-7}$ & $1.4 \times 10^{-5}$ & $1.4 \times 10^{-6}$ & $1.4 \times 10^{-9}$ & $2.8 \times 10^{-8}$ & $-2.8 \times 10^{-9}$ & $10^{-7}$ & $2 \times 10^{-6}$ \\
\hline\hline
Case & (1) & (2) & (3) & (4) & (5) & (6) & (7) & (8) & (9) & (10) & (11) & (12) \\
\hline
$(1)$ & $ 0$ & $49.9$ & $64.7$ & $35.0$ & $-0.224$ & $6.67$ & $-0.311$ & $-0.0267$ & $-0.117$ & $0.0379$ & $ 0.211$ & $1.15$ \\[\defaultaddspace]
$(2)$ & $47.7$ & $0$ & $15.6$ & $13.4$ & $47.7$ & $49.9$ & $47.8$ & $47.7$ & $47.7$ & $47.7$ & $47.6$ & $47.4$ \\[\defaultaddspace]
$(3)$ & $62.4$ & $13.3$ & $0$ & $28.4$ & $62.4$ & $64.5$ & $62.6$ & $62.4$ & $62.4$ & $62.4$ & $62.4$ & $62.2$ \\[\defaultaddspace]
$(4)$ & $32.7$ & $15.8$ & $30.6$ & $0$ & $32.7$ & $35.2$ & $32.9$ & $32.7$ & $32.7$ & $32.7$ & $32.7$ & $32.5$ \\[\defaultaddspace]
$(5)$ & $0.250$ & $49.9$ & $64.7$ & $35.0$ & $0$ & $6.61$ & $-0.324$ & $ 0.248$ & $ 0.214$ & $ 0.253$ & $ 0.314$ & $1.21$ \\[\defaultaddspace]
$(6)$ & $8.51$ & $52.3$ & $66.9$ & $37.7$ & $8.44$ & $0$ & $7.71$ & $8.51$ & $8.48$ & $8.51$ & $8.54$ & $9.31$ \\[\defaultaddspace]
$(7)$ & $1.15$ & $50.1$ & $64.9$ & $35.2$ & $1.08$ & $5.84$ & $0$ & $1.15$ & $1.13$ & $1.15$ & $1.20$ & $2.06$ \\[\defaultaddspace]
$(8)$ & $ 0.0268$ & $49.9$ & $64.7$ & $35.0$ & $-0.223$ & $6.67$ & $-0.310$ & $0$ & $-0.114$ & $0.0465$ & $ 0.212$ & $1.15$ \\[\defaultaddspace]
$(9)$ & $0.121$ & $49.9$ & $64.7$ & $35.0$ & $-0.196$ & $6.66$ & $-0.314$ & $ 0.118$ & $0$ & $ 0.127$ & $ 0.239$ & $1.17$ \\[\defaultaddspace]
$(10)$ & $-0.0378$ & $49.9$ & $64.7$ & $35.0$ & $-0.227$ & $6.67$ & $-0.310$ & $-0.0463$ & $-0.122$ & $0$ & $ 0.208$ & $1.15$ \\[\defaultaddspace]
$(11)$ & $-0.206$ & $49.9$ & $64.7$ & $35.0$ & $-0.289$ & $6.71$ & $-0.322$ & $-0.207$ & $-0.230$ & $-0.203$ & $0$ & $1.11$ \\[\defaultaddspace]
$(12)$ & $-0.710$ & $49.6$ & $64.4$ & $34.7$ & $-0.706$ & $7.40$ & $-0.391$ & $-0.709$ & $-0.708$ & $-0.711$ & $-0.703$ & $0$ \\[\defaultaddspace]
\hline
\end{tabular}
\end{adjustbox}
\caption{The same as in Table~\ref{MC_corr_avg},
but considering all 171 independent combinations of pairs of different frequency channels.}
\label{MC_cross_avg}
\end{table}
\begin{table}[ht!]
\centering
\begin{adjustbox}{width=1\textwidth}
\begin{tabular}{|c|c|ccc|cccccc|cc|}
\hline
{\bf $\sigma$ level} & \multicolumn{1}{c|}{Current}
& \multicolumn{3}{c|}{FIRAS CIB amplitude}
& \multicolumn{6}{c|}{$(\Delta \varepsilon / \varepsilon_{\rm i})_{z_1}$}
& \multicolumn{2}{c|}{$(\Delta \varepsilon / \varepsilon_{\rm i})_{\rm late}$}\\[\defaultaddspace]
{\bf significance} & blackbody  &  &  (units $10^{-5}$) & & $8 \times 10^{-8}$ & $10^{-5}$ & $10^{-6}$ & $10^{-9}$ & $2 \times 10^{-8}$ & $-2 \times 10^{-9}$ & $4 \times 10^{-7}$ & $8 \times 10^{-6}$ \\
\hline\hline
& \multicolumn{1}{c|}{} &
$I_0^{\rm bf}$ & $+ 1\,\sigma$ & $ - 1\,\sigma$
& \multicolumn{6}{c|}{$\mu_0$}
& \multicolumn{2}{c|}{$u$}\\[\defaultaddspace]
&  & $1.3$ & $1.7$ & $0.9$ & $1.12 \times 10^{-7}$ & $1.4 \times 10^{-5}$ & $1.4 \times 10^{-6}$ & $1.4 \times 10^{-9}$ & $2.8 \times 10^{-8}$ & $-2.8 \times 10^{-9}$ & $10^{-7}$ & $2 \times 10^{-6}$ \\
\hline\hline
Case & (1) & (2) & (3) & (4) & (5) & (6) & (7) & (8) & (9) & (10) & (11) & (12) \\
\hline
$(1)$ & $0$ & $51.8$ & $67.0$ & $36.3$ & $-0.191$ & $12.4$ & $ 0.142$ & $-0.0230$ & $-0.101$ & $0.0327$ & $ 0.309$ & $5.10$ \\[\defaultaddspace]
$(2)$ & $49.4$ & $0$ & $16.2$ & $13.9$ & $49.4$ & $52.6$ & $49.5$ & $49.4$ & $49.4$ & $49.4$ & $49.3$ & $49.2$ \\[\defaultaddspace]
$(3)$ & $64.6$ & $13.7$ & $0$ & $29.4$ & $64.7$ & $67.5$ & $64.8$ & $64.7$ & $64.7$ & $64.6$ & $64.6$ & $64.4$ \\[\defaultaddspace]
$(4)$ & $33.9$ & $16.4$ & $31.8$ & $0$ & $33.9$ & $37.7$ & $34.1$ & $33.9$ & $33.9$ & $33.9$ & $33.9$ & $33.9$ \\[\defaultaddspace]
$(5)$ & $0.222$ & $51.8$ & $67.1$ & $36.3$ & $0$ & $12.3$ & $0.0689$ & $ 0.220$ & $ 0.189$ & $ 0.225$ & $ 0.402$ & $5.05$ \\[\defaultaddspace]
$(6)$ & $13.5$ & $55.1$ & $70.0$ & $40.2$ & $13.4$ & $0$ & $12.2$ & $13.5$ & $13.5$ & $13.5$ & $13.4$ & $11.4$ \\[\defaultaddspace]
$(7)$ & $1.48$ & $52.0$ & $67.3$ & $36.5$ & $1.38$ & $11.2$ & $0$ & $1.48$ & $1.46$ & $1.49$ & $1.43$ & $4.57$ \\[\defaultaddspace]
$(8)$ & $0.0231$ & $51.8$ & $67.0$ & $36.3$ & $-0.190$ & $12.4$ & $ 0.142$ & $0$ & $-0.0987$ & $0.0401$ & $ 0.309$ & $5.10$ \\[\defaultaddspace]
$(9)$ & $0.105$ & $51.8$ & $67.0$ & $36.3$ & $-0.169$ & $12.4$ & $ 0.123$ & $ 0.102$ & $0$ & $ 0.110$ & $ 0.338$ & $5.09$ \\[\defaultaddspace]
$(10)$ & $-0.0326$ & $51.8$ & $67.0$ & $36.3$ & $-0.193$ & $12.4$ & $ 0.145$ & $-0.0399$ & $-0.106$ & $0$ & $ 0.308$ & $5.10$ \\[\defaultaddspace]
$(11)$ & $-0.143$ & $51.7$ & $67.0$ & $36.3$ & $-0.324$ & $12.3$ & $-0.0738$ & $-0.147$ & $-0.200$ & $-0.136$ & $0$ & $4.86$ \\[\defaultaddspace]
$(12)$ & $4.35$ & $51.5$ & $66.8$ & $36.2$ & $4.26$ & $9.77$ & $3.06$ & $4.35$ & $4.33$ & $4.35$ & $4.10$ & $0$ \\[\defaultaddspace]
\hline
\end{tabular}
\end{adjustbox}
\caption{The same as in Table~\ref{MC_corr_avg},
but considering each of the 19 frequency channels independently and all 171 independent combinations of pairs of different frequency channels.}
\label{MC_corr_cross_avg}
\end{table}

\subsubsection{Monte Carlo results at about $1^\circ$ resolution}

To understand the typical implications of different assumptions,
we first perform a series of Monte
Carlo simulations, identical to that described in Sect.~\ref{sec:MC_ideal},
but including potential foreground and calibration residuals,
modelled according to Sect.~\ref{sec:res_mod}, assuming $E_{\rm for} = 10^{-2}$ and $E_{\rm cal} = 10^{-4}$.
The main results (the average values of $\sqrt{|\Delta \chi^2|} \; {\rm sign}(\Delta \chi^2)$) are presented in
Tables~\ref{MC_corr_avg},~\ref{MC_cross_avg}, and~\ref{MC_corr_cross_avg}, while the corresponding rms values are reported in Appendix~\ref{app_rms_res} (see
Tables~\ref{MC_corr_rms},~\ref{MC_cross_rms}, and~\ref{MC_corr_cross_rms}).

With these levels of potential residuals, the improvement with respect to FIRAS in the recovery of the CIB amplitude ranges
from a factor of approximately 4 (with an rms of about 1 in the estimate of this improvement factor)
for approach (a) to a factor of about 15 or 20 
(with an rms of about 3) for approaches (b) and (c), respectively.

The improvement found for the recovery of CMB spectral distortion parameters is also very promising.
The sensitivity to $u$ improves with respect to FIRAS by a factor of $20$ (except for the less stable approach (b)), which is suitable for detecting reionization imprints (of the sort predicted in astrophysical reionization models)
at about $5\,\sigma$, while the improvement on
BE distortions is about a factor of 40 (approach (c)).
Note that these results are derived considering the full sky, and thus we could expect to obtain improvements by applying masks to avoid regions with significant potential contamination, as discussed in the next section.

\subsubsection{Application of masks}

We performed some additional tests assuming $E_{\rm for} = 10^{-2}$ and $E_{\rm cal} = 10^{-4}$, but applying appropriate masks to the sky.
Clearly, in this way we reduce the available statistical information (as we verified
through tests carried out under ideal conditions of perfect foreground subtraction and calibration), but in realistic cases we may expect to improve the quality of results by
reducing the impact of potential residuals.

\begin{table}[ht!]
\centering
\begin{adjustbox}{width=1\textwidth}
\begin{tabular}{|c|c|ccc|cccccc|cc|}
\hline
{\bf $\sigma$ level} & \multicolumn{1}{c|}{Current}
& \multicolumn{3}{c|}{FIRAS CIB amplitude}
& \multicolumn{6}{c|}{$(\Delta \varepsilon / \varepsilon_{\rm i})_{z_1}$}
& \multicolumn{2}{c|}{$(\Delta \varepsilon / \varepsilon_{\rm i})_{\rm late}$}\\[\defaultaddspace]
{\bf significance} & blackbody  &  &  (units $10^{-5}$) & & $8 \times 10^{-8}$ & $10^{-5}$ & $10^{-6}$ & $10^{-9}$ & $2 \times 10^{-8}$ & $-2 \times 10^{-9}$ & $4 \times 10^{-7}$ & $8 \times 10^{-6}$ \\
\hline\hline
& \multicolumn{1}{c|}{} &
$I_0^{\rm bf}$ & $+ 1\,\sigma$ & $ - 1\,\sigma$
& \multicolumn{6}{c|}{$\mu_0$}
& \multicolumn{2}{c|}{$u$}\\[\defaultaddspace]
&  & $1.3$ & $1.7$ & $0.9$ & $1.12 \times 10^{-7}$ & $1.4 \times 10^{-5}$ & $1.4 \times 10^{-6}$ & $1.4 \times 10^{-9}$ & $2.8 \times 10^{-8}$ & $-2.8 \times 10^{-9}$ & $10^{-7}$ & $2 \times 10^{-6}$ \\
\hline\hline
Case & (1) & (2) & (3) & (4) & (5) & (6) & (7) & (8) & (9) & (10) & (11) & (12) \\
\hline
$(1)$ & $0$   & $73.5$ & $95.1$ & $51.8$ & $0.139$ & $14.8$ & $1.50$ & $0.00844$ & $ 0.0473$ & $-0.0113$ & $ -0.385$ & $5.29$ \\[\defaultaddspace]
$(2)$ & $69.1$ & $0$ & $23.7$ & $19.4$ & $69.1$ & $72.7$ & $69.3$ & $69.1$ & $69.1$ & $69.1$ & $69.0$ & $68.7$ \\[\defaultaddspace]
$(3)$ & $90.6$ & $19.2$ & $0$ & $41.0$ & $90.6$ & $93.9$ & $90.8$ & $90.6$ & $90.6$ & $90.6$ & $90.6$ & $90.2$ \\[\defaultaddspace]
$(4)$ & $47.3$ & $23.8$ & $45.4$ & $0$ & $47.4$ & $51.6$ & $47.6$ & $47.3$ & $47.3$ & $47.3$ & $47.3$ & $47.0$ \\[\defaultaddspace]
$(5)$ & $0.0920$ & $73.5$ & $95.1$ & $51.8$ & $0$ & $14.7$ & $1.38$ & $ 0.0905$ & $ 0.0611$ & $ 0.0950$ & $ -0.422$ & $5.22$ \\[\defaultaddspace]
$(6)$ & $14.7$ & $76.9$ & $98.2$ & $55.7$ & $14.6$ & $0$ & $13.3$ & $14.7$ & $14.7$ & $14.7$ & $14.6$ & $12.2$ \\[\defaultaddspace]
$(7)$ & $1.45$ & $73.7$ & $95.3$ & $52.0$ & $1.33$ & $13.3$ & $0$ & $1.45$ & $1.42$ & $1.46$ & $1.22$ & $4.53$ \\[\defaultaddspace]
$(8)$ & $-0.00818$ & $73.5$ & $95.1$ & $51.8$ & $0.138$ & $14.8$ & $1.50$ & $0$ & $ 0.0457$ & $-0.0137$ & $ -0.386$ & $5.29$ \\[\defaultaddspace]
$(9)$ & $-0.0223$ & $73.5$ & $95.1$ & $51.8$ & $0.109$ & $14.8$ & $1.47$ & $-0.0227$ & $0$ & $-0.0213$ & $ -0.398$ & $5.27$ \\[\defaultaddspace]
$(10)$ & $0.0121$ & $73.5$ & $95.1$ & $51.8$ & $0.142$ & $14.8$ & $1.50$ & $ 0.0150$ & $ 0.0506$ & $0$ & $ -0.384$ & $5.29$ \\[\defaultaddspace]
$(11)$ & $0.558$ & $73.5$ & $95.0$ & $51.7$ & $0.541$ & $14.6$ & $1.44$ & $0.558$ & $0.551$ & $0.559$ & $0$ & $5.00$ \\[\defaultaddspace]
$(12)$ & $6.10$ & $73.2$ & $94.7$ & $51.6$ & $6.04$ & $12.6$ & $5.46$ & $6.10$ & $6.08$ & $6.10$ & $5.81$ & $0$ \\[\defaultaddspace]
\hline
\end{tabular}
\end{adjustbox}
\caption{Values of $\sqrt{|\Delta \chi^2|} \; {\rm sign}(\Delta \chi^2)$ for a single realization at $N_{\rm side} = 64$, using the {\it Planck\/} mask-76 extended to exclude regions at $|b| \le 30^\circ$.  We 
adopt $E_{\rm for} = 10^{-2}$ and $E_{\rm cal} = 10^{-4}$, and consider each of the 19 frequency channels
and all 171 independent combinations of pairs of different frequencies.}
\label{A_corr_cross_mask76ExtGal}
\end{table}

We use the ``{\it Planck\/} common mask 76'' (in temperature)
and the extension of this mask that excludes all pixels at
$|b| \le 30^\circ$.\footnote{We also considered a mask that excludes also all pixels within $30^\circ$ of the Ecliptic plane, to avoid zodiacal-light contamination,
but the resulting map has considerably less statistical power due to the
low overall sky fraction.}

Having already addressed the rms uncertainty in the $\sqrt{|\Delta \chi^2|} \; {\rm sign}(\Delta \chi^2)$ estimates,
in this test (as well as in the following ones)
we will consider a single realization only, in order to avoid repeating a huge number of unnecessary simulations.
For the sake of simplicity, we omit reporting the results found in the less stable approach (b).

In the case of the extended mask and including also the cross-comparisons between different frequency channels (approach (c)),
we found a significant improvement (see Table~\ref{A_corr_cross_mask76ExtGal}) with respect to the results based on the full sky; the significance of the
CIB amplitude recovery improves by about 50\,\% and that on the BE distortion improves by about 20\,\%.
This indicates the relevance of optimising the selection of the sky region
for which the analysis is applied, and of comparing results obtained with different masks.

\subsubsection{Varying assumptions on potential foreground and calibration residuals}
\label{var_cal_res}

We now consider the implications of different levels of potential residuals, evaluating both better and worse cases with respect to the reference case analysed before.
Given the results obtained in the previous section we will focus on the case of the extended mask. We present here the main outcomes of this analysis,
while the tables with the corresponding numerical results are reported in Appendix \ref{app_var_cal_res} for sake of completeness.

\begin{itemize}

\item

{\bf Improving foreground subtraction}

\noindent
We now evaluate the improvement in component separation of total intensity maps by considering the case of $E_{\rm for} = 10^{-3}$.
The results, summarised in
Table~\ref{Optimistic_corr_cross_mask76ExtGal} (for approach (c)), can be compared with those
of Table~\ref{A_corr_cross_mask76ExtGal}.
We find an improvement by a factor of approximately 10 in the recovery of the
CIB amplitude, in line with that assumed in foreground removal, and by a factor of 5 (or 6) in the recovery of $\mu_0$ (or $u$), implying that
calibration uncertainty is relatively more important for estimating CMB distortion parameters than for estimating the CIB amplitude. In fact, the CIB is better constrained at higher frequencies,
where foregrounds are more relevant.

\item

{\bf The case of poorer calibration}

\noindent
We discuss here the degradation in sensitivity entailed by keeping
$E_{\rm for} = 10^{-2}$, but replacing CORE's calibration-accuracy goal of $E_{\rm cal} = 10^{-4}$ with
 $E_{\rm cal} = 10^{-3}$ at $\nu \le 295$\,GHz and $E_{\rm cal} = 10^{-2}$ at $\nu \ge 340$\,GHz.

The results, summarised in Table~\ref{Pessimistic_corr_cross_mask76ExtGal} (for approach (c)), can be compared with those
of Table~\ref{A_corr_cross_mask76ExtGal}.
In spite of the assumed degradation in calibration accuracy at high frequencies
(particularly relevant for the CIB),
we find that the recovery of the CIB amplitude is very weakly affected, while the significance of the $\mu_0$ (or $u$) determination
degrades by factor of approximately 2 (or 25--30\,\%).
This result strengthens the conclusion of the previous subsection that
calibration uncertainty is relatively more important for estimating CMB distortion parameters than for the CIB amplitude.

For the set of assumptions adopted here, we find an improvement with respect to FIRAS by factor of
around 20 in the recovery of the CIB amplitude, 15 on the constraints on the Comptonization parameter $u$ (or for its detection, at a level of about 3--4$\,\sigma$ for astrophysical reionization models),
and about 24 for the constraints on chemical potential $\mu_0$.

We finally consider a worst case scenario with $E_{\rm for} = E_{\rm cal} = 10^{-2}$. Even in this situation, we find an improvement with respect to FIRAS by a factor
of 17 in the recovery of the CIB amplitude, and a factor of a few for CMB spectral distortion parameters, specifically around
4 for BE distortions and a marginal detection
of astrophysical reionization models for Comptonization distortions.

\item

{\bf Poorer calibration together with improved foreground subtraction}

\noindent
We now consider a combination of the two cases above, i.e.,
a further improvement in component separation of total intensity maps represented by $E_{\rm for} = 10^{-3}$
and a calibration accuracy parameterised by $E_{\rm cal} = 10^{-3}$ at $\nu \le 295$\,GHz and $E_{\rm cal} = 10^{-2}$ at $\nu \ge 340$\,GHz.

The results obtained in approach (c) are summarised in Table~\ref{Intermediate_corr_cross_mask76ExtGal}.
We find that the significance of CIB amplitude recovery is intermediate between the results found in the previous cases,
while the degradation due to the poorer calibration is approximately compensated by the
improvement due to better foreground subtraction in the case of Comptonization distortions, but
only partially compensated in the case of BE distortions.

Overall, our analysis indicates
that the relevance of calibration accuracy increases from CIB amplitude to Comptonization-distortion and to BE-distortion recovery,
while the relevance of the quality of foreground subtraction increases from BE distortions to Comptonization distortions and to CIB amplitude recovery.
This conclusion reflects the increase of the foreground level and of the relative amplitude of the imprints left by the three types of signals at increasing frequencies (for CORE).

\item

{\bf Varying the reference angular scale}

\noindent
We finally consider assumptions of errors in foreground subtraction and in calibration in the range discussed above, but at smaller angular scales, specifically
at $N_{\rm side} = 256$.  The corresponding pixel linear size ($\simeq 13.7$ arcmin) is similar to the FWHM resolution of CORE channels at $\nu \lsim 80$\,GHz that are necessary for the mitigation of
low-frequency foreground emission.

With this adopted set-up and considering the most advantageous approach (i.e., approach (c)),
assuming a foreground mitigation parameterised by $E_{\rm for} = 10^{-3}$,
we find (see Table \ref{standard_corr_cross_mask76ExtGal_ns256}) an improvement with respect to FIRAS by a factor of
80--90 for the recovery of CIB amplitude, around 80 on the constraints for the Comptonization parameter $u$ 
(implying a precise measure of the energy injections associated to astrophysical reionization models),
and about 150 on the constraints on chemical potential $\mu_0$.
Adopting $E_{\rm for} = 10^{-2}$, we find instead an improvement by a factor of
75 for the recovery of CIB amplitude, 50 for the
constraints on the Comptonization parameter $u$, 
and 80 for the constraints on the chemical potential $\mu_0$.

We further consider the same set-up but at $N_{\rm side} = 128$, i.e., with a pixel side 2 times larger. As expected, we find results intermediate between those derived at $N_{\rm side} = 64$ and 256.

\end{itemize}

\begin{table}[t!]
\begin{center}
\begin{adjustbox}{width=1\textwidth}
{\begin{tabular}{|c|c|c|c|c|c|}
    \hline
      & $E_{\rm cal}$ (\%) & $E_{\rm for}$ (\%) &  CIB amplitude &  Bose-Einstein & Comptonization \\
    \hline
    \hline
    Ideal case, all sky & - & - & $\simeq 4.4 \times 10^3$ & $\simeq 10^3$ & $\simeq 6.0  \times 10^2$ \\
        \hline
    All sky & $10^{-4}$ & $10^{-2}$ & $\simeq 15$  & $\simeq 42$ & $\simeq 18$ \\
    P76      & $10^{-4}$ & $10^{-2}$ & $\simeq 19$ & $\simeq 42$ & $\simeq 18$ \\
    P76ext & $10^{-2}$ & $10^{-2}$ & $\simeq 17$ & $\sim 4$ & $\sim 2$ \\
    P76ext & $10^{-4}$ & $10^{-2}$ & $\simeq 22$  & $\simeq 47$ & $\simeq 21$ \\
    P76ext & $10^{-4}$ & $10^{-3}$ & $\simeq 2.1 \times 10^2$ & $\simeq 2.4  \times 10^2$ & $\simeq 1.1  \times 10^2$ \\
    P76ext & $10^{-3}_{(\le 295)}$--$10^{-2}_{(\ge 340)}$ & $10^{-2}$ & $\simeq 19$ & $\simeq 26$ & $\simeq 11$ \\
    P76ext & $10^{-3}_{(\le 295)}$--$10^{-2}_{(\ge 340)}$ & $10^{-3}$ & $\simeq 48$ & $\simeq 35$ & $\simeq 15$ \\
           \hline
    P76ext, $N_{\rm side}=128$ & $10^{-3}_{(\le 295)}$--$10^{-2}_{(\ge 340)}$ & $10^{-2}$ & $\simeq 38$ & $\simeq 51$ & $\simeq 23$ \\
    P76ext,  $N_{\rm side}=128$ & $10^{-3}_{(\le 295)}$--$10^{-2}_{(\ge 340)}$ & $10^{-3}$ & $\simeq 43$ & $\simeq 87$ & $\simeq 39$ \\
    P76ext, $N_{\rm side}=256$ & $10^{-3}_{(\le 295)}$--$10^{-2}_{(\ge 340)}$ & $10^{-2}$ & $\simeq 76$ & $\simeq 98$ & $\simeq 44$ \\
    P76ext, $N_{\rm side}=256$ & $10^{-3}_{(\le 295)}$--$10^{-2}_{(\ge 340)}$ & $10^{-3}$ & $\simeq 85$ & $\simeq 1.6 \times 10^2$ & $\simeq 73$ \\
    \hline
\end{tabular}}
\end{adjustbox}
\end{center}
\caption{\small Predicted improvement in the recovery of the distortion parameters discussed in the text with respect to FIRAS for different calibration and foreground residual assumptions. This table summarizes the
results derived with approach (c). ``P06'' stands for the {\it Planck\/} common mask, while ``P06ext'' is the extended P06 mask. When not explicitly stated, all values refer to $E_{\rm cal}$ and $E_{\rm for}$ at $N_{\rm side}=64$.}
\label{tab:improv_FIRAS}
\end{table}

\subsection{Summary of simulation results}
\label{sec:sum_simdist}

We have presented above a large set of simulations for different choices of the parameters characterising foreground and calibration residual levels.
The main results are summarised in Table~\ref{tab:improv_FIRAS} in terms of improvements with respect to FIRAS, 
in order to parametrically quantify the accuracy required to achieve significant improvements. 
For other values of $E_{\rm for}$ and $E_{\rm cal}$, we find an almost linear dependence on them for the improvement factor in parameter recovery.

\section{Discussion and conclusions}
\label{sec:end}

We have carried out a detailed investigation of three distinct scientific
implications coming from exploitation of the observer's peculiar velocity
effects in CORE maps. The determination of the CMB dipole amplitude and
direction is an important observable in modern cosmology. It provides
information on our velocity with respect to the CMB reference frame, which
is expected to dominate the effect. Related investigations in other
wavebands, which exploit signals from different types of astrophysical
sources, probe different shells in redshift, and together provide an important
test of fundamental principles in cosmology. In particular, the alignment of
the CMB dipole with those independently measured from galaxy and cluster
catalogues is regarded as indirect proof of the kinematic origin of the CMB
dipole.  The specific relation between the amplitudes of the CMB and large-scale
structure dipoles, predicted by the linear perturbation theory, has been used
to obtain estimates of the redshift-space distortion parameter independent of
(but consistent with) those coming from redshift surveys.
It is thus important to look for possible departures from a purely kinematic
character for the CMB dipole. In this
context, surveys from space are clearly appealing, since they represent the
best (and perhaps only) way to precisely measure this large-scale signal.

We performed detailed simulations in the context of a mission like CORE, to
understand the expectations, and potential issues arising from future CMB
surveys
beyond the already excellent results produced by {\it Planck}. The sampling of
the sky turns out to be the main limiting factor for the precise measurement of
the dipole direction and (obviously together with calibration) also a crucial
limiting factor for the precise measurement of dipole amplitude. We found
that the recovered uncertainty scales linearly with the map pixel linear size
(i.e., inversely with $N_{\rm side}$).
Although maps can be oversampled through a proper scanning
strategy and by setting the sampling time of the data acquisition well below
that corresponding to the beam resolution, it is clear that the experimental
resolution plays a crucial role in this respect. Among CMB space missions
proposed for the future, CORE has the best angular resolution. The dipole
direction determination can be averaged over the various frequency channels,
improving accuracy and providing cross-checks for systematics.
With the assumption of a pure blackbody,
the same holds for the amplitude. However, when searching for dipole spectral
signatures, increasing the accuracy at each frequency (which results from a
better sky sampling) turns out to be even more important.

An observer moving with respect to the CMB rest frame will also see boosting
imprints on the CMB at $\ell>1,$ due to Doppler and aberration effects, which
can be measured in harmonic space as correlations between $\ell$ and $\ell+1$
modes (assuming that the CMB is statistically isotropic in its rest frame). Such
a signal can be measured independently in temperature and polarization, which
constitutes a new consistency check, with a signal-to-noise ratio of about 8
for $TT$, 7 for $TE+ET$ and 7 for $EE$. Overall, CORE can achieve a
signal-to-noise ratio of almost 13, which improves on the capabilities of
{\it Planck\/} (about ${\rm S/N}\simeq 4$, only in $TT$) and is essentially
that of an
ideal cosmic-variance-limited experiment up to $\ell \simeq 2000$. We stress
the importance of performing high-sensitivity measurements at close to arcminute
resolution in order to be sensitive to the correlations at high multipoles that
yield most of the signal. Since CORE will also provide good measurements of the
tSZ effect and the CIB, which are also assumed to be statistically isotropic
in the CMB rest frame, we additionally investigated boosting effects
in these maps. However, we found that the aberration effect on tSZ maps and
the boosting effects on the CIB are smaller than in the CMB maps, and that the
predicted signal-to-noise is less than 1 in both cases.

Beyond FIRAS, great hopes are expected for PIXIE, which has been proposed to
NASA to observe CMB polarization and the CMB spectrum with degree resolution
and is designed to have a precision about $10^3$ times better than FIRAS,
mainly relying on the achievement of extreme quality in its absolute
calibration, and a corresponding similar improvement on CMB spectral
distortion parameters \cite{2011JCAP...07..025K}.  Note that even if PIXIE
fails to fully achieve these ambitious goals, an improvement in calibration
precision of even one or two orders of magnitude with respect to FIRAS
calibration, in addition to being intrinsically interesting for strengthening
the limits on CMB distortion parameters, will imply an analogous improvement
for the calibration of other CMB projects. In general, combining results
from experiments like PIXIE and CORE will offer a chance to have maps with
substantialily improved calibration, sensitivity, and resolution.

CMB anisotropy missions will not perform absolute measurements of the CMB
spectrum, but can observe the frequency spectral behaviour of the CMB and CIB dipoles.
We exploit the sensitivity of an experiment like CORE for the recovery of the
parameters $u$ and $\mu_0$ of Comptonization and BE spectral distortions, as well as for the amplitude of the CIB spectrum. Assuming perfect relative calibration and absence of foreground contamination,
the CORE sensitivity and frequency coverage, combined with its resolution (to cope with sampling uncertainty), could allow us to achieve an improvement with respect to FIRAS by a factor of around
1--$4\times 10^3$, 500--600, and 600--1000 in the recovery of the CIB spectrum amplitude, $u$, and $\mu_0$, respectively; the best results are obtained
from the joint information contained in each of the frequency channels independently {\it and} in all the independent combinations of pairs of different frequencies.
Combining pairs of different frequencies turns out to be particularly advantageous for the CIB dipole spectrum, since it exhibits a steeper frequency behaviour.

As expected, foregrounds are critical in both absolute and differential
methods. Relative calibration accuracy is an important limiting factor in CMB
anisotropy experiments in general and even more so for analyses based on the
dipole. In current data analysis pipelines the dipole itself is
in fact typically used for calibration, which raises the issue of a circular
argument.  However, for all-sky mapping experiments (like {\it WMAP\/} and
{\it Planck}), the {\it orbital\/} dipole from the Earth and satellite motion
is ultimately used for calibration, rather than the CMB dipole itself.
Precise calibration is always challenging, and it is unclear what the limiting
step will be for any new experiment.  Nevertheless, in principle it will be
possible to measure the spectrum of the dipole with an anisotropy experiment.
In general, improving and extending calibration methods is crucial for
these analyses. Various approaches can be integrated into the data reduction
design, ranging from a better
instrumental characterization to cross-correlation between different CMB
surveys and substantial refinements in astronomical calibration sources.

We have carried out a large set of simulations, summarised in Table~\ref{tab:improv_FIRAS}, 
to parametrically quantify the accuracy required to achieve significant improvements with respect to FIRAS. 

We find that the importance of the impact of calibration errors decreases from
BE distortions to Comptonization distortions and to the CIB amplitude,
while the opposite holds for
the impact of foreground contamination (in agreement with the increase with
frequency of foreground level and of the relative amplitude of the imprints
left by the three types of signal). Applying suitable masks also yields an
improvement in parameter estimation.

In the case of 1\,\% accuracy (at a reference scale of about $1^\circ$) in
both foreground removal and relative calibration (i.e., $E_{\rm for} = E_{\rm
cal} = 10^{-2}$), CORE will be able to improve the recovery of the CIB spectrum
amplitude of a factor of about 17, to achieve a marginal detection of the
energy release associated with astrophysical reionization models, and to improve
by a factor of approximately 4 the limits on early energy dissipations.
On the other hand, an
improvement of a factor of 20 for CIB amplitude, of 10 for $u$,
and of around 25 for the chemical potential $\mu_0$, is found by improving the
relative calibration error to $\simeq 0.1$\,\%. Any further improvement in
foreground mitigation and calibration will enable still more precise
results to be achieved.

\acknowledgments

Partial support by ASI/INAF Agreement 2014-024-R.1 for the
{\it Planck\/} LFI Activity of Phase E2 and by ASI through the contract I-022-11-0 LSPE is acknowledged.
C.H.-M. acknowledges financial support of the Spanish Ministry of Economy and Competitiveness via I+D project AYA-2015-66211-C2-2-P.
J.G.N. acknowledges financial support from the Spanish MINECO for a {\it Ramon y Cajal} fellowship (RYC-2013-13256) and the I+D 2015 project AYA2015-65887-P (MINECO/FEDER).
C.J.M. is supported by an FCT Research Professorship, contract reference IF/00064/2012, funded by FCT/MCTES (Portugal) and POPH/FSE (EC).
M.Q. is supported by the Brazilian research agencies CNPq and FAPERJ.
We acknowledge the use of the ESA Planck Legacy Archive.
Some of the results in this paper have been derived using the {\tt HEALPix} package.
The use of the computational cluster at INAF-IASF Bologna is acknowledged.
It is a pleasure to thank Arpine Kozmanyan for useful discussions on the
{\tt CosmoMC} sampler.


\appendix


\section{Appendix -- Likelihoods of CMB dipole parameters}\label{sec:Likelihoods}

For the sake of completeness, we report here the likelihoods computed for CMB dipole parameters and the confidence levels in their estimation.
We limit the presentation here to the lowest and highest resolutions among those exploited in this analysis.

\clearpage

\begin{figure}[ht!]
\centering
 \includegraphics[scale=0.32]{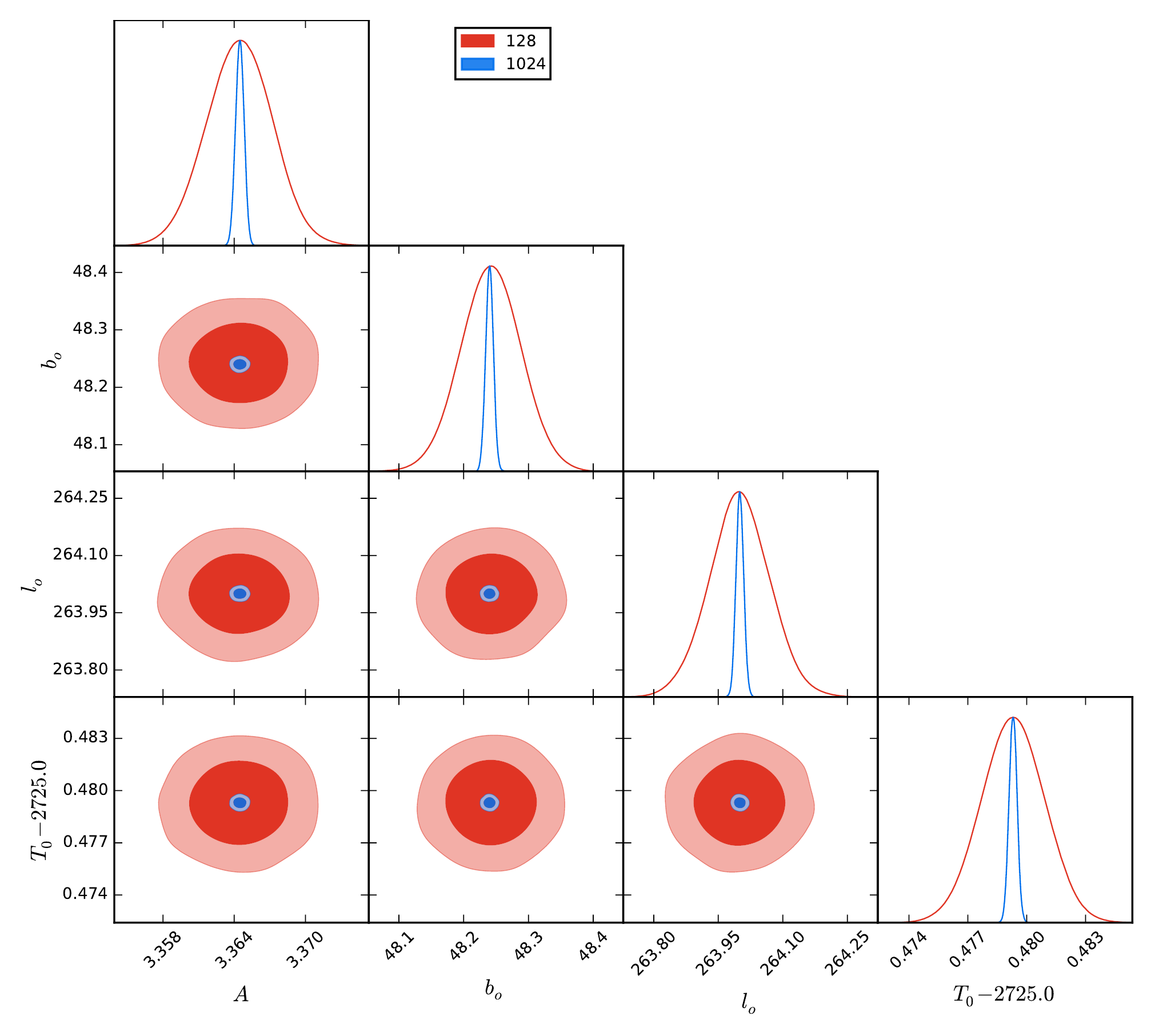}
 \includegraphics[scale=0.32]{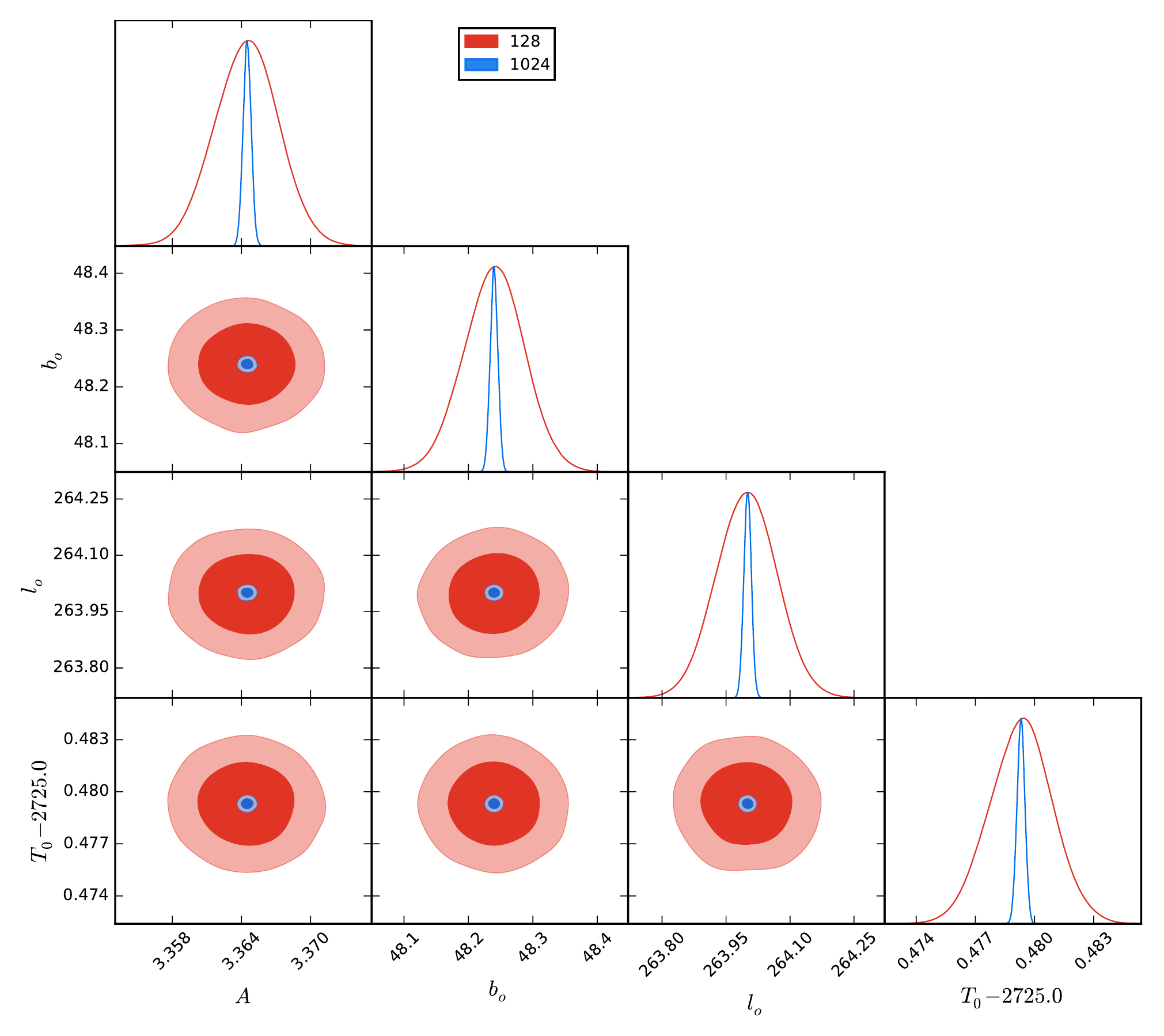}
 \includegraphics[scale=0.32]{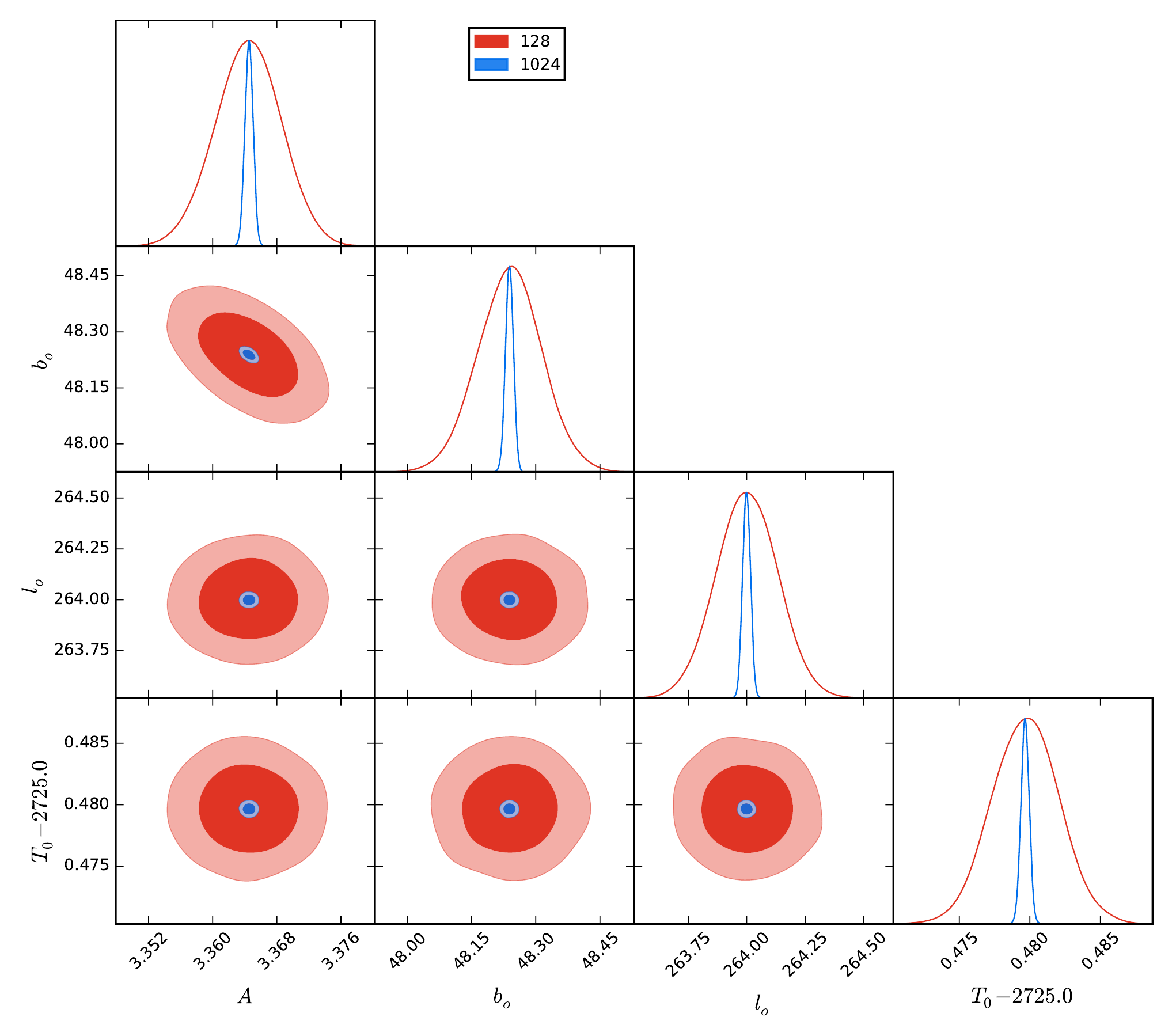}
 \includegraphics[scale=0.32]{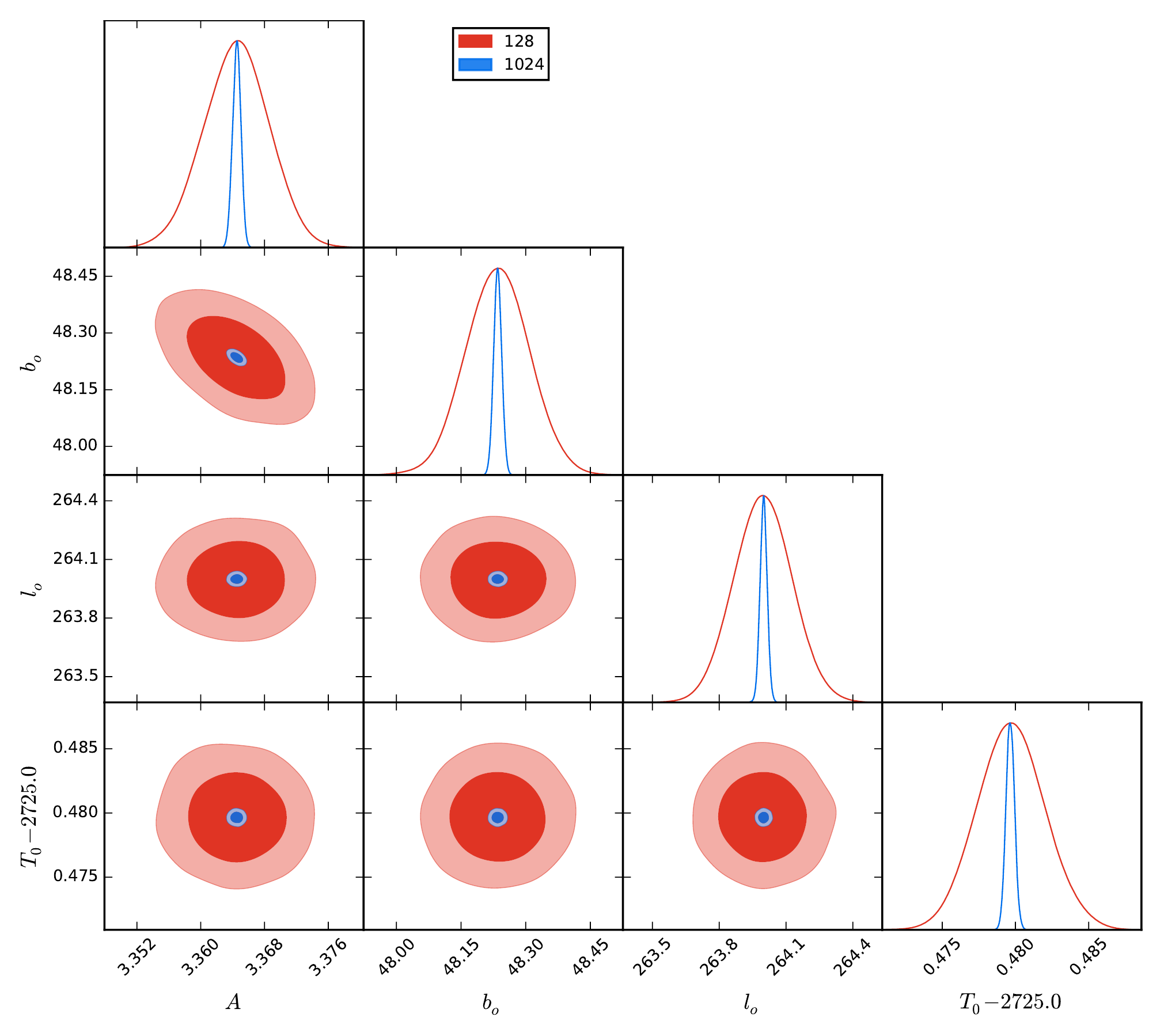}
  \caption{Marginalised likelihoods, and 68\% and 95\% contours of the parameters $A$, $b_0$, $l_0$, and $T_{0}$ at $N_{\rm side}=128$ (red) and at $N_{\rm side}=1024$ (blue): 
top left, dipole only; top right, dipole+noise; bottom left,
dipole+noise+mask; and bottom right, dipole+noise+mask+systematics. The reference frequency channel is
60\,GHz and the noise is 7.5\,$\mu$K.arcmin. The mask used here is the {\it Planck\/} Galactic mask extended to cut out
$\pm30^\circ$ of the Galactic plane. The level of systematics correspond to the pessimistic expectation of calibration errors and sky residuals.}%
\label{fig:conflevel128+1024}
\end{figure}
\begin{table}[ht!]
\begin{adjustbox}{width=1\textwidth}
\begin{tabular}{|c|c|c|c|c|}
\hline
$N_{\rm side}=128$ & $A({\rm mK})$ & $b_0(^\circ)$ & $l_0(^\circ)$ & $T_{0}({\rm mK})$\tabularnewline
\hline
\hline
dipole & $\ensuremath{3.3644\pm0.0028}$ & $48.242\pm0.047$ & $\ensuremath{263.999\pm0.070}$ & $2725.4793\pm0.0016$\tabularnewline
\hline
dip+noi & $\ensuremath{3.3644\pm0.0028}$ & $48.240\pm0.047$ & $\ensuremath{263.998\pm0.071}$ & $2725.4793\pm0.0016$\tabularnewline
\hline
dip+noi+mask & $\ensuremath{3.3644\pm0.0041}$ & $48.240\pm0.075$ & $\ensuremath{264.00\pm0.13}$ & $2725.4797\pm0.0024$\tabularnewline
\hline
dip+noi+mask+sys & $\ensuremath{3.3645\pm0.0041}$ & $48.235\pm0.074$ & $\ensuremath{264.00\pm0.13}$ & $2725.4797\pm0.0023$\tabularnewline
\hline
\hline
$N_{\rm side}=1024$ & $A({\rm mK})$ & $b_0(^\circ)$ & $l_0(^\circ)$ & $T_{0}({\rm mK})$\tabularnewline
\hline
\hline
dipole & $\ensuremath{3.36447\pm0.00036}$ & $48.2399\pm0.0060$ & $\ensuremath{264.0002\pm0.0088}$ & $2725.47930\pm0.00020$\tabularnewline
\hline
dip+noi & $\ensuremath{\ensuremath{3.36450\pm0.00035}}$ & $\ensuremath{48.2398\pm0.0059}$ & $\ensuremath{\ensuremath{264.0005\pm0.0087}}$ & $2725.47931\pm0.00020$\tabularnewline
\hline
dip+noi+mask & $\ensuremath{3.36454\pm0.00051}$ & $\ensuremath{48.2387\pm0.0091}$ & $264.000\pm0.017$ & $2725.47966\pm0.00029$\tabularnewline
\hline
dip+noi+mask+sys & $\ensuremath{3.36451\pm0.00052}$ & $\ensuremath{48.2352\pm0.0092}$ & $\ensuremath{264.000\pm0.016}$ & $\ensuremath{2725.47965\pm0.00029}$\tabularnewline
\hline
\end{tabular}
\end{adjustbox}
\caption{68\,\% confidence levels of the parameters $A$, $b_0$, $l_0$, and $T_{0}$ from Fig.~\ref{fig:conflevel128+1024}.}%
\label{table:conflevel128+1024}
\end{table}


\clearpage{}

\begin{figure}[ht!]
\centering
 \includegraphics[scale=0.39]{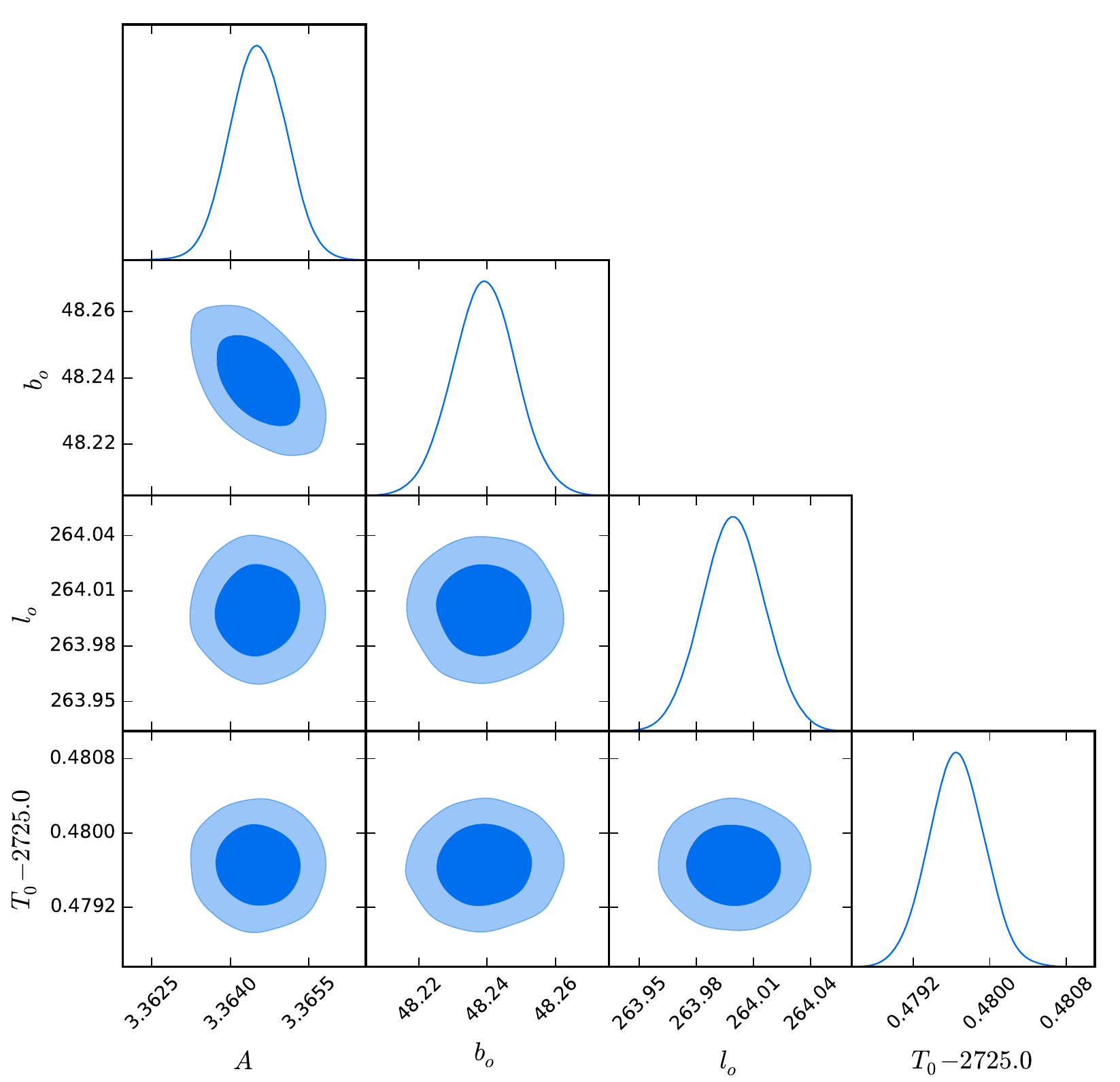}%
 \includegraphics[scale=0.39]{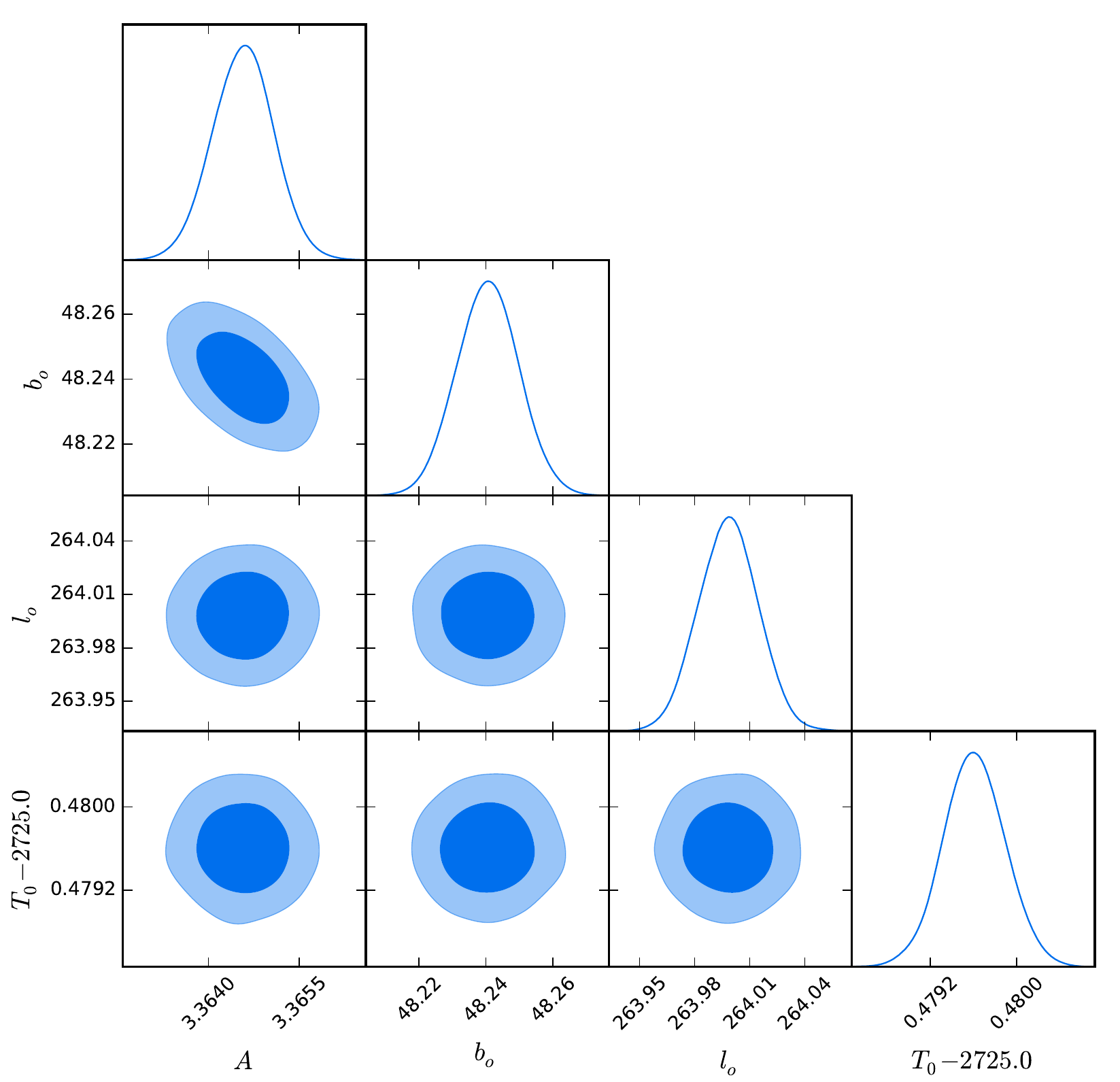}
 \includegraphics[scale=0.39]{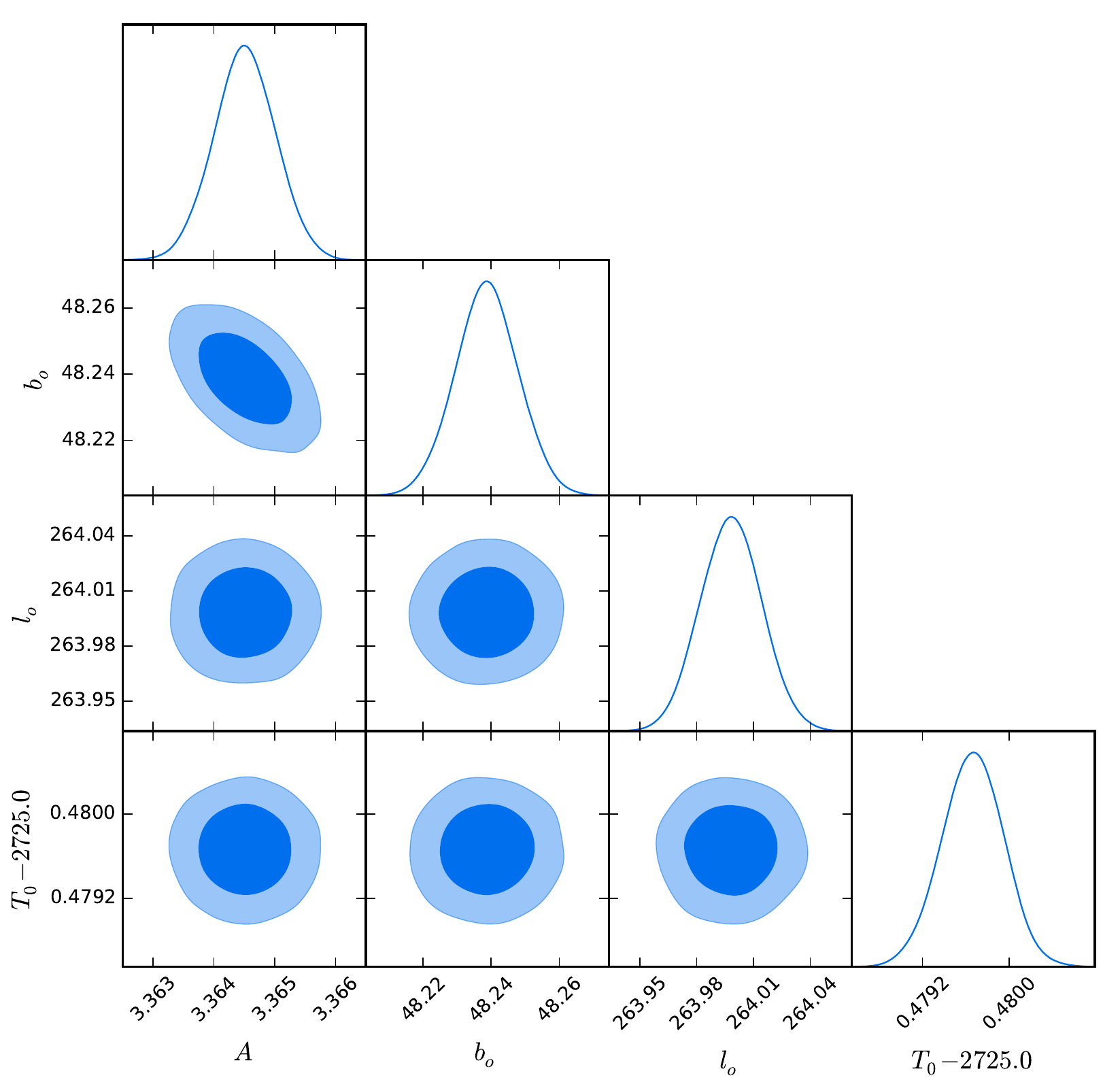}%
 \includegraphics[scale=0.39]{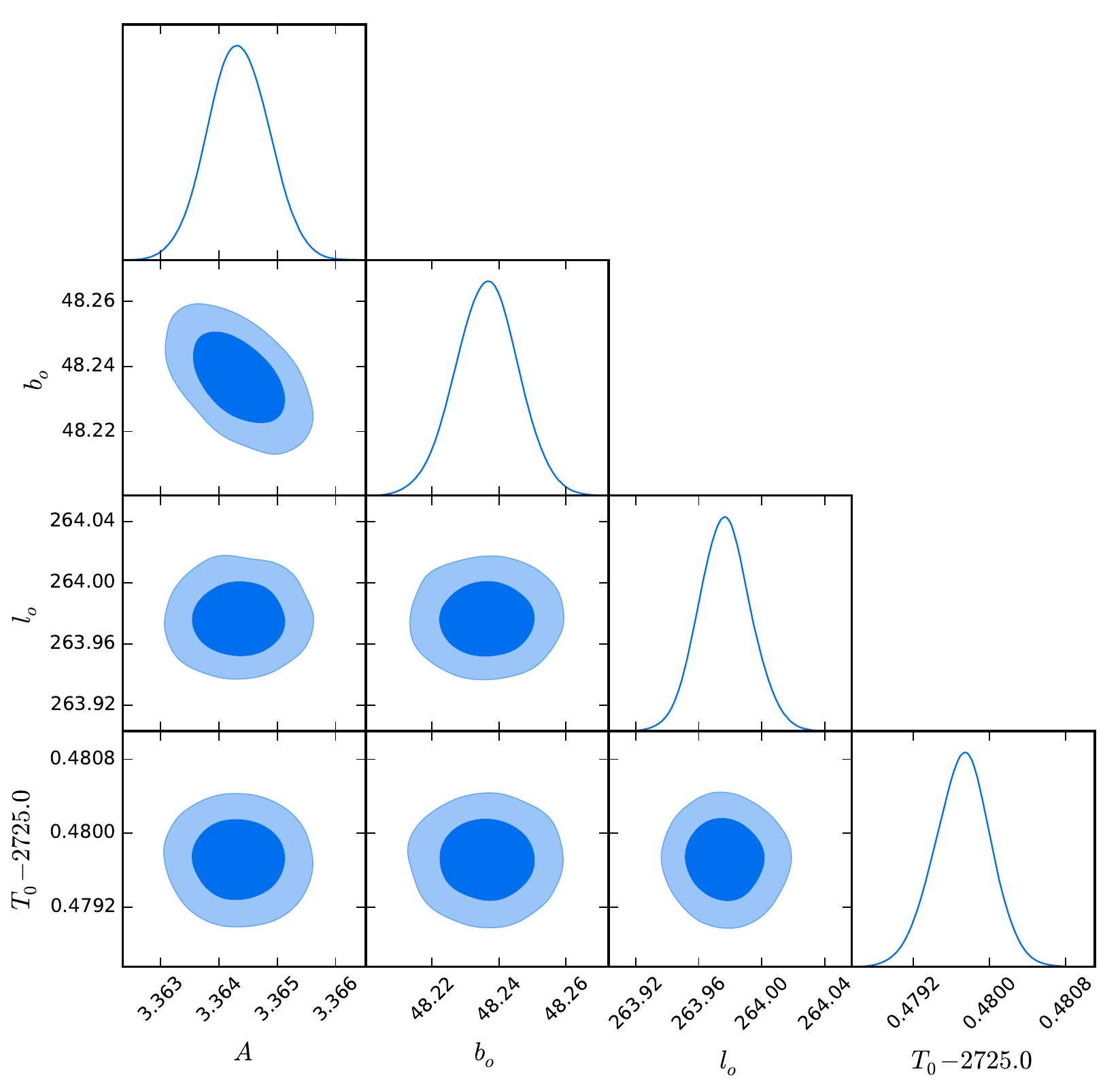}
  \caption{Marginalised likelihoods, and 68\% and 95\% contours of the parameters $A$, $b_0$, $l_0$, and $T_{0}$.
Input maps are at $N_{\rm side}=1024$, with noise, mask and residuals (calibration errors and sky residuals). {\it Top}: 100\,GHz with optimistic (left) and pessimistic (right) systematics. {\it Bottom}: 220\,GHz with optimistic (left) and pessimistic (right) systematics.}%
\label{fig:conflevel100220}
\end{figure}
\begin{table}[ht!]
\begin{adjustbox}{width=1\textwidth}
\begin{tabular}{|c|c|c|c|c|}
\hline
$N_{\rm side}=1024$  & $A({\rm mK})$ & $b_0(^\circ)$ & $l_0(^\circ)$ & $T_{0}({\rm mK})$\tabularnewline
\hline
\hline
60\,GHz, good sys & $\ensuremath{3.36454\pm0.00052}$ & $\ensuremath{48.2387\pm0.0093}$ & $\ensuremath{263.999\pm0.016}$ & $\ensuremath{2725.47965\pm0.00029}$\tabularnewline
\hline
60\,GHz, bad sys & $\ensuremath{3.36451\pm0.00052}$ & $\ensuremath{48.2352\pm0.0092}$ & $\ensuremath{264.000\pm0.016}$ & $\ensuremath{2725.47965\pm0.00029}$\tabularnewline
\hline
100\,GHz, good sys & $\ensuremath{3.36453\pm0.00053}$ & $\ensuremath{48.2393\pm0.0093}$ & $\ensuremath{264.000\pm0.016}$ & $\ensuremath{2725.47965\pm0.00029}$\tabularnewline
\hline
100\,GHz, bad sys & $\ensuremath{3.36457\pm0.00051}$ & $\ensuremath{48.2406\pm0.0093}$ & $\ensuremath{263.998\pm0.016}$ & $\ensuremath{2725.47961\pm0.00029}$\tabularnewline
\hline
145\,GHz, good sys & $\ensuremath{3.36452\pm0.00052}$ & $\ensuremath{48.2391\pm0.0093}$ & $\ensuremath{263.999\pm0.017}$ & $\ensuremath{2725.47967\pm0.00029}$\tabularnewline
\hline
145\,GHz, bad sys & $\ensuremath{3.36434\pm0.00051}$ & $\ensuremath{48.2391\pm0.0091}$ & $\ensuremath{263.996\pm0.017}$ & $\ensuremath{2725.47965\pm0.00029}$\tabularnewline
\hline
220\,GHz, good sys & $\ensuremath{3.36451\pm0.00051}$ & $\ensuremath{48.2387\pm0.0092}$ & $\ensuremath{263.998\pm0.016}$ & $\ensuremath{2725.47966\pm0.00029}$\tabularnewline
\hline
220\,GHz, bad sys & $\ensuremath{3.36434\pm0.00052}$ & $\ensuremath{48.2364\pm0.0094}$ & $\ensuremath{263.977\pm0.016}$ & $\ensuremath{2725.47972\pm0.00029}$\tabularnewline
\hline
\end{tabular}
\end{adjustbox}
\caption{68\,\% confidence level of the parameters $A$, $b_0$, $l_0$, and $T_{0}$ at 60, 100, 145 and 220 GHz (see Fig.~\ref{fig:conflevel100220} for the likelihoods at 100 and 220 GHz).}%
\label{table:conflevel60220}
\end{table}



\begin{figure}[ht!]
\centering
 \includegraphics[scale=0.39]{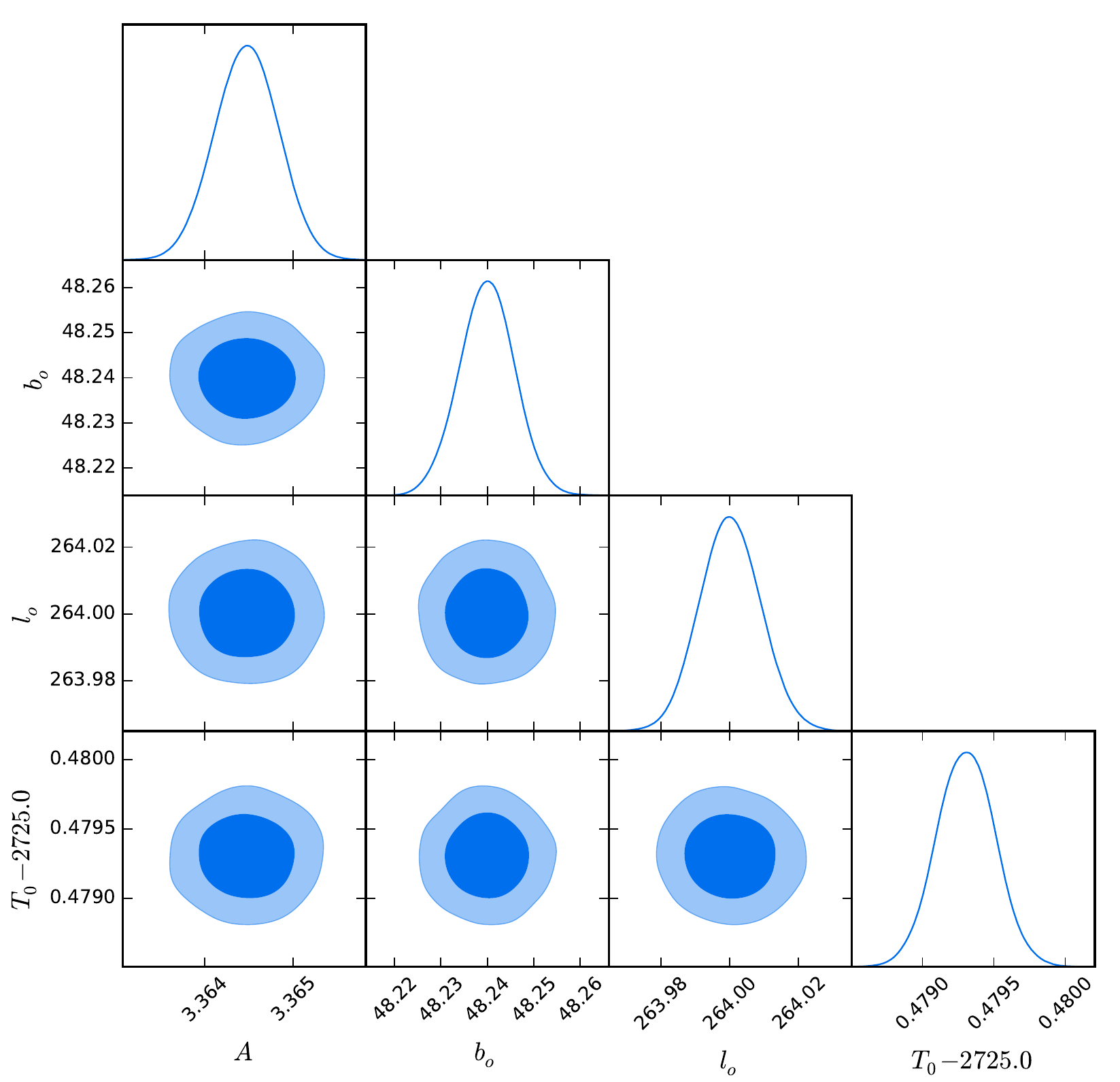}%
 \includegraphics[scale=0.39]{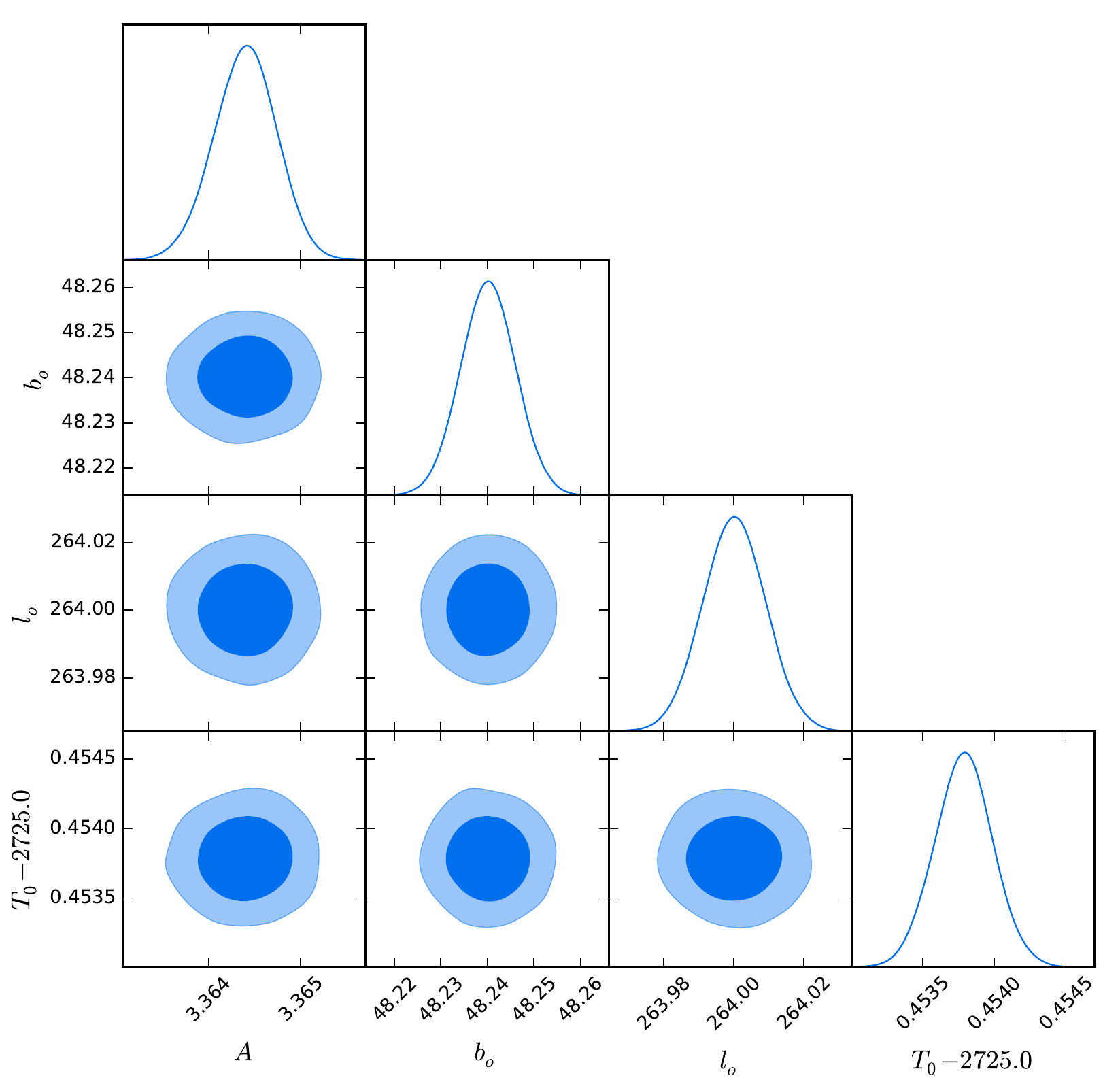}
  \caption{Marginalised likelihoods, and 68\% and 95\% contours of the parameters $A$, $b_0$, $l_0$, and $T_{0}$.
Input maps are dipole-only, at 60\,GHz and at $N_{\rm side}=1024$. On the left: blackbody. On the right: Bose-Einstein (chemical potential $\mu_{0}=1.4\times10^{-5}$).}
\label{fig:conflevelBE}
\end{figure}
\begin{table}[ht!]
\begin{adjustbox}{width=1\textwidth}
\begin{tabular}{|c|c|c|c|c|}
\hline
$N_{\rm side}=1024$  & $A({\rm mK})$ & $b_0(^\circ)$ & $l_0(^\circ)$ & $T_{0}({\rm mK})$\tabularnewline
\hline
\hline
blackbody & $\ensuremath{3.36447\pm0.00036}$ & $48.2399\pm0.0060$ & $\ensuremath{264.0002\pm0.0088}$ & $2725.47930\pm0.00020$\tabularnewline
\hline
Bose-Einstein & $\ensuremath{3.36440\pm0.00034}$ & $\ensuremath{48.2402\pm0.0059}$ & $264.0002\pm0.0090$ & $2725.45379\pm0.00020$\tabularnewline
\hline
\end{tabular}
\end{adjustbox}
\caption{68\,\% confidence level of the parameters $A$, $b_0$, $l_0$, and $T_{0}$ of Fig.~\ref{fig:conflevelBE}.}%
\label{table:conflevelBE}
\end{table}

\clearpage


\section{Appendix -- Rms values from Monte Carlo simulations: ideal case}
\label{app_rms_ideal}

We present here the tables with the estimates of the rms of the $\sqrt{|\Delta \chi^2|} \; {\rm sign}(\Delta \chi^2)$ quoted from a Monte Carlo simulation
at $N_{\rm side} = 64$, using all-sky maps and adopting perfect foreground subtraction and calibration.

\begin{table}[ht!]
\centering
\begin{adjustbox}{width=1\textwidth}
\begin{tabular}{|c|c|ccc|cccccc|cc|}
\hline
{\bf Rms of $\sigma$ level} & \multicolumn{1}{c|}{Current}
& \multicolumn{3}{c|}{FIRAS CIB amplitude}
& \multicolumn{6}{c|}{$(\Delta \varepsilon / \varepsilon_{\rm i})_{z_1}$}
& \multicolumn{2}{c|}{$(\Delta \varepsilon / \varepsilon_{\rm i})_{\rm late}$}\\[\defaultaddspace]
{\bf significance} & blackbody  &  &  (units $10^{-5}$) & & $8 \times 10^{-8}$ & $10^{-5}$ & $10^{-6}$ & $10^{-9}$ & $2 \times 10^{-8}$ & $-2 \times 10^{-9}$ & $4 \times 10^{-7}$ & $8 \times 10^{-6}$ \\
\hline\hline
& \multicolumn{1}{c|}{} &
$I_0^{\rm bf}$ & $+ 1\,\sigma$ & $ - 1\,\sigma$
& \multicolumn{6}{c|}{$\mu_0$}
& \multicolumn{2}{c|}{$u$}\\[\defaultaddspace]
&  & $1.3$ & $1.7$ & $0.9$ & $1.12 \times 10^{-7}$ & $1.4 \times 10^{-5}$ & $1.4 \times 10^{-6}$ & $1.4 \times 10^{-9}$ & $2.8 \times 10^{-8}$ & $-2.8 \times 10^{-9}$ & $10^{-7}$ & $2 \times 10^{-6}$ \\
\hline\hline
Case & (1) & (2) & (3) & (4) & (5) & (6) & (7) & (8) & (9) & (10) & (11) & (12) \\
\hline
$(1)$ & $0$& $4.83$ & $6.99$ & $3.16$ & $0.872$ & $0.568$ & $0.621$ & $0.129$ & $0.562$ & $0.183$ & $0.976$ & $1.06$ \\[\defaultaddspace]
$(2)$ & $4.22$ & $0$& $4.22$ & $0.00$ & $4.22$ & $5.27$ & $5.16$ & $4.22$ & $4.22$ & $4.22$ & $5.16$ & $5.16$ \\[\defaultaddspace]
$(3)$ & $6.99$ & $3.16$ & $0$& $5.16$ & $6.99$ & $4.22$ & $6.99$ & $6.99$ & $6.99$ & $6.99$ & $6.99$ & $6.75$ \\[\defaultaddspace]
$(4)$ & $0.00$ & $3.16$ & $5.16$ & $0$& $0.00$ & $4.22$ & $3.16$ & $0.00$ & $0.00$ & $0.00$ & $3.16$ & $0.00$ \\[\defaultaddspace]
$(5)$ & $1.06$ & $4.83$ & $6.99$ & $3.16$ & $0$& $0.675$ & $0.621$ & $1.06$ & $0.976$ & $1.07$ & $1.05$ & $1.06$ \\[\defaultaddspace]
$(6)$ & $0.667$ & $6.75$ & $6.32$ & $4.22$ & $0.738$ & $0$& $0.823$ & $0.667$ & $0.738$ & $0.568$ & $0.667$ & $1.08$ \\[\defaultaddspace]
$(7)$ & $0.636$ & $4.83$ & $6.75$ & $3.16$ & $0.636$ & $0.675$ & $0$& $0.636$ & $0.632$ & $0.635$ & $0.657$ & $1.14$ \\[\defaultaddspace]
$(8)$ & $0.129$ & $4.83$ & $6.99$ & $3.16$ & $0.871$ & $0.568$ & $0.610$ & $0$& $0.549$ & $0.224$ & $0.976$ & $1.06$ \\[\defaultaddspace]
$(9)$ & $0.578$ & $4.83$ & $6.99$ & $3.16$ & $0.813$ & $0.675$ & $0.619$ & $0.564$ & $0$& $0.606$ & $0.993$ & $1.06$ \\[\defaultaddspace]
$(10)$ & $0.182$ & $4.83$ & $6.99$ & $3.16$ & $0.873$ & $0.667$ & $0.613$ & $0.223$ & $0.587$ & $0$& $0.974$ & $1.06$ \\[\defaultaddspace]
$(11)$ & $1.04$ & $5.16$ & $6.75$ & $3.16$ & $1.14$ & $0.568$ & $0.620$ & $1.05$ & $1.07$ & $1.04$ & $0$& $0.994$ \\[\defaultaddspace]
$(12)$ & $0.994$ & $5.16$ & $6.32$ & $3.16$ & $0.919$ & $1.05$ & $1.08$ & $0.994$ & $1.07$ & $0.994$ & $0.966$ & $0$\\[\defaultaddspace]
\hline
\end{tabular}
\end{adjustbox}
\caption{Rms values of $\sqrt{|\Delta \chi^2|} \; {\rm sign}(\Delta \chi^2)$ from a Monte Carlo simulation at $N_{\rm side} = 64$, full sky,
adopting perfect foreground subtraction and calibration, and considering each of the 19 frequency channels.}
\label{MC_corr_rms_ideal}
\end{table}
\begin{table}
\centering
\begin{adjustbox}{width=1\textwidth}
\begin{tabular}{|c|c|ccc|cccccc|cc|}
\hline
{\bf Rms of $\sigma$ level} & \multicolumn{1}{c|}{Current}
& \multicolumn{3}{c|}{FIRAS CIB amplitude}
& \multicolumn{6}{c|}{$(\Delta \varepsilon / \varepsilon_{\rm i})_{z_1}$}
& \multicolumn{2}{c|}{$(\Delta \varepsilon / \varepsilon_{\rm i})_{\rm late}$}\\[\defaultaddspace]
{\bf significance} & blackbody  &  &  (units $10^{-5}$) & & $8 \times 10^{-8}$ & $10^{-5}$ & $10^{-6}$ & $10^{-9}$ & $2 \times 10^{-8}$ & $-2 \times 10^{-9}$ & $4 \times 10^{-7}$ & $8 \times 10^{-6}$ \\
\hline\hline
& \multicolumn{1}{c|}{} &
$I_0^{\rm bf}$ & $+ 1\,\sigma$ & $ - 1\,\sigma$
& \multicolumn{6}{c|}{$\mu_0$}
& \multicolumn{2}{c|}{$u$}\\[\defaultaddspace]
&  & $1.3$ & $1.7$ & $0.9$ & $1.12 \times 10^{-7}$ & $1.4 \times 10^{-5}$ & $1.4 \times 10^{-6}$ & $1.4 \times 10^{-9}$ & $2.8 \times 10^{-8}$ & $-2.8 \times 10^{-9}$ & $10^{-7}$ & $2 \times 10^{-6}$ \\
\hline\hline
Case & (1) & (2) & (3) & (4) & (5) & (6) & (7) & (8) & (9) & (10) & (11) & (12) \\
\hline
$(1)$ & $ 0.00$ & $31.6$ & $42.2$ & $14.8$ & $2.93$ & $2.67$ & $2.52$ & $0.342$ & $1.52$ & $0.485$ & $2.89$ & $2.18$\\[\defaultaddspace]
$(2)$ & $ 31.6$ & $0$ & $6.75$ & $6.67$ & $0.00$ & $51.6$ & $0.00$ & $31.6$ & $0.00$ & $31.6$ & $31.6$ & $31.6$\\[\defaultaddspace]
$(3)$ & $ 42.2$ & $5.16$ & $0$ & $8.43$ & $42.2$ & $42.2$ & $31.6$ & $42.2$ & $42.2$ & $42.2$ & $42.2$ & $51.6$\\[\defaultaddspace]
$(4)$ & $ 10.8$ & $8.23$ & $11.4$ & $0$ & $10.8$ & $12.9$ & $12.9$ & $10.8$ & $10.8$ & $10.8$ & $11.7$ & $12.9$\\[\defaultaddspace]
$(5)$ & $ 3.11$ & $31.6$ & $42.2$ & $14.8$ & $0$ & $2.59$ & $2.53$ & $3.09$ & $2.70$ & $3.15$ & $4.15$ & $2.20$\\[\defaultaddspace]
$(6)$ & $ 2.30$ & $51.6$ & $42.2$ & $14.5$ & $2.49$ & $0$ & $2.33$ & $2.33$ & $2.67$ & $2.30$ & $2.50$ & $2.46$\\[\defaultaddspace]
$(7)$ & $ 2.59$ & $31.6$ & $31.6$ & $14.5$ & $2.64$ & $2.56$ & $0$ & $2.62$ & $2.60$ & $2.60$ & $2.58$ & $2.31$\\[\defaultaddspace]
$(8)$ & $0.343$ & $31.6$ & $42.2$ & $14.8$ & $2.91$ & $2.56$ & $2.51$ & $0$ & $1.48$ & $0.594$ & $2.91$ & $2.19$\\[\defaultaddspace]
$(9)$ & $ 1.56$ & $31.6$ & $42.2$ & $14.8$ & $2.58$ & $2.59$ & $2.53$ & $1.52$ & $0$ & $1.63$ & $3.29$ & $2.20$\\[\defaultaddspace]
$(10)$ & $0.484$ & $31.6$ & $42.2$ & $14.8$ & $2.96$ & $2.67$ & $2.53$ & $0.593$ & $1.60$ & $0$ & $2.85$ & $2.18$\\[\defaultaddspace]
$(11)$ & $ 2.39$ & $31.6$ & $42.2$ & $14.8$ & $2.93$ & $2.81$ & $2.48$ & $2.40$ & $2.60$ & $2.36$ & $0$ & $2.20$\\[\defaultaddspace]
$(12)$ & $ 2.10$ & $31.6$ & $52.7$ & $14.0$ & $2.13$ & $2.59$ & $2.30$ & $2.09$ & $2.12$ & $2.12$ & $2.10$ & $0$\\[\defaultaddspace]
\hline
\end{tabular}
\end{adjustbox}
\caption{The same as in Table~\ref{MC_corr_rms_ideal}, but
considering all 171 independent combinations of pairs of different frequency channels.}
\label{MC_cross_rms_ideal}
\end{table}
\begin{table}
\centering
\begin{adjustbox}{width=1\textwidth}
\begin{tabular}{|c|c|ccc|cccccc|cc|}
\hline
{\bf Rms of $\sigma$ level} & \multicolumn{1}{c|}{Current}
& \multicolumn{3}{c|}{FIRAS CIB amplitude}
& \multicolumn{6}{c|}{$(\Delta \varepsilon / \varepsilon_{\rm i})_{z_1}$}
& \multicolumn{2}{c|}{$(\Delta \varepsilon / \varepsilon_{\rm i})_{\rm late}$}\\[\defaultaddspace]
{\bf significance} & blackbody  &  &  (units $10^{-5}$) & & $8 \times 10^{-8}$ & $10^{-5}$ & $10^{-6}$ & $10^{-9}$ & $2 \times 10^{-8}$ & $-2 \times 10^{-9}$ & $4 \times 10^{-7}$ & $8 \times 10^{-6}$ \\
\hline\hline
& \multicolumn{1}{c|}{} &
$I_0^{\rm bf}$ & $+ 1\,\sigma$ & $ - 1\,\sigma$
& \multicolumn{6}{c|}{$\mu_0$}
& \multicolumn{2}{c|}{$u$}\\[\defaultaddspace]
&  & $1.3$ & $1.7$ & $0.9$ & $1.12 \times 10^{-7}$ & $1.4 \times 10^{-5}$ & $1.4 \times 10^{-6}$ & $1.4 \times 10^{-9}$ & $2.8 \times 10^{-8}$ & $-2.8 \times 10^{-9}$ & $10^{-7}$ & $2 \times 10^{-6}$ \\
\hline\hline
Case & (1) & (2) & (3) & (4) & (5) & (6) & (7) & (8) & (9) & (10) & (11) & (12) \\
\hline
$(1)$ & $0$ & $52.7$ & $31.6$ & $31.6$ & $2.69$ & $2.17$ & $2.14$ & $0.340$ & $1.50$ & $0.482$ & $1.40$ & $1.57$\\[\defaultaddspace]
$(2)$ & $51.6$ & $0$ & $8.43$ & $6.75$ & $51.6$ & $0.00$ & $51.6$ & $51.6$ & $51.6$ & $51.6$ & $51.6$ & $31.6$\\[\defaultaddspace]
$(3)$ & $31.6$ & $5.16$ & $0$ & $11.0$ & $31.6$ & $52.7$ & $31.6$ & $31.6$ & $31.6$ & $31.6$ & $31.6$ & $51.6$\\[\defaultaddspace]
$(4)$ & $31.6$ & $6.32$ & $12.6$ & $0$ & $31.6$ & $0.00$ & $0.00$ & $31.6$ & $31.6$ & $31.6$ & $31.6$ & $51.6$\\[\defaultaddspace]
$(5)$ & $3.10$ & $52.7$ & $31.6$ & $31.6$ & $0$ & $2.28$ & $2.14$ & $3.08$ & $2.68$ & $3.14$ & $2.15$ & $1.40$\\[\defaultaddspace]
$(6)$ & $2.20$ & $0.00$ & $48.3$ & $31.6$ & $2.21$ & $0$ & $2.17$ & $2.06$ & $2.21$ & $2.20$ & $2.21$ & $2.59$\\[\defaultaddspace]
$(7)$ & $2.23$ & $51.6$ & $31.6$ & $31.6$ & $2.23$ & $2.27$ & $0$ & $2.22$ & $2.22$ & $2.22$ & $2.57$ & $1.65$\\[\defaultaddspace]
$(8)$ & $0.340$ & $52.7$ & $31.6$ & $31.6$ & $2.68$ & $2.17$ & $2.12$ & $0$ & $1.46$ & $0.591$ & $1.40$ & $1.57$\\[\defaultaddspace]
$(9)$ & $1.53$ & $52.7$ & $31.6$ & $31.6$ & $2.40$ & $2.28$ & $2.16$ & $1.49$ & $0$ & $1.61$ & $1.58$ & $1.57$\\[\defaultaddspace]
$(10)$ & $0.481$ & $52.7$ & $31.6$ & $31.6$ & $2.72$ & $2.31$ & $2.14$ & $0.589$ & $1.57$ & $0$ & $1.38$ & $1.57$\\[\defaultaddspace]
$(11)$ & $1.48$ & $51.6$ & $31.6$ & $31.6$ & $2.15$ & $2.18$ & $2.48$ & $1.49$ & $1.65$ & $1.46$ & $0$ & $1.57$\\[\defaultaddspace]
$(12)$ & $1.23$ & $31.6$ & $52.7$ & $51.6$ & $1.58$ & $2.72$ & $1.87$ & $1.23$ & $1.23$ & $1.23$ & $1.23$ & $0$\\[\defaultaddspace]
\hline
\end{tabular}
\end{adjustbox}
\caption{The same as in Table~\ref{MC_corr_rms_ideal}, but
considering each of the 19 frequency channels and all 171 independent
combinations of pairs of different frequency channels.}
\label{MC_corr_cross_rms_ideal}
\end{table}

\clearpage

\section{Appendix -- Ideal case at high resolution}
\label{ideal_highres}

We repeat here the same analysis carried out in the previous section, but now working at $N_{\rm side} = 512$, i.e., at about 7 arcmin resolution, and considering a single realization.
The results are reported in Table~\ref{MC_corr_cross_avg_ideal_512} relative to approach (c).

%
\begin{table}[ht!]
\centering
\begin{adjustbox}{width=1\textwidth}
\begin{tabular}{|c|c|ccc|cccccc|cc|}
\hline
{\bf $\sigma$ level} & \multicolumn{1}{c|}{Current}
& \multicolumn{3}{c|}{FIRAS CIB amplitude}
& \multicolumn{6}{c|}{$(\Delta \varepsilon / \varepsilon_{\rm i})_{z_1}$}
& \multicolumn{2}{c|}{$(\Delta \varepsilon / \varepsilon_{\rm i})_{\rm late}$}\\[\defaultaddspace]
{\bf significance} & blackbody  &  &  (units $10^{-5}$) & & $8 \times 10^{-8}$ & $10^{-5}$ & $10^{-6}$ & $10^{-9}$ & $2 \times 10^{-8}$ & $-2 \times 10^{-9}$ & $4 \times 10^{-7}$ & $8 \times 10^{-6}$ \\
\hline\hline
& \multicolumn{1}{c|}{} &
$I_0^{\rm bf}$ & $+ 1\,\sigma$ & $ - 1\,\sigma$
& \multicolumn{6}{c|}{$\mu_0$}
& \multicolumn{2}{c|}{$u$}\\[\defaultaddspace]
&  & $1.3$ & $1.7$ & $0.9$ & $1.12 \times 10^{-7}$ & $1.4 \times 10^{-5}$ & $1.4 \times 10^{-6}$ & $1.4 \times 10^{-9}$ & $2.8 \times 10^{-8}$ & $-2.8 \times 10^{-9}$ & $10^{-7}$ & $2 \times 10^{-6}$ \\
\hline\hline
Case & (1) & (2) & (3) & (4) & (5) & (6) & (7) & (8) & (9) & (10) & (11) & (12) \\
\hline
$(1)$ & $0$ & $14400$ & $18800$ & $10100$ & $-1.39$ & $320.$ & $30.5$ & $-0.326$ & $-1.31$ & $0.467$ & $8.65$ & $161.$\\[\defaultaddspace]
$(2)$ & $14500$ & $0$ & $4380$ & $4380$ & $14500$ & $14500$ & $14500$ & $14500$ & $14500$ & $14500$ & $14500$ & $14400$\\[\defaultaddspace]
$(3)$ & $18800$ & $4330$ & $0$ & $8710$ & $18800$ & $18800$ & $18800$ & $18800$ & $18800$ & $18800$ & $18800$ & $18800$\\[\defaultaddspace]
$(4)$ & $10100$ & $4380$ & $8710$ & $0$ & $10100$ & $10100$ & $10100$ & $10100$ & $10100$ & $10100$ & $10100$ & $10000$\\[\defaultaddspace]
$(5)$ & $3.90$ & $14400$ & $18800$ & $10100$ & $0$ & $317.$ & $27.9$ & $3.86$ & $3.19$ & $3.97$ & $9.06$ & $160.$\\[\defaultaddspace]
$(6)$ & $323.$ & $14500$ & $10100$ & $10100$ & $321.$ & $0$ & $291.$ & $323.$ & $323.$ & $323.$ & $322.$ & $332.$\\[\defaultaddspace]
$(7)$ & $33.8$ & $18800$ & $18800$ & $10100$ & $31.2$ & $288.$ & $0$ & $33.8$ & $33.1$ & $33.8$ & $33.4$ & $158.$\\[\defaultaddspace]
$(8)$ & $0.329$ & $14400$ & $18800$ & $10100$ & $-1.41$ & $320.$ & $30.4$ & $0$ & $-1.29$ & $0.575$ & $8.66$ & $161.$\\[\defaultaddspace]
$(9)$ & $1.60$ & $14400$ & $18800$ & $10100$ & $-1.64$ & $319.$ & $29.8$ & $1.55$ & $0$ & $1.69$ & $8.69$ & $160.$\\[\defaultaddspace]
$(10)$ & $-0.458$ & $14400$ & $18800$ & $10100$ & $-1.35$ & $320.$ & $30.5$ & $-0.559$ & $-1.36$ & $0$ & $8.65$ & $161.$\\[\defaultaddspace]
$(11)$ & $7.27$ & $14400$ & $18800$ & $10100$ & $6.56$ & $318.$ & $29.7$ & $7.26$ & $7.01$ & $7.30$ & $0$ & $153.$\\[\defaultaddspace]
$(12)$ & $159.$ & $14500$ & $18800$ & $10000$ & $159.$ & $328.$ & $156.$ & $159.$ & $159.$ & $159.$ & $151.$ & $0$\\[\defaultaddspace]
\hline
\end{tabular}
\end{adjustbox}
\caption{Values of $\sqrt{|\Delta \chi^2|} \; {\rm sign}(\Delta \chi^2)$ for a single realization at $N_{\rm side} = 512$, for the full sky,
adopting perfect foreground subtraction and calibration, and considering each of the 19 frequency channels
and all 171 independent combinations of pairs of different frequencies.}
\label{MC_corr_cross_avg_ideal_512}
\end{table}

\clearpage

\section{Appendix -- Rms values from Monte Carlo simulations: including potential
residuals}
\label{app_rms_res}

We report here tables with the estimates of the rms of the $\sqrt{|\Delta \chi^2|} \; {\rm sign}(\Delta \chi^2)$ quoted from a Monte Carlo simulation
at $N_{\rm side} = 64$, with all-sky data, and including potential foreground and calibration residuals.

\begin{table}[ht!]
\centering
\begin{adjustbox}{width=1\textwidth}
\begin{tabular}{|c|c|ccc|cccccc|cc|}
\hline
{\bf Rms of $\sigma$ level} & \multicolumn{1}{c|}{Current}
& \multicolumn{3}{c|}{FIRAS CIB amplitude}
& \multicolumn{6}{c|}{$(\Delta \varepsilon / \varepsilon_{\rm i})_{z_1}$}
& \multicolumn{2}{c|}{$(\Delta \varepsilon / \varepsilon_{\rm i})_{\rm late}$}\\[\defaultaddspace]
{\bf significance} & blackbody  &  &  (units $10^{-5}$) & & $8 \times 10^{-8}$ & $10^{-5}$ & $10^{-6}$ & $10^{-9}$ & $2 \times 10^{-8}$ & $-2 \times 10^{-9}$ & $4 \times 10^{-7}$ & $8 \times 10^{-6}$ \\
\hline\hline
& \multicolumn{1}{c|}{} &
$I_0^{\rm bf}$ & $+ 1\,\sigma$ & $ - 1\,\sigma$
& \multicolumn{6}{c|}{$\mu_0$}
& \multicolumn{2}{c|}{$u$}\\[\defaultaddspace]
&  & $1.3$ & $1.7$ & $0.9$ & $1.12 \times 10^{-7}$ & $1.4 \times 10^{-5}$ & $1.4 \times 10^{-6}$ & $1.4 \times 10^{-9}$ & $2.8 \times 10^{-8}$ & $-2.8 \times 10^{-9}$ & $10^{-7}$ & $2 \times 10^{-6}$ \\
\hline\hline
Case & (1) & (2) & (3) & (4) & (5) & (6) & (7) & (8) & (9) & (10) & (11) & (12) \\
\hline
$(1)$ & $0$ & $0.798$ & $0.804$ & $0.802$ & $0.274$ & $0.471$ & $0.673$ & $0.0308$ & $0.137$ & $0.0435$ & $0.416$ & $0.469$ \\[\defaultaddspace]
$(2)$ & $0.838$ & $ 0$ & $0.792$ & $0.973$ & $0.846$ & $0.637$ & $0.819$ & $0.838$ & $0.838$ & $0.838$ & $0.838$ & $0.701$ \\[\defaultaddspace]
$(3)$ & $0.829$ & $0.981$ & $ 0$ & $0.871$ & $0.829$ & $0.682$ & $0.806$ & $0.829$ & $0.829$ & $0.829$ & $0.823$ & $0.753$ \\[\defaultaddspace]
$(4)$ & $0.859$ & $0.788$ & $0.799$ & $ 0$ & $0.857$ & $0.578$ & $0.837$ & $0.859$ & $0.860$ & $0.859$ & $0.841$ & $0.667$ \\[\defaultaddspace]
$(5)$ & $0.276$ & $0.798$ & $0.804$ & $0.800$ & $ 0$ & $0.480$ & $0.746$ & $0.274$ & $0.239$ & $0.280$ & $0.359$ & $0.472$ \\[\defaultaddspace]
$(6)$ & $0.490$ & $0.605$ & $0.657$ & $0.529$ & $0.495$ & $ 0$ & $0.486$ & $0.490$ & $0.480$ & $0.481$ & $0.484$ & $0.509$ \\[\defaultaddspace]
$(7)$ & $0.788$ & $0.795$ & $0.808$ & $0.768$ & $0.775$ & $0.477$ & $ 0$ & $0.788$ & $0.784$ & $0.789$ & $0.775$ & $0.474$ \\[\defaultaddspace]
$(8)$ & $ 0.0308$ & $0.798$ & $0.804$ & $0.802$ & $0.272$ & $0.471$ & $0.673$ & $ 0$ & $0.134$ & $0.0533$ & $0.415$ & $0.470$ \\[\defaultaddspace]
$(9)$ & $0.138$ & $0.798$ & $0.804$ & $0.802$ & $0.237$ & $0.478$ & $0.702$ & $0.134$ & $ 0$ & $0.145$ & $0.399$ & $0.469$ \\[\defaultaddspace]
$(10)$ & $ 0.0435$ & $0.798$ & $0.804$ & $0.802$ & $0.277$ & $0.486$ & $0.670$ & $0.0533$ & $0.144$ & $ 0$ & $0.418$ & $0.469$ \\[\defaultaddspace]
$(11)$ & $0.455$ & $0.798$ & $0.798$ & $0.784$ & $0.383$ & $0.482$ & $0.768$ & $0.454$ & $0.437$ & $0.457$ & $ 0$ & $0.469$ \\[\defaultaddspace]
$(12)$ & $0.539$ & $0.709$ & $0.733$ & $0.624$ & $0.542$ & $0.525$ & $0.591$ & $0.539$ & $0.539$ & $0.538$ & $0.543$ & $0$ \\[\defaultaddspace]
\hline
\end{tabular}
\end{adjustbox}
\caption{Rms values of $\sqrt{|\Delta \chi^2|} \; {\rm sign}(\Delta \chi^2)$ from a Monte Carlo simulation at $N_{\rm side} = 64$, full sky,
adopting $E_{\rm for} = 10^{-2}$ and $E_{\rm cal} = 10^{-4}$,  and considering  each of the 19 frequency channels.}
\label{MC_corr_rms}
\end{table}
\begin{table}[ht!]
\centering
\begin{adjustbox}{width=1\textwidth}
\begin{tabular}{|c|c|ccc|cccccc|cc|}
\hline
{\bf Rms of $\sigma$ level} & \multicolumn{1}{c|}{Current}
& \multicolumn{3}{c|}{FIRAS CIB amplitude}
& \multicolumn{6}{c|}{$(\Delta \varepsilon / \varepsilon_{\rm i})_{z_1}$}
& \multicolumn{2}{c|}{$(\Delta \varepsilon / \varepsilon_{\rm i})_{\rm late}$}\\[\defaultaddspace]
{\bf significance} & blackbody  &  &  (units $10^{-5}$) & & $8 \times 10^{-8}$ & $10^{-5}$ & $10^{-6}$ & $10^{-9}$ & $2 \times 10^{-8}$ & $-2 \times 10^{-9}$ & $4 \times 10^{-7}$ & $8 \times 10^{-6}$ \\
\hline\hline
& \multicolumn{1}{c|}{} &
$I_0^{\rm bf}$ & $+ 1\,\sigma$ & $ - 1\,\sigma$
& \multicolumn{6}{c|}{$\mu_0$}
& \multicolumn{2}{c|}{$u$}\\[\defaultaddspace]
&  & $1.3$ & $1.7$ & $0.9$ & $1.12 \times 10^{-7}$ & $1.4 \times 10^{-5}$ & $1.4 \times 10^{-6}$ & $1.4 \times 10^{-9}$ & $2.8 \times 10^{-8}$ & $-2.8 \times 10^{-9}$ & $10^{-7}$ & $2 \times 10^{-6}$ \\
\hline\hline
Case & (1) & (2) & (3) & (4) & (5) & (6) & (7) & (8) & (9) & (10) & (11) & (12) \\
\hline
$(1)$ & $0$ & $2.69$ & $2.69$ & $2.68$ & $0.332$ & $1.77$ & $1.34$ & $0.0362$ & $0.164$ & $0.0512$ & $0.359$ & $1.54$ \\[\defaultaddspace]
$(2)$ & $2.67$ & $0$ & $2.77$ & $2.88$ & $2.67$ & $2.69$ & $2.67$ & $2.67$ & $2.67$ & $2.67$ & $2.67$ & $2.66$ \\[\defaultaddspace]
$(3)$ & $2.65$ & $2.91$ & $0$ & $2.70$ & $2.64$ & $2.68$ & $2.67$ & $2.65$ & $2.65$ & $2.65$ & $2.65$ & $2.65$ \\[\defaultaddspace]
$(4)$ & $2.69$ & $2.76$ & $2.70$ & $0$ & $2.69$ & $2.72$ & $2.71$ & $2.69$ & $2.69$ & $2.69$ & $2.69$ & $2.68$ \\[\defaultaddspace]
$(5)$ & $0.320$ & $2.69$ & $2.69$ & $2.68$ & $0$ & $1.78$ & $1.28$ & $0.318$ & $0.278$ & $0.324$ & $0.483$ & $1.56$ \\[\defaultaddspace]
$(6)$ & $1.39$ & $2.69$ & $2.71$ & $2.64$ & $1.38$ & $0$ & $1.38$ & $1.39$ & $1.37$ & $1.39$ & $1.38$ & $1.40$ \\[\defaultaddspace]
$(7)$ & $1.08$ & $2.69$ & $2.69$ & $2.68$ & $1.04$ & $1.85$ & $0$ & $1.08$ & $1.08$ & $1.09$ & $1.12$ & $1.59$ \\[\defaultaddspace]
$(8)$ & $ 0.0362$ & $2.69$ & $2.69$ & $2.68$ & $0.329$ & $1.77$ & $1.34$ & $0$ & $0.160$ & $0.0627$ & $0.361$ & $1.54$ \\[\defaultaddspace]
$(9)$ & $0.161$ & $2.69$ & $2.69$ & $2.68$ & $0.286$ & $1.78$ & $1.33$ & $0.157$ & $0$ & $0.169$ & $0.395$ & $1.55$ \\[\defaultaddspace]
$(10)$ & $ 0.0512$ & $2.69$ & $2.69$ & $2.68$ & $0.336$ & $1.77$ & $1.35$ & $0.0629$ & $0.172$ & $0$ & $0.355$ & $1.54$ \\[\defaultaddspace]
$(11)$ & $0.360$ & $2.69$ & $2.72$ & $2.67$ & $0.491$ & $1.77$ & $1.38$ & $0.362$ & $0.397$ & $0.356$ & $0$ & $1.51$ \\[\defaultaddspace]
$(12)$ & $1.64$ & $2.69$ & $2.69$ & $2.66$ & $1.68$ & $1.79$ & $2.08$ & $1.64$ & $1.65$ & $1.64$ & $1.60$ & $0$ \\[\defaultaddspace]
\hline
\end{tabular}
\end{adjustbox}
\caption{The same as in Table~\ref{MC_corr_rms},
but considering all 171 independent combinations of pairs of different frequency channels.}
\label{MC_cross_rms}
\end{table}
\begin{table}[ht!]
\centering
\begin{adjustbox}{width=1\textwidth}
\begin{tabular}{|c|c|ccc|cccccc|cc|}
\hline
{\bf Rms of $\sigma$ level} & \multicolumn{1}{c|}{Current}
& \multicolumn{3}{c|}{FIRAS CIB amplitude}
& \multicolumn{6}{c|}{$(\Delta \varepsilon / \varepsilon_{\rm i})_{z_1}$}
& \multicolumn{2}{c|}{$(\Delta \varepsilon / \varepsilon_{\rm i})_{\rm late}$}\\[\defaultaddspace]
{\bf significance} & blackbody  &  &  (units $10^{-5}$) & & $8 \times 10^{-8}$ & $10^{-5}$ & $10^{-6}$ & $10^{-9}$ & $2 \times 10^{-8}$ & $-2 \times 10^{-9}$ & $4 \times 10^{-7}$ & $8 \times 10^{-6}$ \\
\hline\hline
& \multicolumn{1}{c|}{} &
$I_0^{\rm bf}$ & $+ 1\,\sigma$ & $ - 1\,\sigma$
& \multicolumn{6}{c|}{$\mu_0$}
& \multicolumn{2}{c|}{$u$}\\[\defaultaddspace]
&  & $1.3$ & $1.7$ & $0.9$ & $1.12 \times 10^{-7}$ & $1.4 \times 10^{-5}$ & $1.4 \times 10^{-6}$ & $1.4 \times 10^{-9}$ & $2.8 \times 10^{-8}$ & $-2.8 \times 10^{-9}$ & $10^{-7}$ & $2 \times 10^{-6}$ \\
\hline\hline
Case & (1) & (2) & (3) & (4) & (5) & (6) & (7) & (8) & (9) & (10) & (11) & (12) \\
\hline
$(1)$ & $ 0$ & $2.79$ & $2.82$ & $2.79$ & $0.411$ & $1.10$ & $1.48$ & $0.0458$ & $0.205$ & $0.0647$ & $0.469$ & $0.638$ \\[\defaultaddspace]
$(2)$ & $ 2.79$ & $0$ & $2.87$ & $3.03$ & $2.79$ & $2.74$ & $2.78$ & $2.79$ & $2.79$ & $2.79$ & $2.79$ & $2.75$ \\[\defaultaddspace]
$(3)$ & $2.77$ & $3.04$ & $0$ & $2.84$ & $2.76$ & $2.75$ & $2.79$ & $2.76$ & $2.76$ & $2.77$ & $2.77$ & $2.75$ \\[\defaultaddspace]
$(4)$ & $2.82$ & $2.86$ & $2.81$ & $0$ & $2.82$ & $2.73$ & $2.82$ & $2.82$ & $2.82$ & $2.82$ & $2.80$ & $2.74$ \\[\defaultaddspace]
$(5)$ & $0.406$ & $2.79$ & $2.79$ & $2.78$ & $0$ & $1.10$ & $1.43$ & $0.404$ & $0.352$ & $0.411$ & $0.534$ & $0.649$ \\[\defaultaddspace]
$(6)$ & $1.02$ & $2.71$ & $2.76$ & $2.62$ & $1.02$ & $0$ & $1.01$ & $1.02$ & $1.02$ & $1.02$ & $1.02$ & $1.34$ \\[\defaultaddspace]
$(7)$ & $1.18$ & $2.79$ & $2.79$ & $2.78$ & $1.14$ & $1.10$ & $0$ & $1.18$ & $1.17$ & $1.18$ & $1.21$ & $0.853$ \\[\defaultaddspace]
$(8)$ & $0.0457$ & $2.79$ & $2.82$ & $2.79$ & $0.409$ & $1.10$ & $1.48$ & $0$ & $0.199$ & $0.0794$ & $0.470$ & $0.638$ \\[\defaultaddspace]
$(9)$ & $0.204$ & $2.79$ & $2.82$ & $2.79$ & $0.356$ & $1.08$ & $1.47$ & $0.199$ & $0$ & $0.214$ & $0.475$ & $0.642$ \\[\defaultaddspace]
$(10)$ & $ 0.0647$ & $2.79$ & $2.82$ & $2.79$ & $0.417$ & $1.10$ & $1.48$ & $0.0794$ & $0.215$ & $0$ & $0.467$ & $0.637$ \\[\defaultaddspace]
$(11)$ & $ 0.503$ & $2.81$ & $2.82$ & $2.78$ & $0.547$ & $1.11$ & $1.54$ & $0.503$ & $0.513$ & $0.502$ & $0$ & $0.636$ \\[\defaultaddspace]
$(12)$ & $ 0.842$ & $2.76$ & $2.78$ & $2.74$ & $0.876$ & $1.55$ & $1.87$ & $0.843$ & $0.849$ & $0.840$ & $0.859$ & $0$ \\[\defaultaddspace]
\hline
\end{tabular}
\end{adjustbox}
\caption{The same as in Table~\ref{MC_corr_rms},
but considering each of the 19 frequency channels 
and all 171 independent combinations of pairs of different frequency channels.}
\label{MC_corr_cross_rms}
\end{table}


\section{Appendix -- Results for different assumptions on potential foreground and calibration residuals} 
\label{app_var_cal_res}

We report here some of the results discussed in Sect.~\ref{var_cal_res}.

\begin{table}[ht!]
\centering
\begin{adjustbox}{width=1\textwidth}
\begin{tabular}{|c|c|ccc|cccccc|cc|}
\hline
{\bf $\sigma$ level} & \multicolumn{1}{c|}{Current}
& \multicolumn{3}{c|}{FIRAS CIB amplitude}
& \multicolumn{6}{c|}{$(\Delta \varepsilon / \varepsilon_{\rm i})_{z_1}$}
& \multicolumn{2}{c|}{$(\Delta \varepsilon / \varepsilon_{\rm i})_{\rm late}$}\\[\defaultaddspace]
{\bf significance} & blackbody  &  &  (units $10^{-5}$) & & $8 \times 10^{-8}$ & $10^{-5}$ & $10^{-6}$ & $10^{-9}$ & $2 \times 10^{-8}$ & $-2 \times 10^{-9}$ & $4 \times 10^{-7}$ & $8 \times 10^{-6}$ \\
\hline\hline
& \multicolumn{1}{c|}{} &
$I_0^{\rm bf}$ & $+ 1\,\sigma$ & $ - 1\,\sigma$
& \multicolumn{6}{c|}{$\mu_0$}
& \multicolumn{2}{c|}{$u$}\\[\defaultaddspace]
&  & $1.3$ & $1.7$ & $0.9$ & $1.12 \times 10^{-7}$ & $1.4 \times 10^{-5}$ & $1.4 \times 10^{-6}$ & $1.4 \times 10^{-9}$ & $2.8 \times 10^{-8}$ & $-2.8 \times 10^{-9}$ & $10^{-7}$ & $2 \times 10^{-6}$ \\
\hline\hline
Case & (1) & (2) & (3) & (4) & (5) & (6) & (7) & (8) & (9) & (10) & (11) & (12) \\
\hline
$(1)$ & $0.00$ & $700.$ & $911.$ & $487.$ & $0.859$ & $75.7$ & $7.85$ & $0.0689$ & $0.341$ & $-0.0956$ & $2.35$ & $30.7$ \\[\defaultaddspace]
$(2)$ & $699.$ & $0.00$ & $211.$ & $213.$ & $699.$ & $715.$ & $700.$ & $699.$ & $699.$ & $699.$ & $699.$ & $696.$ \\[\defaultaddspace]
$(3)$ & $910.$ & $211.$ & $0.00$ & $424.$ & $910.$ & $926.$ & $911.$ & $910.$ & $910.$ & $910.$ & $910.$ & $907.$ \\[\defaultaddspace]
$(4)$ & $486.$ & $213.$ & $424.$ & $0.00$ & $486.$ & $504.$ & $487.$ & $486.$ & $486.$ & $486.$ & $486.$ & $483.$ \\[\defaultaddspace]
$(5)$ & $-0.103$ & $700.$ & $911.$ & $487.$ & $0.00$ & $75.1$ & $7.24$ & $-0.122$ & $-0.276$ & $-0.0379$ & $2.19$ & $30.5$ \\[\defaultaddspace]
$(6)$ & $75.1$ & $716.$ & $926.$ & $505.$ & $74.5$ & $0.00$ & $67.6$ & $75.1$ & $75.0$ & $75.1$ & $74.6$ & $69.5$ \\[\defaultaddspace]
$(7)$ & $7.23$ & $701.$ & $912.$ & $488.$ & $6.62$ & $68.2$ & $0.00$ & $7.22$ & $7.07$ & $7.24$ & $7.00$ & $28.7$ \\[\defaultaddspace]
$(8)$ & $-0.0680$ & $700.$ & $911.$ & $487.$ & $0.851$ & $75.7$ & $7.84$ & $0.00$ & $0.331$ & $-0.116$ & $2.35$ & $30.7$ \\[\defaultaddspace]
$(9)$ & $-0.266$ & $700.$ & $911.$ & $487.$ & $0.697$ & $75.6$ & $7.70$ & $-0.262$ & $0.00$ & $-0.275$ & $2.30$ & $30.7$ \\[\defaultaddspace]
$(10)$ & $0.0979$ & $700.$ & $911.$ & $487.$ & $0.875$ & $75.7$ & $7.86$ & $0.121$ & $0.361$ & $0.00$ & $2.35$ & $30.8$ \\[\defaultaddspace]
$(11)$ & $-1.06$ & $700.$ & $911.$ & $487.$ & $-1.04$ & $75.1$ & $7.19$ & $-1.06$ & $-1.09$ & $-1.05$ & $0.00$ & $29.3$ \\[\defaultaddspace]
$(12)$ & $28.5$ & $697.$ & $907.$ & $484.$ & $28.3$ & $69.3$ & $26.4$ & $28.5$ & $28.5$ & $28.5$ & $27.0$ & $0.00$ \\[\defaultaddspace]
\hline
\end{tabular}
\end{adjustbox}
\caption{Values of $\sqrt{|\Delta \chi^2|} \; {\rm sign}(\Delta \chi^2)$ for a single realization at $N_{\rm side} = 64$, using {\it Planck\/} mask-76 extended to exclude regions at $|b| \le 30^\circ$.  We adopt
$E_{\rm for} = 10^{-3}$ and $E_{\rm cal} = 10^{-4}$, and consider each of the 19 frequency channels
and all 171 independent combinations of pairs of different frequencies.}
\label{Optimistic_corr_cross_mask76ExtGal}
\end{table}

\begin{table}[ht!]
\centering
\begin{adjustbox}{width=1\textwidth}
\begin{tabular}{|c|c|ccc|cccccc|cc|}
\hline
{\bf $\sigma$ level} & \multicolumn{1}{c|}{Current}
& \multicolumn{3}{c|}{FIRAS CIB amplitude}
& \multicolumn{6}{c|}{$(\Delta \varepsilon / \varepsilon_{\rm i})_{z_1}$}
& \multicolumn{2}{c|}{$(\Delta \varepsilon / \varepsilon_{\rm i})_{\rm late}$}\\[\defaultaddspace]
{\bf significance} & blackbody  &  &  (units $10^{-5}$) & & $8 \times 10^{-8}$ & $10^{-5}$ & $10^{-6}$ & $10^{-9}$ & $2 \times 10^{-8}$ & $-2 \times 10^{-9}$ & $4 \times 10^{-7}$ & $8 \times 10^{-6}$ \\
\hline\hline
& \multicolumn{1}{c|}{} &
$I_0^{\rm bf}$ & $+ 1\,\sigma$ & $ - 1\,\sigma$
& \multicolumn{6}{c|}{$\mu_0$}
& \multicolumn{2}{c|}{$u$}\\[\defaultaddspace]
&  & $1.3$ & $1.7$ & $0.9$ & $1.12 \times 10^{-7}$ & $1.4 \times 10^{-5}$ & $1.4 \times 10^{-6}$ & $1.4 \times 10^{-9}$ & $2.8 \times 10^{-8}$ & $-2.8 \times 10^{-9}$ & $10^{-7}$ & $2 \times 10^{-6}$ \\
\hline\hline
Case & (1) & (2) & (3) & (4) & (5) & (6) & (7) & (8) & (9) & (10) & (11) & (12) \\
\hline
$(1)$ & $0$ & $65.2$ & $84.3$ & $46.0$ & $-0.193$ & $7.80$ & $ 0.376$ & $-0.0227$ & $-0.101$ & $0.0322$ & $ 0.563$ & $3.96$ \\[\defaultaddspace]
$(2)$ & $61.2$ & $0$ & $20.9$ & $17.1$ & $61.2$ & $63.0$ & $61.3$ & $61.2$ & $61.2$ & $61.2$ & $61.2$ & $60.9$ \\[\defaultaddspace]
$(3)$ & $80.3$ & $17.0$ & $0$ & $36.3$ & $80.3$ & $82.0$ & $80.4$ & $80.3$ & $80.3$ & $80.3$ & $80.2$ & $79.9$ \\[\defaultaddspace]
$(4)$ & $41.9$ & $21.1$ & $40.2$ & $0$ & $41.9$ & $44.0$ & $42.1$ & $41.9$ & $41.9$ & $41.9$ & $41.9$ & $41.7$ \\[\defaultaddspace]
$(5)$ & $0.214$ & $65.2$ & $84.3$ & $46.0$ & $0$ & $7.73$ & $ 0.286$ & $ 0.212$ & $ 0.183$ & $ 0.217$ & $ 0.595$ & $3.94$ \\[\defaultaddspace]
$(6)$ & $8.44$ & $67.0$ & $86.0$ & $48.0$ & $8.37$ & $0$ & $7.62$ & $8.44$ & $8.42$ & $8.44$ & $8.39$ & $8.11$ \\[\defaultaddspace]
$(7)$ & $1.09$ & $65.3$ & $84.4$ & $46.1$ & $1.02$ & $6.98$ & $0$ & $1.08$ & $1.07$ & $1.09$ & $1.18$ & $3.84$ \\[\defaultaddspace]
$(8)$ & $0.0228$ & $65.2$ & $84.3$ & $46.0$ & $-0.192$ & $7.80$ & $ 0.375$ & $0$ & $-0.0980$ & $0.0395$ & $ 0.563$ & $3.96$ \\[\defaultaddspace]
$(9)$ & $0.103$ & $65.2$ & $84.3$ & $46.0$ & $-0.169$ & $7.78$ & $ 0.355$ & $ 0.100$ & $0$ & $ 0.108$ & $ 0.570$ & $3.95$ \\[\defaultaddspace]
$(10)$ & $-0.0322$ & $65.2$ & $84.3$ & $46.0$ & $-0.195$ & $7.80$ & $ 0.378$ & $-0.0394$ & $-0.105$ & $0$ & $ 0.562$ & $3.96$ \\[\defaultaddspace]
$(11)$ & $-0.517$ & $65.2$ & $84.2$ & $45.9$ & $-0.560$ & $7.71$ & $-0.481$ & $-0.518$ & $-0.529$ & $-0.516$ & $0$ & $3.80$ \\[\defaultaddspace]
$(12)$ & $2.00$ & $64.8$ & $83.9$ & $45.6$ & $1.94$ & $6.61$ & $1.42$ & $1.99$ & $1.98$ & $2.00$ & $1.82$ & $0$ \\[\defaultaddspace]
\hline
\end{tabular}
\end{adjustbox}
\caption{Values of $\sqrt{|\Delta \chi^2|} \; {\rm sign}(\Delta \chi^2)$ for a single realization at $N_{\rm side} = 64$, using {\it Planck\/} mask-76 extended to exclude regions at $|b| \le 30^\circ$.  We
adopt $E_{\rm for} = 10^{-2}$ and $E_{\rm cal} = 10^{-3}$ at $\nu \le 295$\,GHz and $E_{\rm cal} = 10^{-2}$ at $\nu \ge 340$\,GHz, and consider
each of the 19 frequency channels
and all 171 independent combinations of pairs of different frequencies.}
\label{Pessimistic_corr_cross_mask76ExtGal}
\end{table}

\begin{table}[ht!]
\centering
\begin{adjustbox}{width=1\textwidth}
\begin{tabular}{|c|c|ccc|cccccc|cc|}
\hline
{\bf $\sigma$ level} & \multicolumn{1}{c|}{Current}
& \multicolumn{3}{c|}{FIRAS CIB amplitude}
& \multicolumn{6}{c|}{$(\Delta \varepsilon / \varepsilon_{\rm i})_{z_1}$}
& \multicolumn{2}{c|}{$(\Delta \varepsilon / \varepsilon_{\rm i})_{\rm late}$}\\[\defaultaddspace]
{\bf significance} & blackbody  &  &  (units $10^{-5}$) & & $8 \times 10^{-8}$ & $10^{-5}$ & $10^{-6}$ & $10^{-9}$ & $2 \times 10^{-8}$ & $-2 \times 10^{-9}$ & $4 \times 10^{-7}$ & $8 \times 10^{-6}$ \\
\hline\hline
& \multicolumn{1}{c|}{} &
$I_0^{\rm bf}$ & $+ 1\,\sigma$ & $ - 1\,\sigma$
& \multicolumn{6}{c|}{$\mu_0$}
& \multicolumn{2}{c|}{$u$}\\[\defaultaddspace]
&  & $1.3$ & $1.7$ & $0.9$ & $1.12 \times 10^{-7}$ & $1.4 \times 10^{-5}$ & $1.4 \times 10^{-6}$ & $1.4 \times 10^{-9}$ & $2.8 \times 10^{-8}$ & $-2.8 \times 10^{-9}$ & $10^{-7}$ & $2 \times 10^{-6}$ \\
\hline\hline
Case & (1) & (2) & (3) & (4) & (5) & (6) & (7) & (8) & (9) & (10) & (11) & (12) \\
\hline
$(1)$ & $0$ & $161.$ & $209.$ & $112.$ & $-0.601$ & $8.97$ & $-1.83$ & $-0.0680$ & $-0.303$ & $0.0962$ & $0.963$ & $6.15$ \\[\defaultaddspace]
$(2)$ & $157.$ & $0$ & $49.8$ & $46.4$ & $157.$ & $160.$ & $157.$ & $157.$ & $157.$ & $157.$ & $157.$ & $156.$ \\[\defaultaddspace]
$(3)$ & $205.$ & $45.9$ & $0$ & $94.4$ & $205.$ & $208.$ & $205.$ & $205.$ & $205.$ & $205.$ & $205.$ & $204.$ \\[\defaultaddspace]
$(4)$ & $109.$ & $50.2$ & $98.2$ & $0$ & $109.$ & $112.$ & $109.$ & $109.$ & $109.$ & $109.$ & $109.$ & $108.$ \\[\defaultaddspace]
$(5)$ & $ 0.615$ & $161.$ & $209.$ & $112.$ & $0$ & $8.88$ & $-1.78$ & $0.611$ & $0.531$ & $0.622$ & $1.14$ & $6.17$ \\[\defaultaddspace]
$(6)$ & $13.1$ & $164.$ & $212.$ & $116.$ & $13.1$ & $0$ & $12.0$ & $13.1$ & $13.1$ & $13.2$ & $13.1$ & $13.8$ \\[\defaultaddspace]
$(7)$ & $2.43$ & $161.$ & $209.$ & $113.$ & $2.31$ & $7.81$ & $0$ & $2.43$ & $2.40$ & $2.43$ & $2.59$ & $6.46$ \\[\defaultaddspace]
$(8)$ & $0.0680$ & $161.$ & $209.$ & $112.$ & $-0.598$ & $8.97$ & $-1.83$ & $0$ & $-0.296$ & $0.118$ & $0.966$ & $6.15$ \\[\defaultaddspace]
$(9)$ & $ 0.305$ & $161.$ & $209.$ & $112.$ & $-0.522$ & $8.95$ & $-1.82$ & $0.297$ & $0$ & $0.320$ & $1.01$ & $6.15$ \\[\defaultaddspace]
$(10)$ & $-0.0961$ & $161.$ & $209.$ & $112.$ & $-0.609$ & $8.97$ & $-1.83$ & $-0.118$ & $-0.318$ & $0$ & $0.959$ & $6.15$ \\[\defaultaddspace]
$(11)$ & $-0.909$ & $161.$ & $209.$ & $112.$ & $-1.09$ & $8.87$ & $-2.07$ & $-0.912$ & $-0.959$ & $-0.904$ & $0$ & $5.91$ \\[\defaultaddspace]
$(12)$ & $1.65$ & $160.$ & $208.$ & $111.$ & $1.48$ & $8.00$ & $-1.60$ & $1.64$ & $1.61$ & $1.65$ & $1.27$ & $0$ \\[\defaultaddspace]
\hline
\end{tabular}
\end{adjustbox}
\caption{Values of $\sqrt{|\Delta \chi^2|} \; {\rm sign}(\Delta \chi^2)$ for a single realization at $N_{\rm side} = 64$, using {\it Planck\/} mask-76 extended to exclude regions at $|b| \le 30^\circ$.  We
adopt $E_{\rm for} = 10^{-3}$ and $E_{\rm cal} = 10^{-3}$ at $\nu \le 295$\,GHz and $E_{\rm cal} = 10^{-2}$ at $\nu \ge 340$\,GHz, and consider
each of the 19 frequency channels
and all 171 independent combinations of pairs of different frequencies.}
\label{Intermediate_corr_cross_mask76ExtGal}
\end{table}

\begin{table}[ht!]
\centering
\begin{adjustbox}{width=1\textwidth}
\begin{tabular}{|c|c|ccc|cccccc|cc|}
\hline
{\bf $\sigma$ level} & \multicolumn{1}{c|}{Current}
& \multicolumn{3}{c|}{FIRAS CIB amplitude}
& \multicolumn{6}{c|}{$(\Delta \varepsilon / \varepsilon_{\rm i})_{z_1}$}
& \multicolumn{2}{c|}{$(\Delta \varepsilon / \varepsilon_{\rm i})_{\rm late}$}\\[\defaultaddspace]
{\bf significance} & blackbody  &  &  (units $10^{-5}$) & & $8 \times 10^{-8}$ & $10^{-5}$ & $10^{-6}$ & $10^{-9}$ & $2 \times 10^{-8}$ & $-2 \times 10^{-9}$ & $4 \times 10^{-7}$ & $8 \times 10^{-6}$ \\
\hline\hline
& \multicolumn{1}{c|}{} &
$I_0^{\rm bf}$ & $+ 1\,\sigma$ & $ - 1\,\sigma$
& \multicolumn{6}{c|}{$\mu_0$}
& \multicolumn{2}{c|}{$u$}\\[\defaultaddspace]
&  & $1.3$ & $1.7$ & $0.9$ & $1.12 \times 10^{-7}$ & $1.4 \times 10^{-5}$ & $1.4 \times 10^{-6}$ & $1.4 \times 10^{-9}$ & $2.8 \times 10^{-8}$ & $-2.8 \times 10^{-9}$ & $10^{-7}$ & $2 \times 10^{-6}$ \\
\hline\hline
Case & (1) & (2) & (3) & (4) & (5) & (6) & (7) & (8) & (9) & (10) & (11) & (12) \\
\hline
$(1)$ & $0$ & $286.$ & $371.$ & $200.$ & $-1.12$ & $48.6$ & $2.74$ & $-0.133$ & $-0.589$ & $0.189$ & $2.12$ & $21.3$\\[\defaultaddspace]
$(2)$ & $278.$ & $0$ & $88.9$ & $81.7$ & $278.$ & $289.$ & $278.$ & $278.$ & $278.$ & $278.$ & $278.$ & $276.$\\[\defaultaddspace]
$(3)$ & $363.$ & $81.0$ & $0$ & $167.$ & $363.$ & $373.$ & $364.$ & $363.$ & $363.$ & $363.$ & $363.$ & $361.$\\[\defaultaddspace]
$(4)$ & $192.$ & $89.7$ & $175.$ & $0$ & $192.$ & $205.$ & $193.$ & $192.$ & $192.$ & $192.$ & $192.$ & $191.$\\[\defaultaddspace]
$(5)$ & $1.26$ & $286.$ & $371.$ & $200.$ & $0$ & $48.2$ & $2.25$ & $1.25$ & $1.08$ & $1.28$ & $2.38$ & $21.2$\\[\defaultaddspace]
$(6)$ & $52.1$ & $297.$ & $381.$ & $213.$ & $51.7$ & $0$ & $47.1$ & $52.1$ & $52.0$ & $52.1$ & $51.6$ & $46.2$\\[\defaultaddspace]
$(7)$ & $6.57$ & $286.$ & $371.$ & $201.$ & $6.15$ & $43.5$ & $0$ & $6.57$ & $6.47$ & $6.58$ & $6.52$ & $19.9$\\[\defaultaddspace]
$(8)$ & $ 0.134$ & $286.$ & $371.$ & $200.$ & $-1.12$ & $48.5$ & $2.74$ & $0$ & $-0.574$ & $0.232$ & $2.12$ & $21.3$\\[\defaultaddspace]
$(9)$ & $ 0.606$ & $286.$ & $371.$ & $200.$ & $-0.989$ & $48.5$ & $2.62$ & $0.590$ & $0$ & $0.636$ & $2.18$ & $21.3$\\[\defaultaddspace]
$(10)$ & $-0.189$ & $286.$ & $371.$ & $200.$ & $-1.14$ & $48.6$ & $2.75$ & $-0.231$ & $-0.617$ & $0$ & $2.11$ & $21.3$\\[\defaultaddspace]
$(11)$ & $ -1.60$ & $286.$ & $371.$ & $200.$ & $-2.06$ & $48.0$ & $-0.489$ & $-1.61$ & $-1.74$ & $-1.59$ & $0$ & $20.3$\\[\defaultaddspace]
$(12)$ & $17.7$ & $284.$ & $369.$ & $199.$ & $17.4$ & $40.4$ & $14.8$ & $17.7$ & $17.6$ & $17.7$ & $16.7$ & $0$\\[\defaultaddspace]
\hline
\end{tabular}
\end{adjustbox}
\caption{Values of $\sqrt{|\Delta \chi^2|} \; {\rm sign}(\Delta \chi^2)$ for a single realization at $N_{\rm side} = 256$, using {\it Planck\/} mask-76 extended to exclude regions at $|b| \le 30^\circ$.  We
adopt $E_{\rm for} = 10^{-3}$ and $E_{\rm cal} = 10^{-3}$ at $\nu \le 295$\,GHz and $E_{\rm cal} = 10^{-2}$ at $\nu \ge 340$\,GHz, and consider
each of the 19 frequency channels 
and all 171 independent combinations of pairs of different frequencies.}
\label{standard_corr_cross_mask76ExtGal_ns256}
\end{table}

\clearpage


\bibliography{pecmotreferences}{}
\bibliographystyle{JHEP2015}  

\end{document}